\documentclass[12pt,preprint]{aastex}
\pdfoutput=1
\begin{document}
\title{Comparing binary systems from rotating parent gas structures with different total masses}

\author{Guillermo Arreaga-Garc\'{\i}a\altaffilmark{1}}
\affil{Departamento de Investigaci\'on en F\'{\i}sica, Universidad de Sonora, \\
Apdo. Postal 14740, C.P. 83000, Hermosillo, Sonora, Mexico.}

\begin{abstract}
In this paper we continue the investigation reported by \cite{RMAA}
concerning the morphology of binary configurations obtained via the
collapse of rotating parent gas structures with total masses in the
range of M$_T$= 1 to 5 M$_{\odot}$. Here we extend the mass
range and consider the collapse of two uniform gas clumps of M$_T$=
50 and 400 M$_{\odot}$, so that they also rotates rigidly in such a
way that its approximate virial parameter takes the values of 0.5,
1.5, and 2.5 and their collapse is induced initially by implementing
an azimuthal mass perturbation. To assess the effects of the total
mass of the parent gas structure on the nature of the resulting
binary configurations, we also consider the collapse of two cores of
M$_T$= 1 and 5 M$_{\odot}$. We calculate the collapse of all these
parent gas structures using three values of the ratio of thermal
energy to potential energy and for two values of the mass
perturbation amplitude. We next calculate the binary separations,
masses and integral properties of the binary fragments and present
them in terms of the total mass of the parent structure. For most of
our models, we finally calculate the $\beta$ extreme value, so that
a model with a slightly higher $\beta$ value would no longer
collapse.
\end{abstract}


\section{Introduction}
\label{intro}

Stars are formed in gravitationally collapsed clouds made of molecular
hydrogen gas.
These gas clouds appear to be composed of many well-defined
gas substructures, so that these assembled clouds can be very
large and massive. Thus, all gas structures have been
characterized according to \cite{bergin} by
means of their size and mass. For instance, a gas clump is a gas
structure with mass and size in the range of 50-500 M$_{\odot}$ and
0.3-3 pc, while for a gas core these values are in the range of 0.5-5 M$_{\odot}$
and 0.03-0.2 pc.

Numerical simulations of the collapse of a rotating, self-gravitating,
isolated cloud cores, began to be performed many years
ago; see  \cite{bodentohlineblack}, \cite{boden}. One of the
classic models of binary formation is based on the
collapse of a rotating spherical core of 1 M$_{\odot}$, in which an
azimuthal symmetric mass perturbation was initially
implemented such that a binary system was formed by prompt
fragmentation; see the so-called "standard isothermal test case" calculated by
\cite{boss1979}, \cite{boss1991}, \cite{truelove98},
\cite{klein99}, \cite{boss2000}, and \cite{kitsionas}, among
others.

It has always been very important to determine whether or not a gas structure
is stable against gravitational collapse. As early as two decades ago,
theorists proposed collapse and fragmentation criteria by constructing
configuration diagrams, the axes of which are usually
the ratio of thermal energy to potential energy, denoted by $\alpha$,
versus the ratio of rotational energy to gravitational energy, denoted by
$\beta$; see for instance \cite{miyama},
\cite{hachisu1}, \cite{hachisu2}, and \cite{tsuribe1}.

It must be emphasized that numerical simulations are still necessary to
conclusively show
the final configuration obtained from the collapse of a particular
model of gas structure, as all these $\alpha$ versus $\beta$ configuration diagrams
mentioned are, above all, mostly indicative, due to the fact that the parameter
space that determines the final simulation outcome is very large.

We mention that the role played by the total mass of the parent core on the
collapse results has been considered for a long time ago. For instance, \cite{Tscharnuter}
chose initial conditions in order
to study the collapse of a cloud of M=60 M$_\odot$ and compared it to the collapse
of three models of 1 M$_\odot$. \cite{Rozyczka} investigated the collapse
and fragmentation in four cloud models, of which the first three are for 1 M$_\odot$ and the
last one is a model of a very massive uniform cloud of 5000 M$_\odot$.
Furthermore, \cite{boss1986} studied in depth the collapse of
rotating, uniform non-isothermal clouds by considering a four
dimensional parameter space composed by $\alpha$, $\beta$, $T$ and
the total mass M. The upper limit of M was 2 M$_{\odot}$ while the
range of $\alpha$ and $\beta$ varied within 0.01-0.50 and 0-0.33,
respectively. He found four different types of configurations and
noted that as the mass $M$ decreased, cloud fragmentation is less
favored.

More recently, \cite{RMAA} reported numerical simulations focusing on only one
type of binary configuration, consisting
of a well-defined pair of mass condensations that approach
each other, achieve rotational
speed, swing past each other, and finally separate to form the
desired binary system, in which the mass condensations orbit around one
another at the end of the
simulation. A schematic diagram was
also presented in which this particular binary configuration can be located
among others. Similar configurations were also obtained by
\cite{hennebelle} by increasing the external pressure on a rotating
core. The configuration that interested \cite{RMAA}
corresponds to the disk-bar type fragmentation studied
by \cite{matsumoto} and \cite{tsuribe2}.

The schematic diagram reported by \cite{RMAA} was constructed with
the total mass of the parent core M$_T$ as the vertical axis in the range
of 0.75 to 5 M$_{\odot}$, versus
the dimensionless ratio of rotational energy to gravitational energy $\beta$,
as the horizontal axis in the range of 0.1 to 0.21. The dimensionless ratio
of thermal energy to potential energy, $\alpha$, was kept fixed for all
simulations. It was also noted that the increase in the total mass
of the parent core diminished the formation of the desired binary
configuration and instead favored the formation of a single central mass
condensation surrounded by a disk as the final result.

Observationally, much effort has been put forth for
many years in order to determine some physical property that could be
useful in deciding whether or not a particular gas structure will
collapse. Particularly, the virial parameter, denoted here by
$\alpha_{vir}$, has recently been
measured in order to characterize
the dynamical state of gas structures. For instance,
\cite{kauffmann} recently compiled a catalog of 1325
molecular gas clouds of very different sizes, including estimates of
their virial parameters.

There exists a critical virial parameter, denoted here by $\alpha_{vir}^{crit}$, which
comes from a stability study in which perturbations both in pressure and
density gradients were considered; it has been mathematically determined
that $\alpha_{vir}^{crit} \approx 2$. The common belief in the recent past
has been that most gas structures have $\alpha_{vir} > \alpha_{vir}^{crit}$, so
that either they do not collapse or will even be diffused into the interstellar
medium. It is important to note that \cite{kauffmann} recently observed
low values of $\alpha_{vir}$ for regions of high-mass star formation; that is,
these gas structures have $\alpha_{vir} < \alpha_{vir}^{crit}$, so they are expected
to collapse.

The virial parameter is defined
as the ratio of virial mass M$_{vir}$ to the total mass M$_T$ of a gas
structure, so that in
the case of a spherical model of radius $R$, it is given by
$\alpha_{vir}=\frac{5 \, \sigma_v \, R}{G\, M_T}$, where $\sigma_v$ is the
velocity dispersion and $G$ is the Newton's gravitational constant. It should
be noted that the velocity dispersion entering in the
calculation of $\alpha_{vir}$ includes both the thermal
and the non-thermal velocity components.

When the velocity distribution of a particular model
of a gas structure is considered to be composed only of a thermal
component, it is possible to relate the
$\alpha_{vir}$ to the $\beta$ ratio by means of
$\alpha_{vir}= f \, 2 \beta$, where $f$ is a form factor, which is
empirically included to take into account modifications for
non-homogeneous and non-spherical gas distributions.

As it seems that the dimensionless critical virial parameter, which separates the gas
clouds that collapse from those that do not, is close to 2, we here prepare
initial conditions of the SPH particles to have a rigidly rotating gas structure,
such that its $\beta$ is given by 0.1, 0.3, and 0.48.
A possible approximation would be
that of using the average velocity $<v>$ of the particle distribution
instead of the $\sigma_v$. Thus, for a numerical simulation, we would have a
virial parameter approximated by $\alpha_{vir}=\frac{5 \, <v>^2 \, R}{G\, M_T}$.
So that the values of $\beta$ given above will correspond to simulations
with their approximated $\alpha_{vir}$ values given
by 0.5, 1.5, and 2.5, respectively.

Thus, the question raised by the virial parameter in the
observational field, of finding where is the transition between the
cores that collapse to those that do not, in this paper we
translated this question to the numerical simulations field, so that
we here locate those models that are near to the non-collapsing
regime, such that we call them "the last collapsing configuration".
As expected, the occurrence of this last collapsing configuration
depends on initial values of $\alpha$ and $\beta$ and on the total
mass of the parent structure.

With regard to the theoretical aspect, we mention that self-similar
solutions for the collapse of isothermal spheres were first obtained
by \cite{larson} and \cite{penston}, and later by \cite{shu}.
\cite{hunter} performed a detailed mathematical analysis of the
collapse problem and re-discovered previous solutions and also
obtained new solutions. Other self-similar solutions were also found
for the collapse of adiabatic spheres, in which a polytropic
equation of state was assumed. All these solutions provide an
accurate description of the early collapse stage, which mostly
proceeds isothermally, as the gas is optically thin to its own
radiation. A radial density profile was obtained from these
self-similar solutions. In the case of a rotating collapse with
pressure, \cite{tohline2} was able to get also the mathematical
expression of a radial density profile by using an approach based on
the virial theorem. The physical parameters of the parent core are
included in the density profile,  so that it can be simply scaled
for a different set of parameters of a second parent core.

Motivated by the work of \cite{RMAA} mentioned above, in
this paper we want to study the formation of binary systems via the
collapse of a more massive gas structure; thus, we
carry out a fully three-dimensional set of numerical hydrodynamical
simulations aimed to model the gravitational collapse of
a M$_T \approx $ 50 and 400 $\,$ M$_{\odot} $ clumps, by using the SPH particle
technique, such that a finite number of SPH particles is used to sample
the entire gas structure. The formation of two antipode mass seeds
during the early collapse stage of the clumps is here enforced by
implementing a mass perturbation with the same mathematical structure
of the density perturbation successfully used
in the classic collapse calculation of \cite{boss2000} for an isothermal
one solar mass core.

When the peak density of the collapsing core is high enough,
the central region of the core become opaque to its own radiation;
thus the isothermal regime breaks down so that an adiabatic regime
begins. In our case, this transition is taken into account by using
the barotropic equation of state, first proposed by \cite{boss2000},
which is characterized by a critical density.

Now, the value of this critical density depends on how much mass
surrounds and obscures the central region of the core.  In general,
it is expected that the more massive the parent cores are, the
critical density at which the obscuration occurs may be lower.
Strictly speaking, this would imply to change the value of the
critical density of the barotropic equation of state in our suite of
simulations.

In this paper, we were able to follow the collapse of the clumps up
to three orders of magnitude in density within this adiabatic
regime, where there are no scaling solutions known, out of which new
solutions can be simply scaled from the initial mass of the parent
core, as can be done in the isothermal collapse. Because of this,
the treatment of this advanced collapse stage must be investigated
by means of numerical simulations. As expected, for the first
collapse stage, no significance differences in the collapse of a
higher mass structure are seen. However, as we will see later, the
influence of a higher mass parent structure on the collapse get
manifested only in the non-isothermal collapse stage; see
\cite{sterzik}.

In this paper we re-simulate two core models of 1 and 5 M$_{\odot}$ with
$\beta$ also given by 0.1, 0.3, and 0.48, respectively. These models are similar
but not identical to those already reported by \cite{RMAA}. We consider here all those
collapse models with values of
$\alpha$ given by 0.1, 0.2, and 0.3, in order to make a comparison of
our results with those of \cite{riaz}, who studied the thermal sensitivity
of binary formation via the collapse of cores. To complement the qualitative
study of binary configurations that we carry out in the first part of this paper, we next present a
quantitative analysis of some of the resulting binary configurations, including the calculation of the
binary separation and masses of the fragments of some
particular binary configurations, and we present them in terms of the total mass of the parent structure.

It should be mentioned that we have not considered here the
collapse of a more massive gas structure, such as a gas cloud with mass
and size of the order of 10$^3$-10$^4 \,$ M$_{\odot}$ and 2-15 pc.
This is due to the fact that it has been suggested by several authors that
the observed masses and core separations in regions of high-mass
star formation cannot be explained by invoking only fragmentation via
gravitational instability; see \cite{liu} among others. Thus, the
formation process of very massive dense cores in high-mass star
formation regions needs other mechanisms to be taken into account
other than that of thermal fragmentation (also called Jeans fragmentation),
for instance, turbulent fragmentation or magneto-hydrodynamical induced fragmentation.

However, for both the cores and the intermediate mass clumps, such as those considered
in this paper, it is expected that only thermal fragmentation is relevant to
explain their fragmentation properties; see \cite{palau14}, \cite{palau15}
and \cite{busquet}.

We finally mention that the collapse of cores to model isolated or binary star formation is an
active field of research, for example, the evolution of rotating cloud cores including turbulent
velocity initial distributions has been investigated by many authors; see for
instance:  \cite{bate2002}, \cite{bate2003},
\cite{delgadonante2004a}, \cite{delgadonante2004b}, \cite{batebonnell2005} and \cite{goodwin}.
Collapse calculations of an isothermal cloud core rotating in a uniform magnetic
field have been presented, among others, by \cite{price}, \cite{hennnebelleteyssier},
\cite{machida2} and \cite{bosskeiser}.

%%%%%%%%%%%%%%%%%%%%%%%%%%%%%%%%%%%%%
%%%%%%%%%%%%%%%%%%%%%%%%%%%%%%%%%%
%%%%%%%%%%%%%%%%%%%%%%%%%%%%%%%%%%%
\section{The physical state of the parent gas structures}
\label{sec:parent}

All the parent gas structures considered in this paper are
uniform spheres, which are rigidly rotating around the $z$ axis with an angular
velocity $\Omega$, so the initial velocity
of the $i-th$ SPH particle is given by $(-\Omega\, y_i,\Omega\, x_i,0)\;$.

The time needed for a test particle to reach
the center of a gas sphere, when gravity is the only force acting on it,
is defined as the free fall time $t_{ff}$, and it is given by
$t_{ff} \approx \sqrt{ \frac{3\, \pi}{32 \, G \, \rho_0}}$, where $\rho_0$ is the average
density. We will use $\rho_0$ and $t_{ff}$ to normalize the figures that appear in the coming
sections.

Following \cite{bodeniv}, the dynamical state of a general gas structure is
usually characterized by the values of the dimensionless ratios
$\alpha$ and $\beta$, which are given by

\begin{equation}
\begin{array}{l}
\alpha \equiv \frac{E_{therm}}{\left|E_{grav}\right|} \vspace{0.25 cm}\\
\beta \equiv \frac{E_{rot}}{\left|E_{grav}\right|}
\end{array}
\label{defalphabeta}
\end{equation}
\noindent For a spherical model of a gas structure of total mass $M_T$, the
total gravitational potential energy
is approximated by $<E_{grav}> \approx - \frac{3}{5} \; \frac{G\, M_T^2}{R}$, where
$R$ is the sphere radius. The
average total thermal energy $<E_{therm}>$ (kinetic plus potential
interaction terms of the molecules) is
$ <E_{therm}> \approx \frac{3}{2} {\cal N} \, k \, T = \frac{3}{2} M_T\, c_0^2$,
where $k$ is the Boltzmann constant, $T$ is the equilibrium temperature,
${\cal N}$ is the total number of molecules in the gas, and $c_0$ is the
speed of sound. The rotational energy of the clump is approximately given by
$E_{rot}=\frac{1}{2} \; I\,\Omega^2=\frac{1}{2}\,\frac{J^2}{I}
\approx \frac{1}{5}\, M_T\, R^2 \, \Omega^2$, where $I\approx \frac{2}{5} \, M_T\,R^2$
is the moment
of inertia and $J=I\,\Omega_0$ is the total angular momentum.

If we consider that the velocity dispersion $\sigma_v$ of a gas
structure is directly related only to the thermal velocity component, then
$\sigma_v = \sqrt{\frac{k\, T_{kin} }{17 \, m_H} }$ where $m_H$ is
the molecular mass of the main gas component and $T_{kin}$ is the
kinetic temperature. In this case, an approximation to be used in this paper
would be using the average velocity $<v>$ of the simulation particles instead
of the $\sigma_v$. Thus, we would have a virial parameter
approximated by $\alpha_{vir}=\frac{5 \, <v>^2 \, R}{G\, M_T}$.

In this paper, we have carefully selected the values of
$c_0$ and $\Omega$, so the energy ratios obtained by
considering all the simulation particles of an initial configuration
(the snapshot zero of a simulation) are initially given
by a pairs of values taken from the following sets:

\begin{equation}
\begin{array}{l}
\alpha \equiv 0.3, 0.2, 0.1 \vspace{0.25 cm}\\
\beta \equiv 0.1, 0.3 , 0.48
\end{array}
\label{valuesalphaybeta}
\end{equation}
\noindent so that the corresponding values of the virial parameter for
all our models are approximately given, respectively, by
\begin{equation}
\alpha_{vir} \equiv 0.5,1.5, 2.5
\label{valuesalphavir}
\end{equation}
%%%%%%%%%%%%%%%%%%%%%%%%%%%%%%%%%%%%%%
\subsection{The cores}
\label{subsec:core}

We will consider the gravitational collapse of two cores: the first
is a variant of the so-called "standard isothermal test case," which was
first calculated by \cite{boss1979} and later calculated by
\cite{burkertboden93} and \cite{bateburkert97}; the main
outcome of this classic model was a protostellar binary system. In this paper, the
core radius is R=4.99 $\times 10^{16} \,$ cm $\equiv$ 0.016 pc
and its mass is M$_T$=1 M$_{\odot}$. Thus, the average density and the
corresponding free fall time of this core are
$\rho_0=$3.8 $\times 10^{-18}\, $ g cm$^{-3}$ and
$t_{ff} \approx $1.0 $\times 10^{12} \,$ s $\equiv $ 34077 yr, respectively.

The radius and mass of the second core are
R=6.16 $\times 10^{17} \,$ cm $\equiv $0.2 pc and
M$_T$=5 M$_{\odot}$, respectively. This core has an average density of
$\rho_0=$1.0 $\times 10^{-20}\, $ g cm$^{-3}$ and a
$t_{ff} \approx $2.0 $\times 10^{13} \,$ s $\equiv 660997 \,$ yr, respectively.

Despite the size and mass differences of these gas structures,
both of them are still cores, as defined statistically by \cite{bergin}.
%%%%%%%%%%%%%%%%%%%%%%%%%%%%%%%%%%%%%%%%%%%
\subsection{The clumps}
\label{subsec:clump}

The first clump structure considered in this paper has a radius and mass given
by R=0.3 pc $\equiv$ 9.24 $\times 10^{17}$ cm and M$_T$=50 M$_{\odot} $, respectively. Its
average density is given by $\varrho_1=$3.0 $\times 10^{-20}$ g cm$^{-3}$. 

The radius and mass of the second clump are given
by R=1 pc $\equiv$ 3.08 $\times 10^{18}\,$cm and
M$_T$=400 M$_{\odot} $, respectively. Its average density is
$\varrho_2=$6.5 $\times 10^{-21}\, $ g cm$^{-3}$ and its corresponding free fall
time is $t_{ff} \approx 2.6 \times 10^{13} \,$ s $\equiv $826 247 yr.

We emphasize that the radius and mass of the clumps have been changed
with respect to those of the cores. In spite of this, it is still
possible to make a comparison of the collapse results, as  the focus
of this paper is to study the effects of changing the values of
$\alpha$ and $\beta$ on the collapse of different parent gas
structures. Besides, if we had kept unchanged some physical property
of the clump structures, so that we had used the core radius, while
the total parent mass is different to that of the cores, then the
resulting clump structure would be un-physical, in the statistical
sense defined by \cite{bergin}.
%%%%%%%%%%%%%%%%%%%%%%%%%%%%%%%%%%%%%%%%%
%%%%%%%%%%%%%%%%%%%%%%%%%%%%%%%%%%%%%%%%%
\subsection{The mass perturbation}
\label{subs:initialmasspert}

For every SPH particle i of a simulation with total number of particles N$_p$,
that is, i=1..N$_p$, the particle mass is given by
$m_0=\frac{M_T}{N_p}$; then we implement a mass perturbation
of the type

\begin{equation}
m_i=m_0+m_0*a \cos\left(m\, \phi_i \right)
\label{masspert}
\end{equation}
\noindent where the perturbation amplitude is set to the values
of a=0.1 and a=0.25; the mode is fixed to m=2 and $\phi$ is the
azimuthal spherical coordinate.

This mass perturbation scheme was
successfully applied in many papers on collapse; see for instance
\cite{NuestroApJ}, \cite{NuestroRMAA}, \cite{NuestroPlummer}, and
notably \cite{springel}, when the Gadget2 code was proven,  among
other tests, with the calculation of the isothermal collapse, where  the unequal
mass particles method was validated.

%%%%%%%%%%%%%%%%%%%%%%%%%%%%%%%%%%%%%%%%%%%%%%%%%%%%%%%%%%%%%%%%%%%%%%%%%%%%%%%%%%%%
\subsection{Resolution}
\label{subs:resol}

\cite{truelove} demonstrated the need of fulfill appropriate spatial resolution
requirements in order to avoid the occurrence of artificial fragmentation in a collapse simulation.

Following~\citet{bateburkert97}, the
smallest mass that a $SPH$ calculation can resolve, $m_r$, is given by
$m_r \approx 2 \, N_{neigh} \, M_J$ where $N_{neigh}$ is the number of
neighboring particles included in the $SPH$ kernel and $M_J$ is the spherical
Jeans mass $M_J$, which is defined by

\begin{equation}
M_J \equiv \frac{4}{3}\pi \; \rho \left(\frac{ \lambda_J}{2} \right)^3
= \frac{ \pi^\frac{5}{2} }{6} \frac{c^3}{ \sqrt{G^3 \, \rho_m} } \;.
\label{mjeans}
\end{equation}
\noindent where $\lambda_J$ is Jeans wavelength and $\rho_m$ is the peak density
reached in a simulation. Therefore, the smallest mass particle $m_p$ in our simulations must
at least be such that $\frac{m_p}{m_r}<1$.

Let us assume that $N_{neigh}=40$, $\rho_m$=1.0 $\times \, 10^{-11}$ g cm$^{-3}$ and that
the sound speed $c$ varies within the values 10 369 and 17 868  cm s$^{-1}$, which
correspond to runs with $\alpha$=0.1 and 0.3 of the 1 M$_{\odot}$ model, respectively. In
this case, the $m_r$ is within 3.0 $\times 10^{27}$ and 1.5 $\times 10^{28}$ g. In the simulations
of a 1 M$_{\odot}$ core, we
used N$_p$=2 000 000 of SPH particles, so that $m_p$ is
$m_p=9.9 \times 10^{26}$ g, and then we have that the ratio of masses is
$m_p/m_r$ 0.06 and 0.3. Therefore The number of particles is high enough to fulfill
the resolution requirements described by \cite{truelove}.

For the 5 M$_{\odot}$ core, the particle mass is $m_p$=4.9 $\times 10^{27}$ g while the
sound speed varies within 9500.0 and 12400.0 cm s$^{-1}$, so that by using again N$_p$=2 000 000 of SPH
particles, the ratios $m_p/m_r$ are appropriate up to peak densities smaller
than $\rho_m$=5.0 $\times \, 10^{-12}$.

For the 50 M$_{\odot}$ clump, we used 12 000 000 of SPH particles, so the particle mass
is $m_p=8.2 \times 10^{27}$ g. In this case, the sound speed varies within 17 497 and
29 576 cm s$^{-1}$, and for a peak density of 1.0 $\times 10^{11}$, $m_r$ are 0.5 and 0.1,
respectively.

For the 400 M$_{\odot}$ clump, a very large number of particles is
needed for the simulations to fulfill the resolution requirements.
Because computational limitations, we could not achieve the desired
resolution and despite of this, we evolved these simulations using
only 2 million of SPH particles; for this lack of reliability, we
consider them useful only for comparison with the lower mass
simulations.
%%%%%%%%%%%%%%%%%%%%%%%%%%%%%%%%%%%%%%%%%%%%%%%%%%%%%%%%%%%%%%%%%%%%%%%%%%%%%
\subsection{The barotropic equation of state}
\label{subs:thermo}

In order to ensure the change in the thermodynamic regime
from isothermal to adiabatic, we here implement the barotropic
equation of state proposed by~\cite{boss2000}:

\begin{equation}
p= c_0^2 \rho \left[ 1 +
\left( \frac{\rho}{\rho_{crit}}\right)^{\gamma -1 }
\right]
\label{beos}
\end{equation}
\noindent where $\gamma$=5/3 and c$_0$ is the sound speed. The
critical density $\rho_{crit}$ has only been given here the value
$\rho_{crit}=5.0 \times 10^{-14} \,$ g cm$^{-3}$. As we will
show in the following sections, in the simulations considered in this paper, the
average peak density increases up to 3 orders of magnitude within the adiabatic regime.
%%%%%%%%%%%%%%%%%%%%%%%%%%%%%%%%%%%%%%%%%%
%%%%%%%%%%%%%%%%%%%%%%%%%%%%%%%%%%%%%%%%%%
\subsection{Evolution Code}
\label{subs:code}

To follow the gravitational collapse of our models, in this paper we
use the fully parallelized particle-based code Gadget2, which is
based on the tree-PM method for computing the gravitational forces
and on the standard SPH method for solving the Euler equations of
hydrodynamics; see~\cite{springel} and also \cite{gadgets}. The Gadget2 code has implemented
a Monaghan-Balsara form for the artificial viscosity;
see~\cite{mona1983} and \cite{balsara1995}. The strength of the
viscosity is regulated by setting the parameter $\alpha_{\nu} = 0.75$ and
$\beta_{\nu}=\frac{1}{2}\, \times \alpha_v$; see Equations 11 and 14
in~\cite{springel}. We have fixed the Courant factor to $0.1$.
%%%%%%%%%%%%%%%%%%%%%%%%%%%%%%%%%%%%%%%%%%%
%%%%%%%%%%%%%%%%%%%%%%%%%%%%%%%%%%%%%%%%%%%%
\subsection{The initial setup}
\label{subs:setup}

Herein we used a radial mesh with spherical geometry, such that a
set of concentric shells was created and populated with SPH particles by means
of a Monte Carlo scheme to set the initial particle
configuration.

Thus, all the SPH particles of each simulation, were
located randomly in all the available surfaces of each spherical
shell.  The total mass in each shell is kept constant, so that the global density of
the gas structure is also constant. In order to have  a constant density
distribution in a local sense, we next applied a radial
perturbation to all the particles of a given shell such that any particle
could be randomly displaced radially outward or inward, but preventing
a perturbed particle from reaching another shell; see section 2.1 of \cite{RMAA}.
%%%%%%%%%%%%%%%%%%%%%%%%%%%%%%%%%%%%%%%%%%%%%%%%
%%%%%%%%%%%%%%%%%%%%%%%%%%%%%%%%%%%%%%%%%%%%%%%%%
\subsection{The collapse models}
\label{sec:modelos}

In Table~\ref{tab:modelos} we summarize the models considered in this paper and the
main
configurations obtained. Column 1 shows the number of the model;  column 3
shows the total mass of the parent structure; column 4 shows the
value of the mass perturbation amplitude, defined by Eq.\ref{masspert}; column 5 and 6
show the values of the dimensionless
ratios $\alpha$ and $\beta$, respectively;
column 7 shows the type of configuration obtained and column 8 shows
the number of the figure and panel, in which the solution configuration is
shown\footnote{The letters on the right side of the
number of the figure indicate the panel of the mosaic where the configuration is located}.

We now emphasize five types of solution configurations that appear frequently, which
may be termed as follows:
(i) a face-to-face configuration, characterized by the presence of two
fragments located one in front of
the other, so that the connecting filament has been entirely accreted; (ii) a
connecting filament configuration, in which the filament still plays a relevant role as a gas
bridge between the two fragments; (iii) a central primary configuration,
characterized by a rotationally
supported solution surrounded by spiral arms; (iv) a filament configuration,
characterized by a thin and dense longitudinal collapsed structure without
fragments present; (v) a binary configuration, characterized by two fragments, so that
they are
orbiting one about the other.

All of these configurations have already been
observed before in the literature; in the next section we describe
how these configurations are transformed one into another because the change in
parameters of the parent gas structure.
%%%%%%%%%%%%%%%%%%%%%%%%%%%%%%%%%%%%%%%%%%
\section{Results}
\label{sec:resultados}

The main results of this paper are shown in color iso-density
plots for a slice of particles around the
equatorial plane of the
spherical gas structure.

As we want to compare several collapse models in this paper, we use
only one panel per model, which usually will correspond to the last
available snapshot, and we use three panels to compose a mosaic in order
to illustrate the resulting configurations obtained from a parent gas structure
of total mass M$_T$.

It should be mentioned that the vertical and horizontal axes of the iso-density plots indicate length
in terms of the sphere radius, which varies from zero to one in all the plots for the initial
snapshot, irrespective of the model; therefore the Cartesian axes X and Y vary initially
from -1 to 1.  In order to facilitate the comparison of the results, in each mosaic at least, we use
the same length scale for all the panels.

A comparison between models is still possible even at slightly different output times because
most of the configurations have already entered a
stable stage. Otherwise, the time evolution should be done
with the sink technique introduced by \cite{batebonnellprice95}, so that the system
can be evolved further in time.

The panels of a mosaic have the same value of the virial parameter
$\alpha_{vir}$ (equivalently, with the same $\beta$ values of 0.1,0.3 and 0.48) for a
given value of $\alpha$. We observed that all the models with the highest $\alpha$ and $\beta$, given by
0.3 and 0.48, respectively, do not collapse but get dispersed. As we want
to include only collapsing configurations, these models have been
skipped and for this reason some mosaics are formed only by two panels.
The last collapsing configurations have been grouped in other figures, because they are more likely to 
make better sense in mathematical terms, as their $\beta$ values may be
too high to keep physical significance in the theory of star formation.

Finally, for each model with M$_T$, $\alpha$ and $\beta$ given, we build two mosaics, each
corresponding to a perturbation amplitude $a$ with values of $a=$0.1 and $a=$0.25,
respectively; see Eq.~\ref{masspert}; see Table \ref{tab:modelos}.
%%%%%%%%%%%%%%%%%%%%%%%%%%%%%%%%%%%%%%%%%
\subsection{The models with M$_T$= 1 M$_{\odot}$}
\label{subsec:rescore1}

\begin{figure}
\begin{center}
\begin{tabular}{ccc}
\includegraphics[width=2.2in]{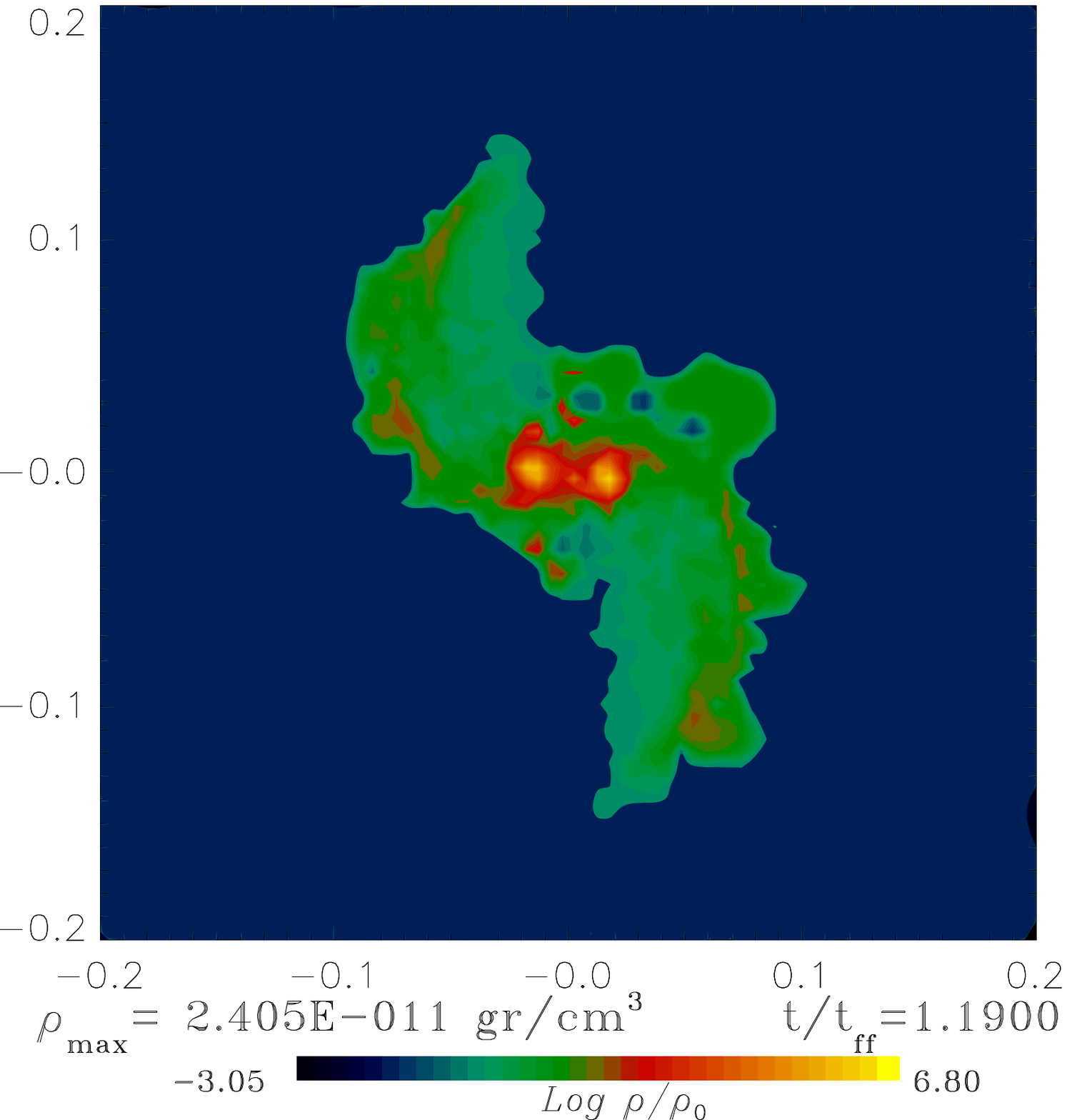} & \includegraphics[width=2.2in]{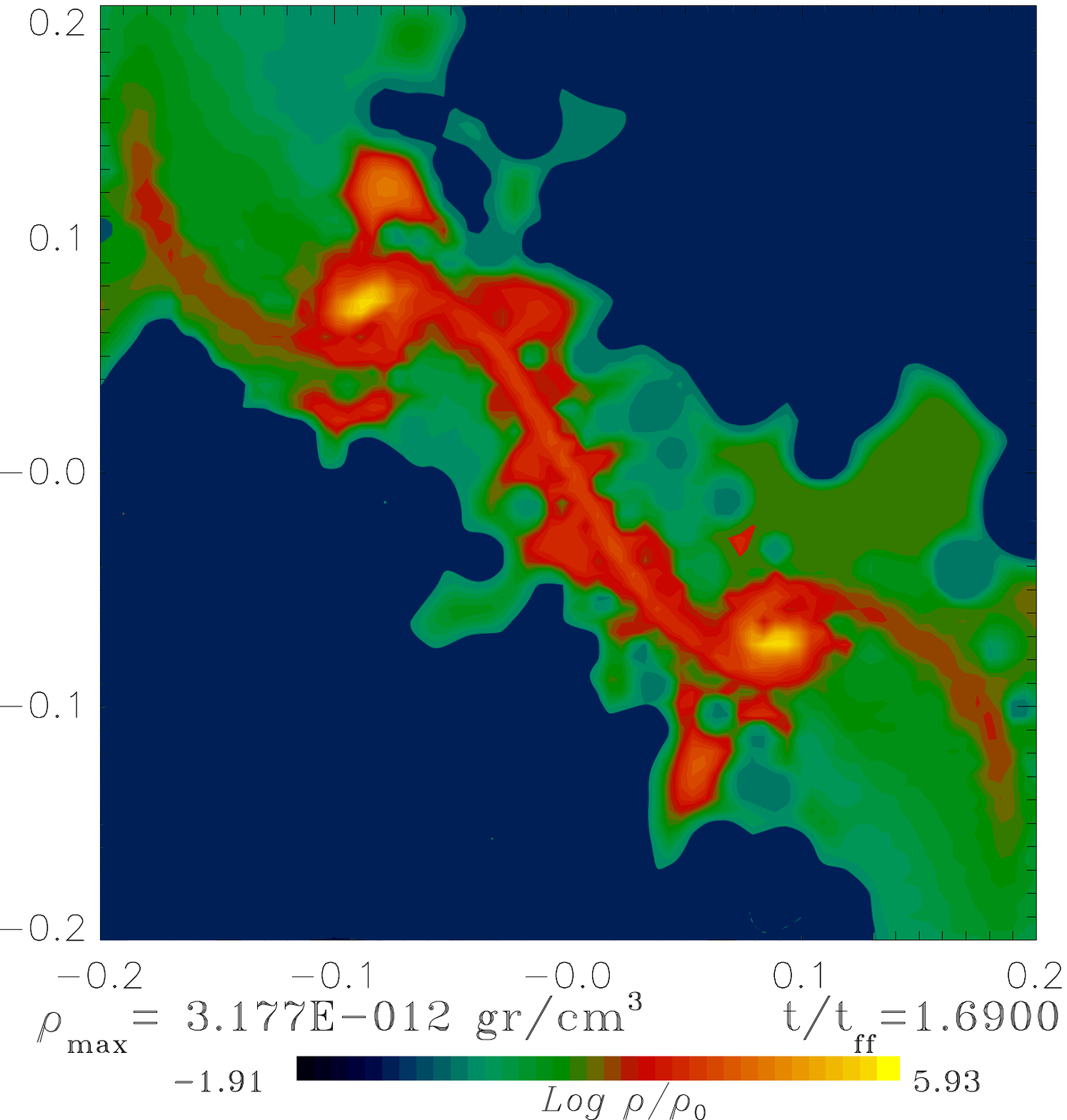}&
\includegraphics[width=2.2in]{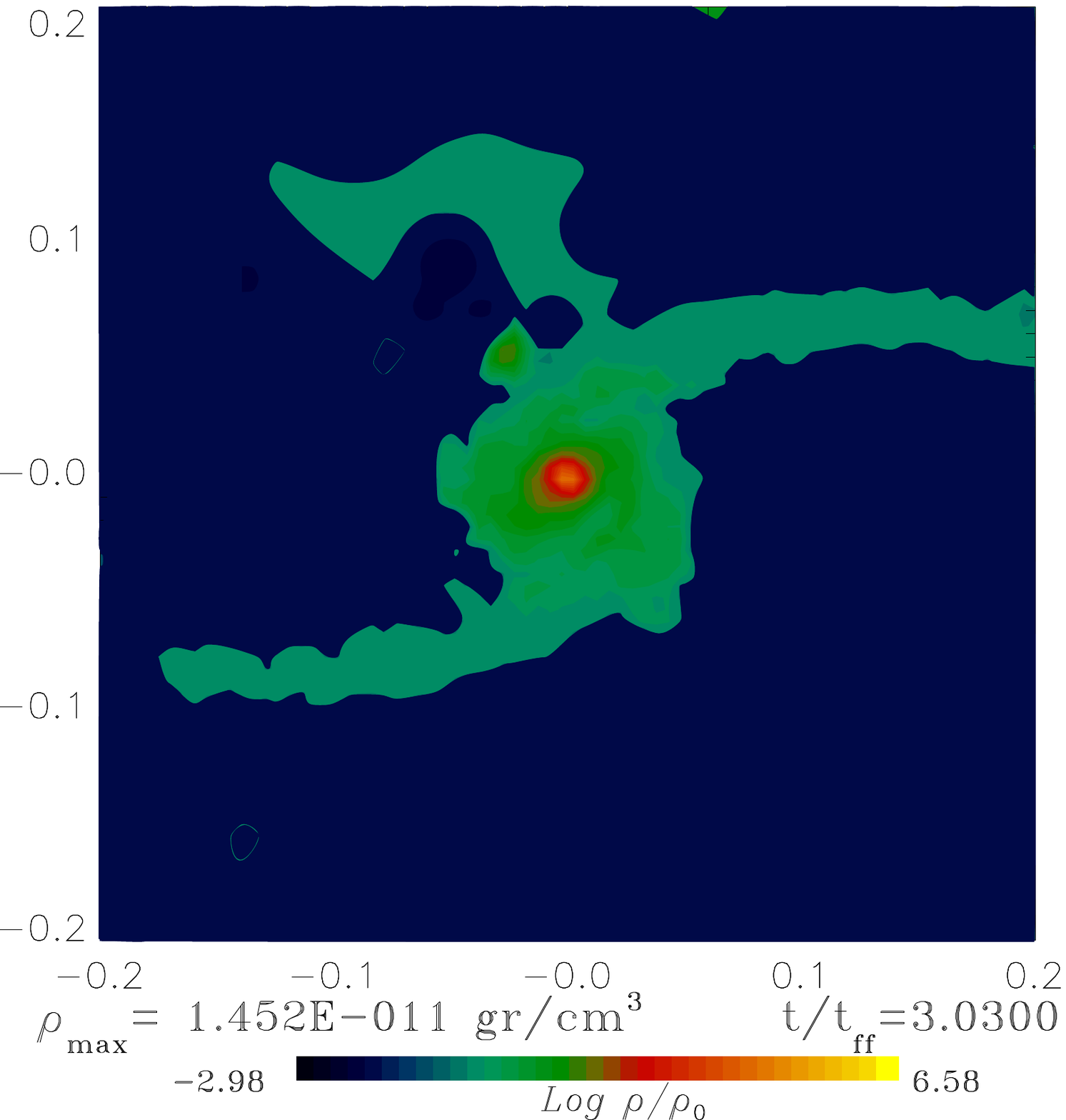}
\end{tabular}
\caption{\label{CPrueba60p2} Core models with M$_T$= 1 M$_{\odot}$, $a=$0.1 and $\alpha=$0.2;
the corresponding $\beta$ are
(left) 0.1
(middle) 0.3
(right) 0.48.}
\end{center}
\end{figure}

\begin{figure}
\begin{center}
\begin{tabular}{cc}
\includegraphics[width=2.2in]{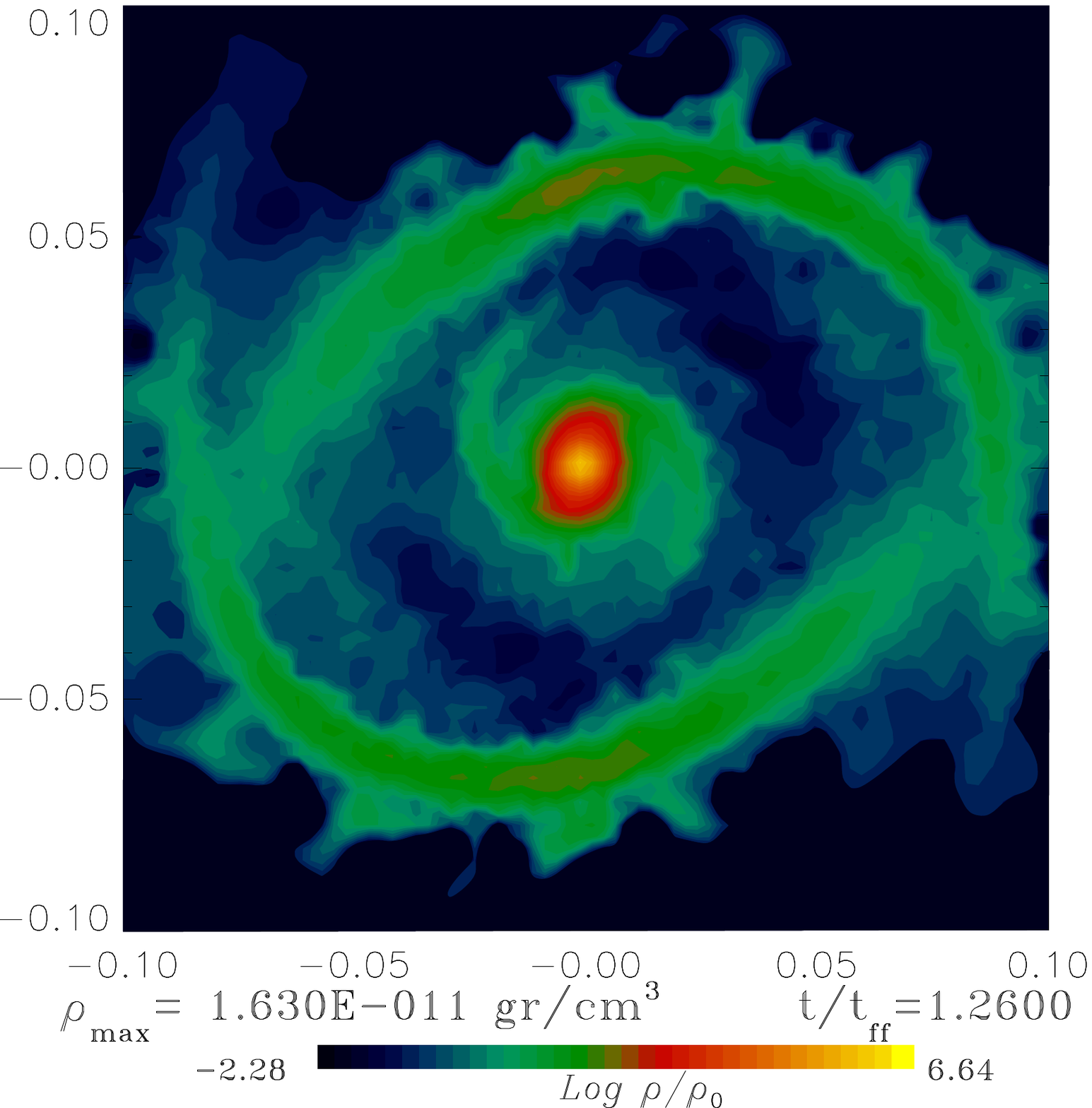} & \includegraphics[width=2.2in]{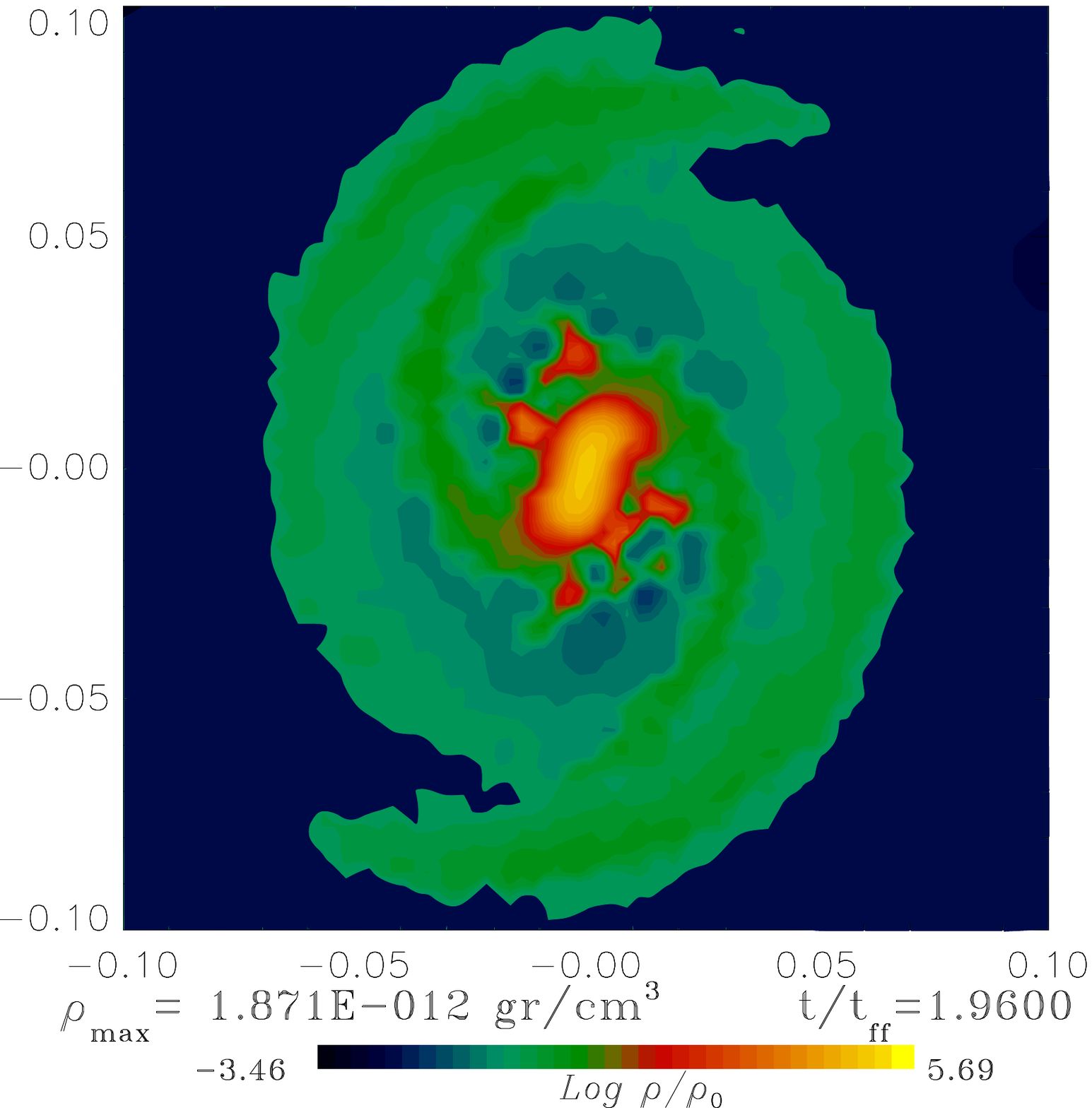} \\
\end{tabular}
\caption{\label{CPrueba60p3} Core models with M$_T$= 1 M$_{\odot}$, $a=$0.1 and $\alpha=$0.3;
the corresponding $\beta$ are
(left) 0.1
(right) 0.3.}
\end{center}
\end{figure}

\begin{figure}
\begin{center}
\begin{tabular}{ccc}
\includegraphics[width=2.2in,height=2.2in]{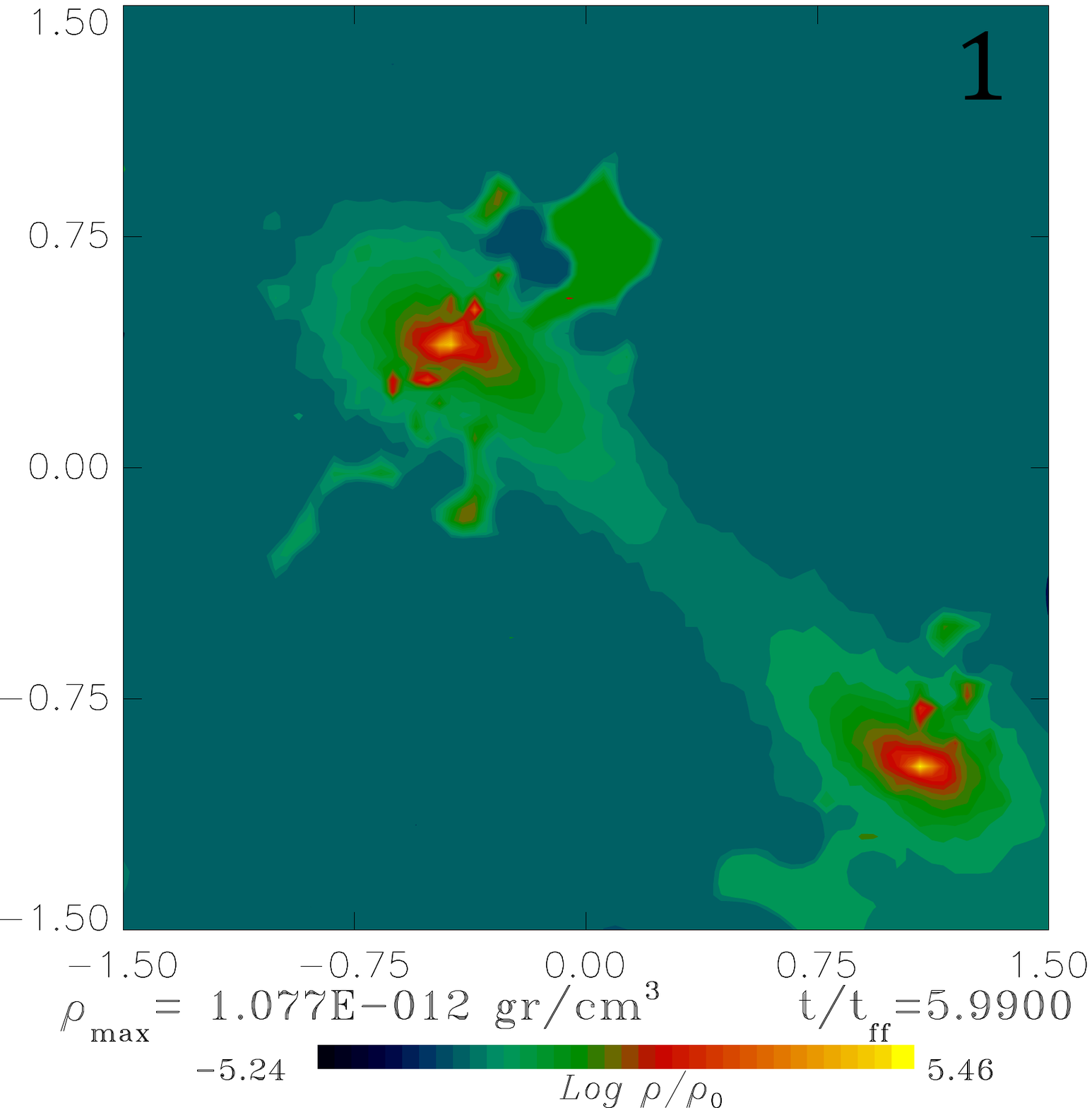} &
\includegraphics[width=2.2in,height=2.2in]{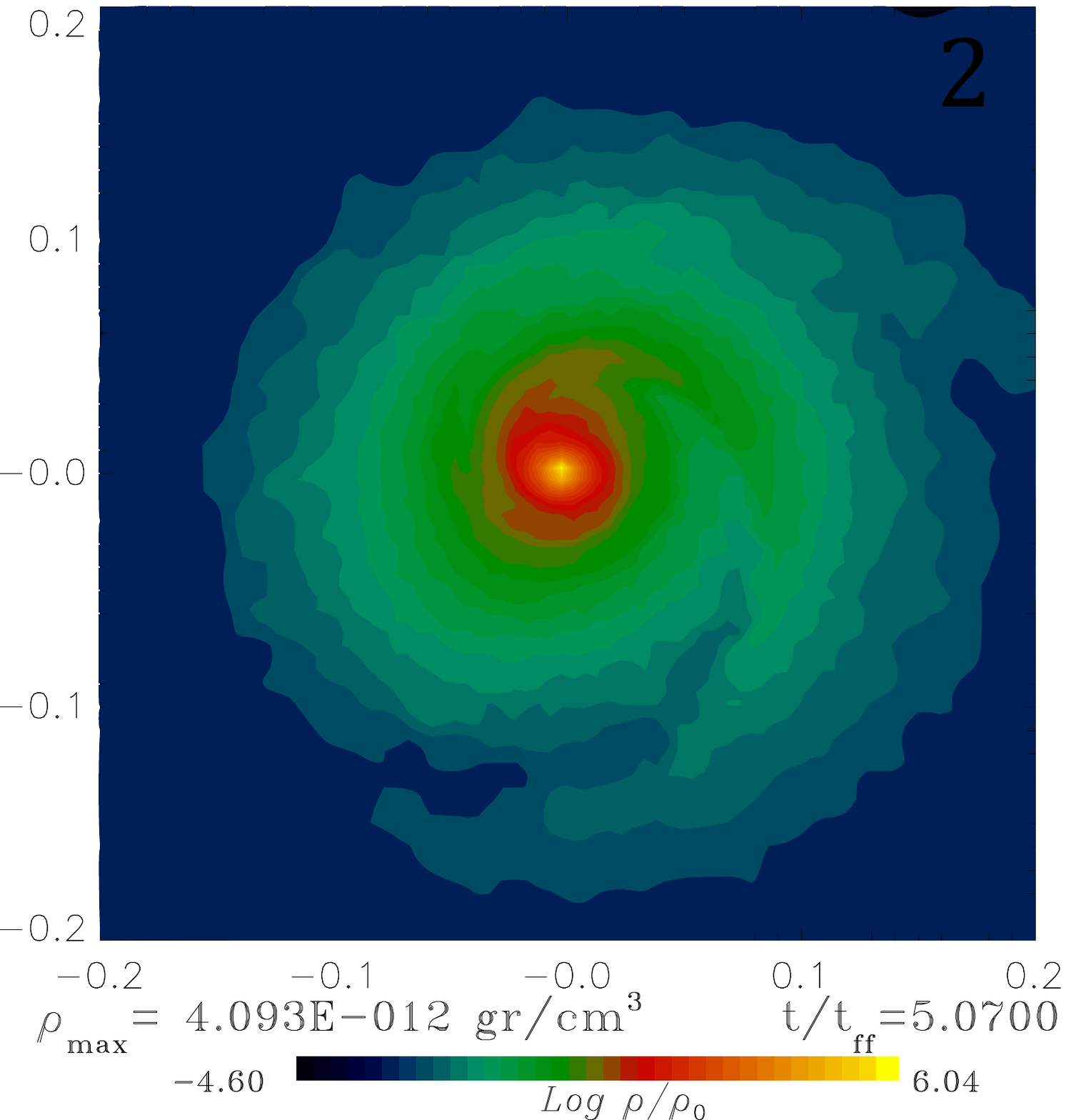} &
\includegraphics[width=2.2in,height=2.2in]{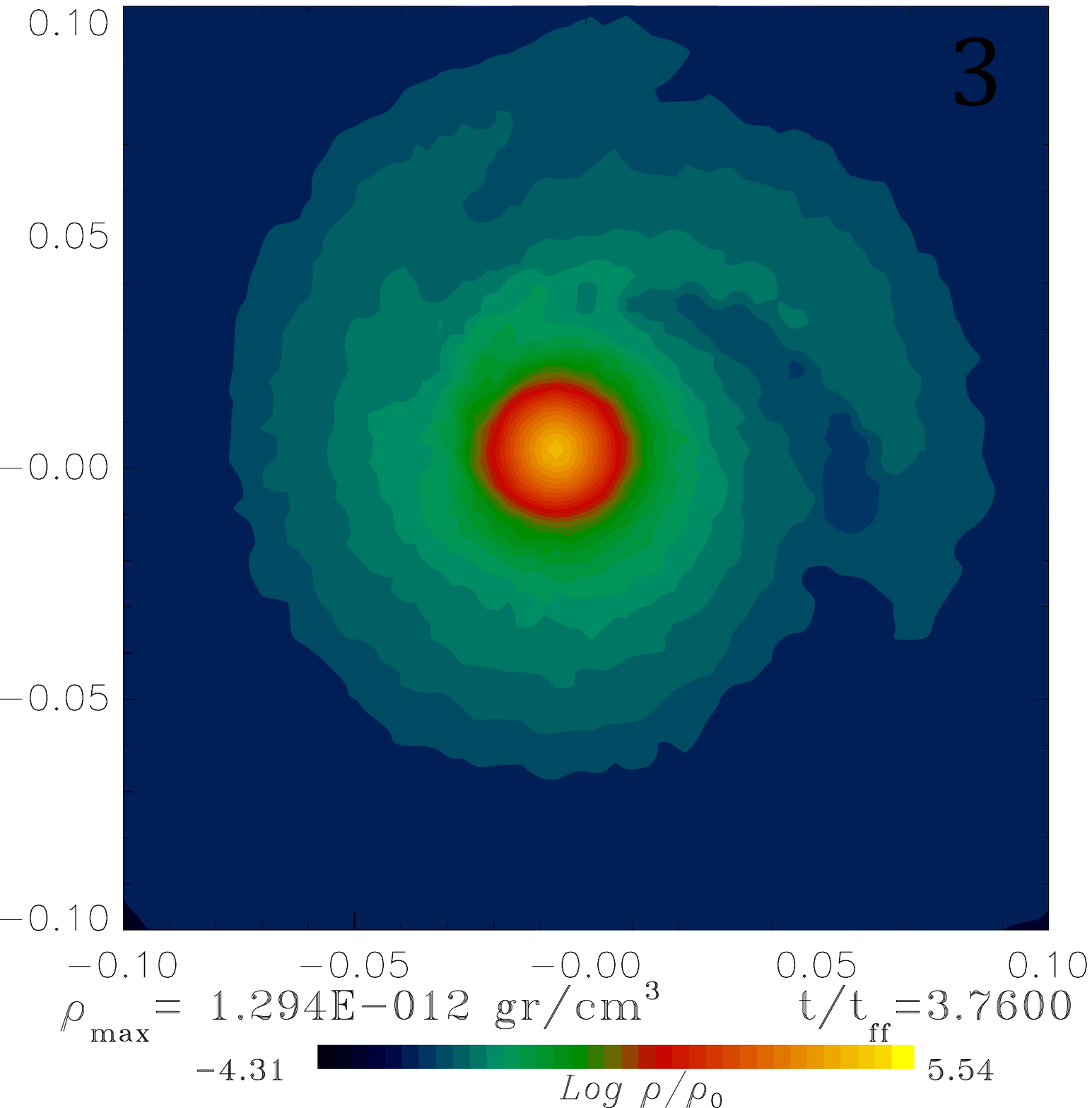}\\
\includegraphics[width=2.2in,height=2.2in]{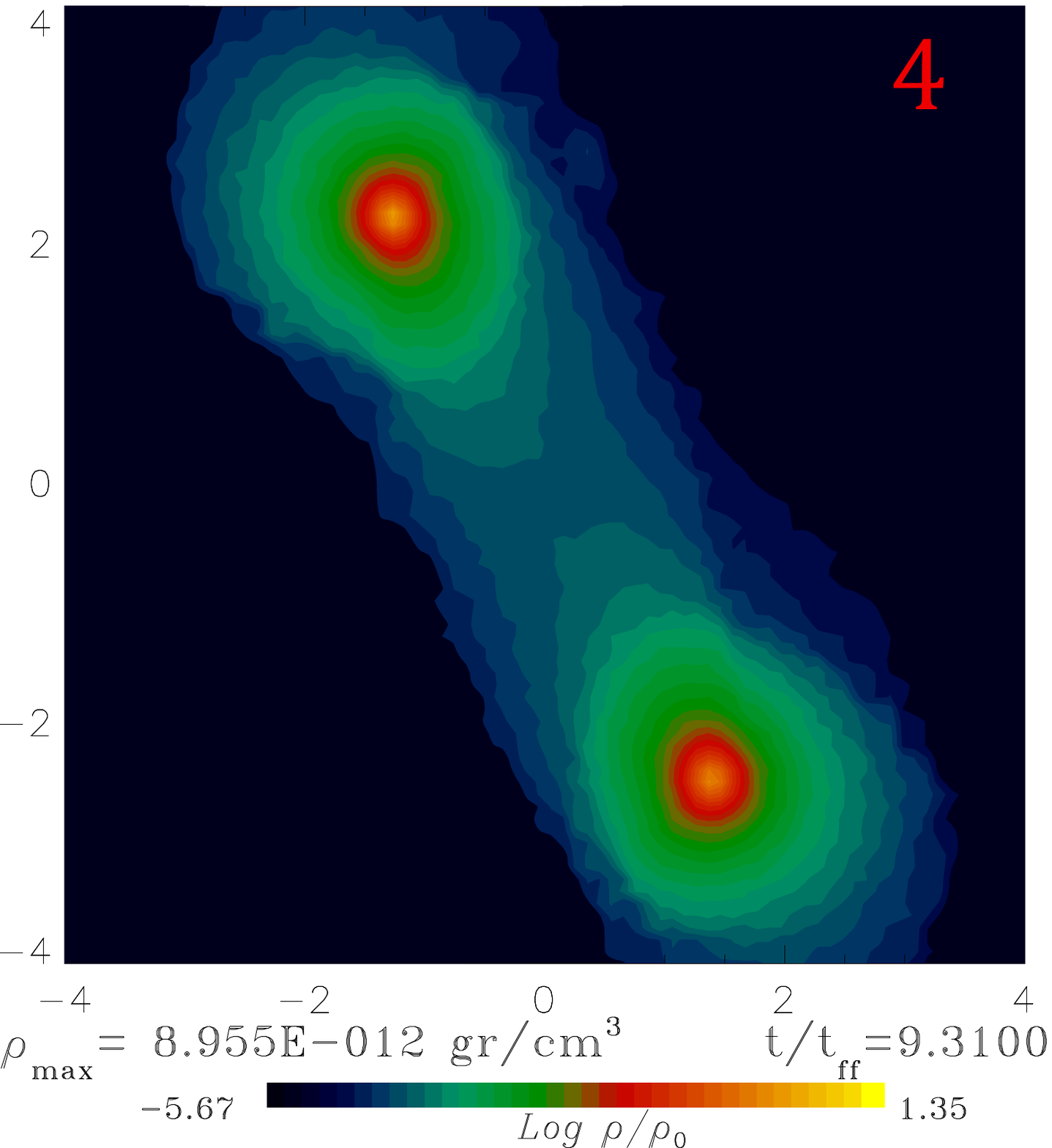} &
\includegraphics[width=2.2in,height=2.2in]{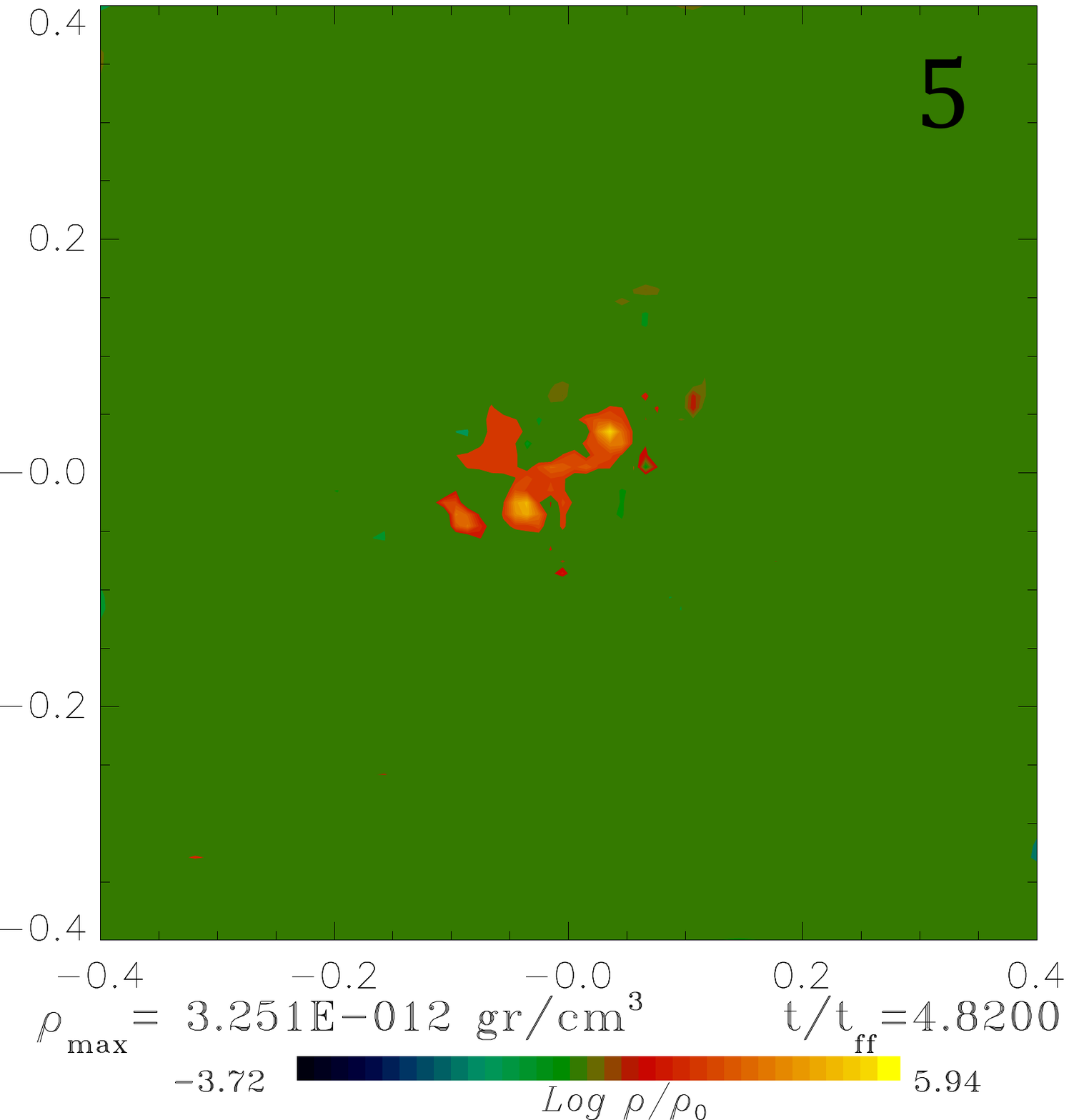} &
\includegraphics[width=2.2in,height=2.2in]{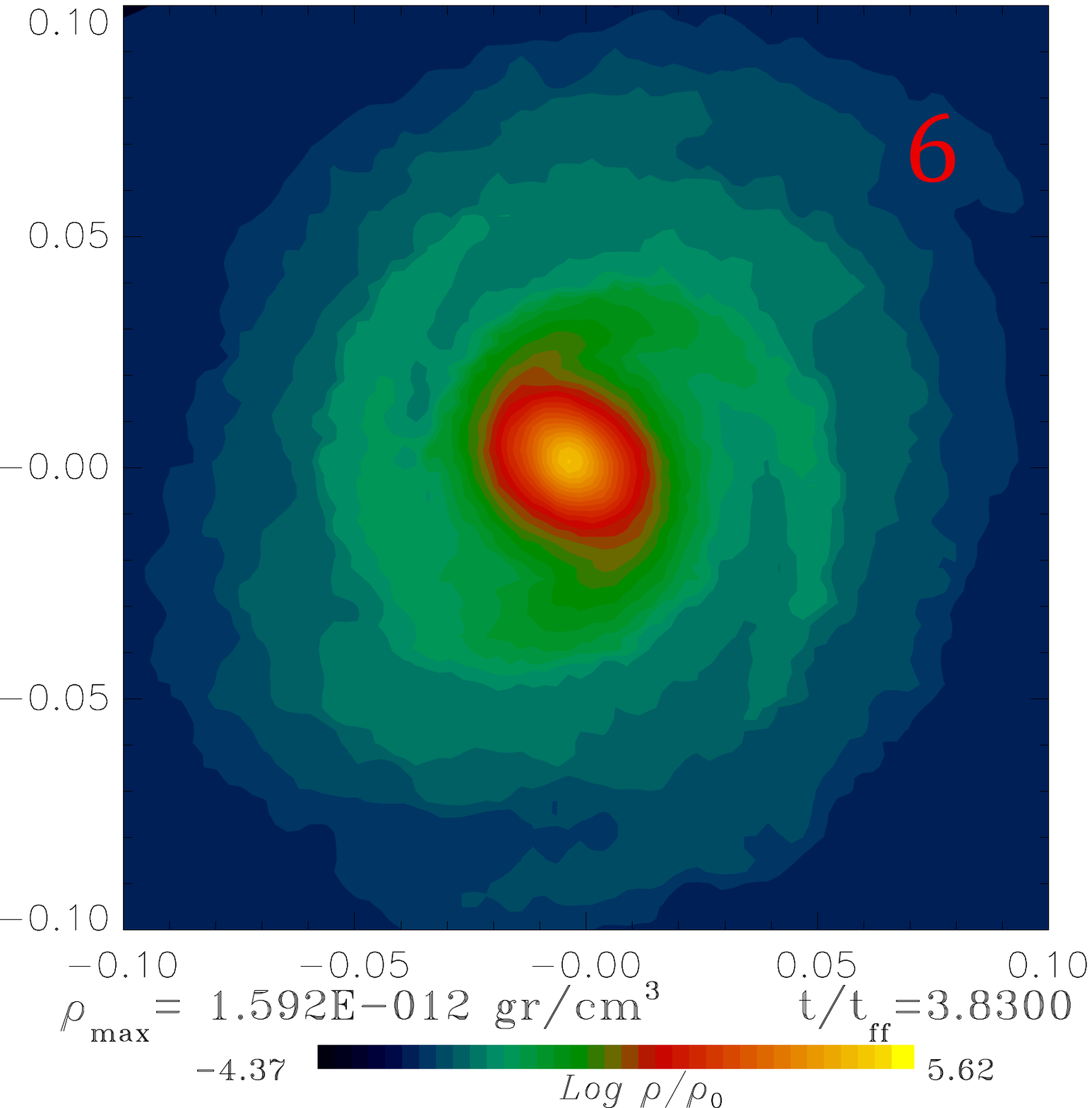}\\
\end{tabular}
\caption{\label{lastcolconfM1}
(1) M$_T$=1 M$_{\odot}$, $a=$0.1, $\alpha=$0.1, $\beta=$0.758;
(2) M$_T$=1 M$_{\odot}$, $a=$0.1, $\alpha=$0.2, $\beta=$0.5344;
(3) M$_T$=1 M$_{\odot}$, $a=$0.1, $\alpha=$0.3, $\beta=$0.3983;
(4) M$_T$=1 M$_{\odot}$, $a=$0.25,$\alpha=$0.1, $\beta=$0.8871;
(5) M$_T$=1 M$_{\odot}$, $a=$0.25,$\alpha=$0.2, $\beta=$0.5411;
(6) M$_T$=1 M$_{\odot}$, $a=$0.25,$\alpha=$0.3, $\beta=$0.3983;
}
\end{center}
\end{figure}

The figures \ref{CPrueba60p1}, \ref{CPrueba60p2}, and \ref{CPrueba60p3} show
the resulting systems obtained from the gravitational collapse of a
M$_T$= 1 M$_{\odot}$ core, similar to "the standard test case".  In Fig.\ref{CPrueba60p1}
all the panels show a well-defined pair of mass condensations connected by a filament.
When $\beta$ increases, the filament disappears so that the last collapsing
configuration shown in panel 1 of Fig.\ref{lastcolconfM1} is a well defined face-to-face
configuration. By observing the axes of the panels it can be seen that the separation of the resulting
fragments has significantly been increased for this last collapsing configuration.

When $\alpha$ is increased from 0.1 to 0.2, the pair of fragments can move one towards the
other, so that an orbiting binary system is formed
for lowest value of $\beta$,
and for higher values of $\beta$, the configurations
end with a central mass condensation surrounded
by short spiral arms, as can be seen in the left and right panels of Fig.\ref{CPrueba60p2} and also in
the panel 2 of Fig.\ref{lastcolconfM1}.

In Fig.\ref{CPrueba60p3}, one can see that even for the lowest value of $\beta$, the
resulting configuration is a central primary, which is rotationally supported, such
that the gravitational collapse slows down.

It is interesting to observe the change in the outcome of models with the same
$\beta$, but in which the $\alpha$ has been increased: from a face-to-face
configuration type to a central primary type. This behavior was also
observed by \cite{riaz}.

\begin{figure}
\begin{center}
\begin{tabular}{ccc}
\includegraphics[width=2.2in]{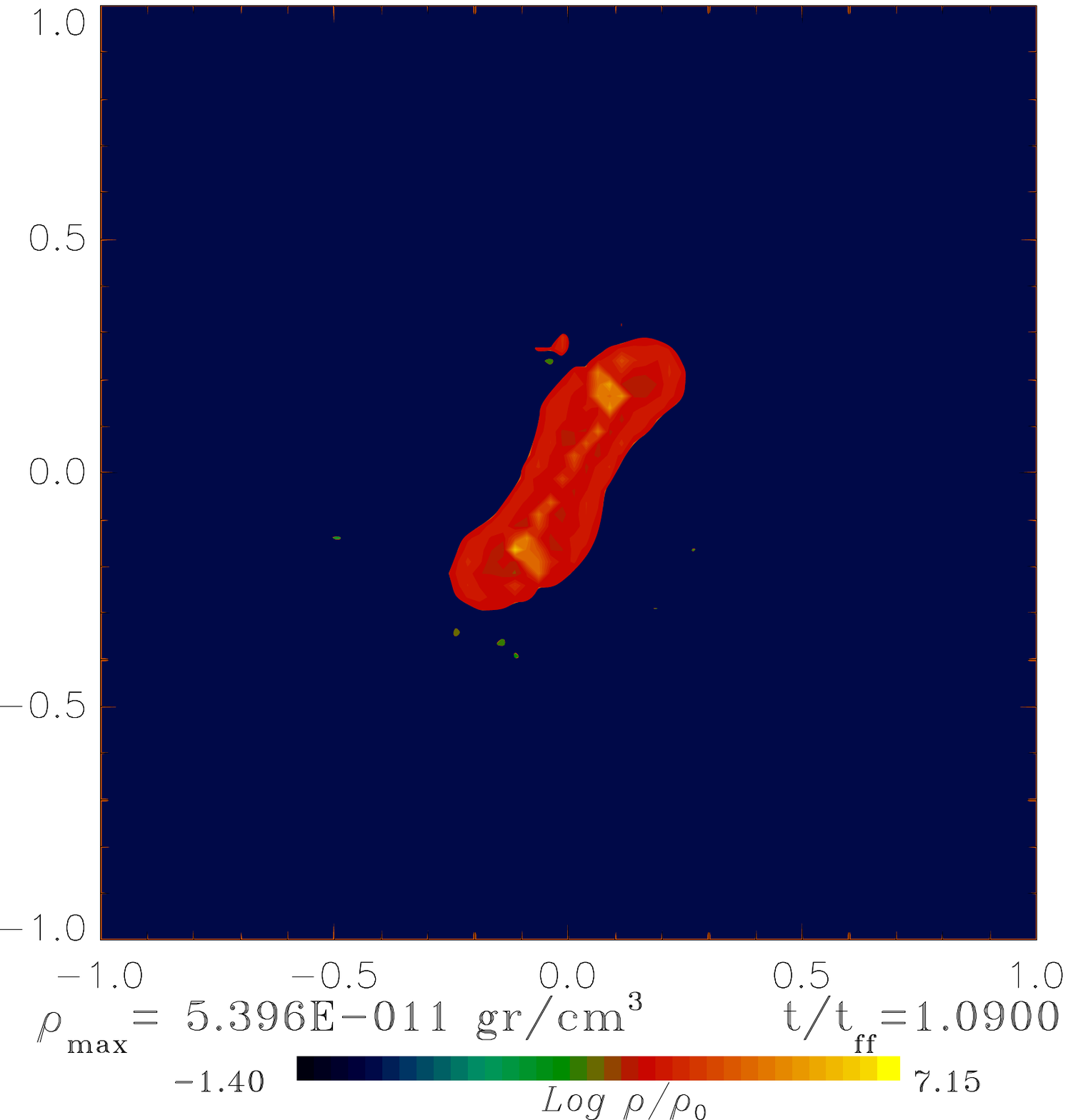} &\includegraphics[width=2.0in,height=1.75in]{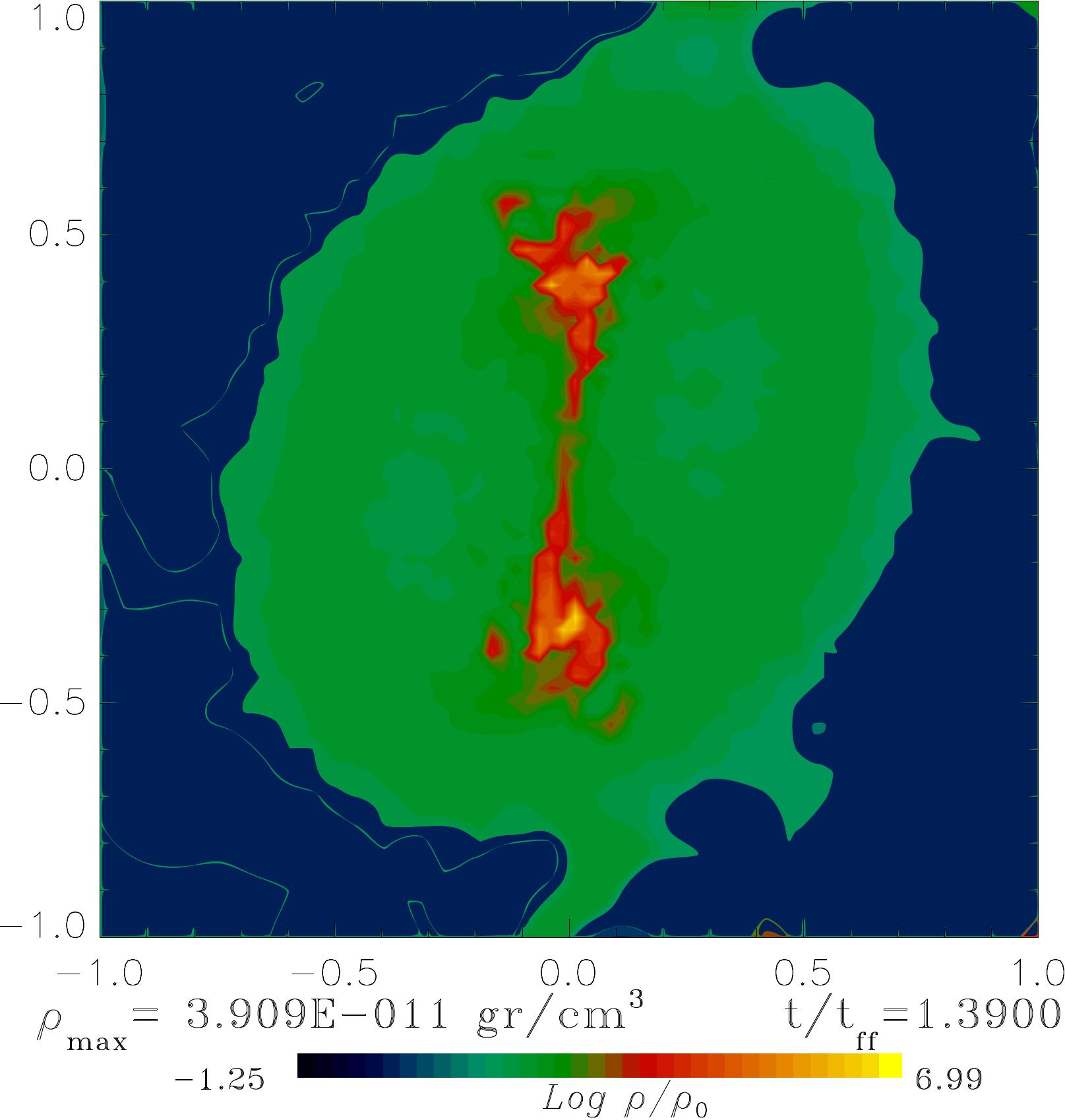} &
\includegraphics[width=2.2in]{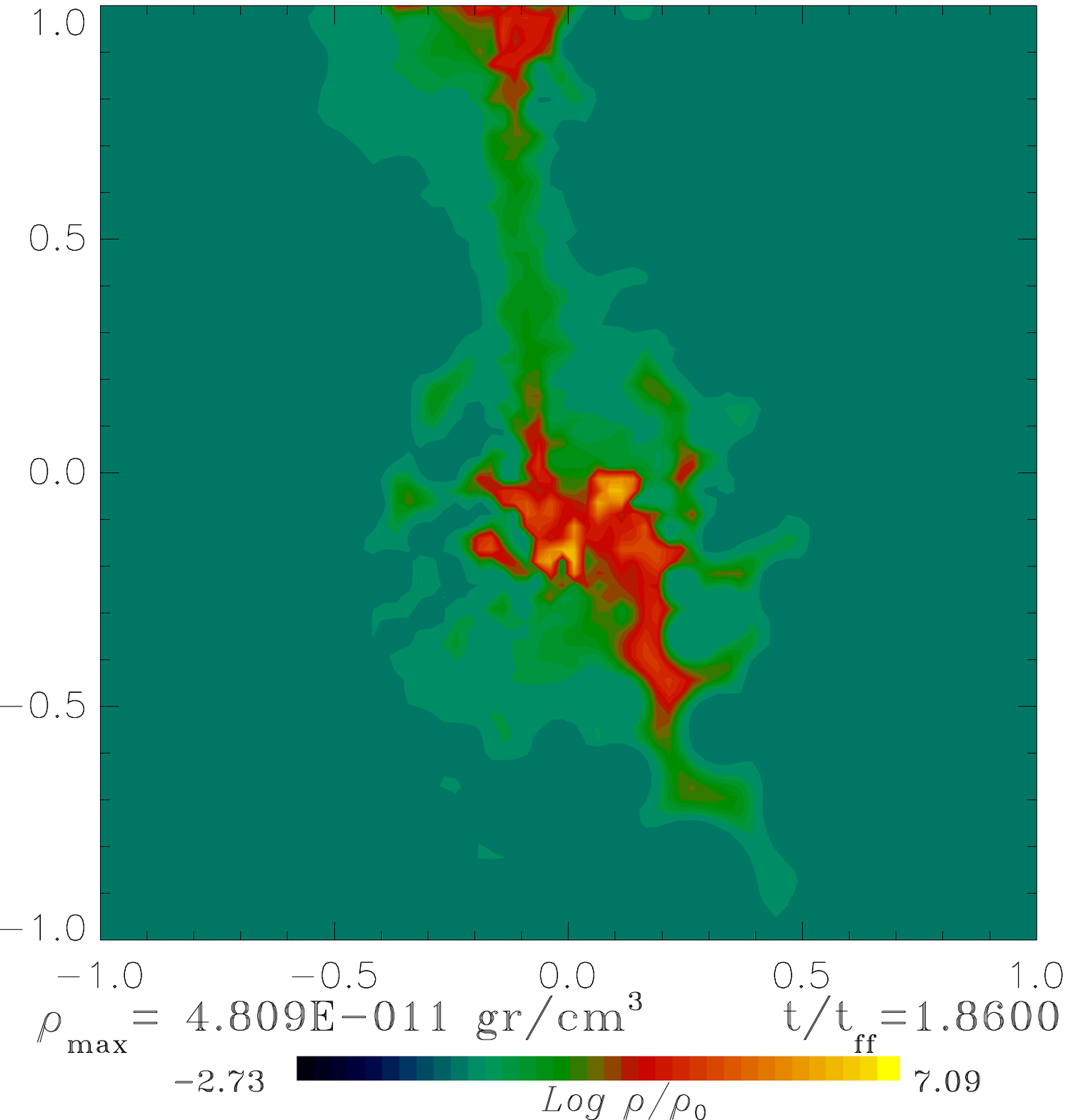}
\end{tabular}
\caption{\label{CPrueba70p1} Core models with M$_T$= 1 M$_{\odot}$, $a=$0.25 and $\alpha=$0.1;
the corresponding $\beta$ are
(left) 0.1
(middle) 0.3
(right) 0.48.}
\end{center}
\end{figure}

When the mass perturbation is increased from $a=$0.1 to $a=$0.25,
we can see in Fig.\ref{CPrueba70p1} and in panel 4 of Fig.\ref{lastcolconfM1}
that a similar behavior is
obtained to that which was seen in Fig. \ref{CPrueba60p1}, but
the binary separations are quite larger than that seen in the
corresponding panels of Fig.\ref{CPrueba60p1}.

We also noted that the number of orbiting binary systems is greatly
increased, as can be seen in Figs.\ref{CPrueba70p1} and \ref{CPrueba70p2}; only the
last collapsing configuration finishes as a central primary, as can be seen
in panel 6 of Fig.\ref{lastcolconfM1}.

%%%%%%%%%%%%%%%%%%%%%%%%%%%%%%%%%%%%%
%%%%%%%%%%%%%%%%%%%%%%%%%%%%%%%%%%%%%
%%%%%%%%%%%%%%%%%%%%%%%%%%%%%%%%%%%%%%
\subsection{The models with M$_T$= 5 M$_{\odot}$}
\label{subsec:rescore2}

\begin{figure}
\begin{center}
\begin{tabular}{ccc}
\includegraphics[width=2.2in]{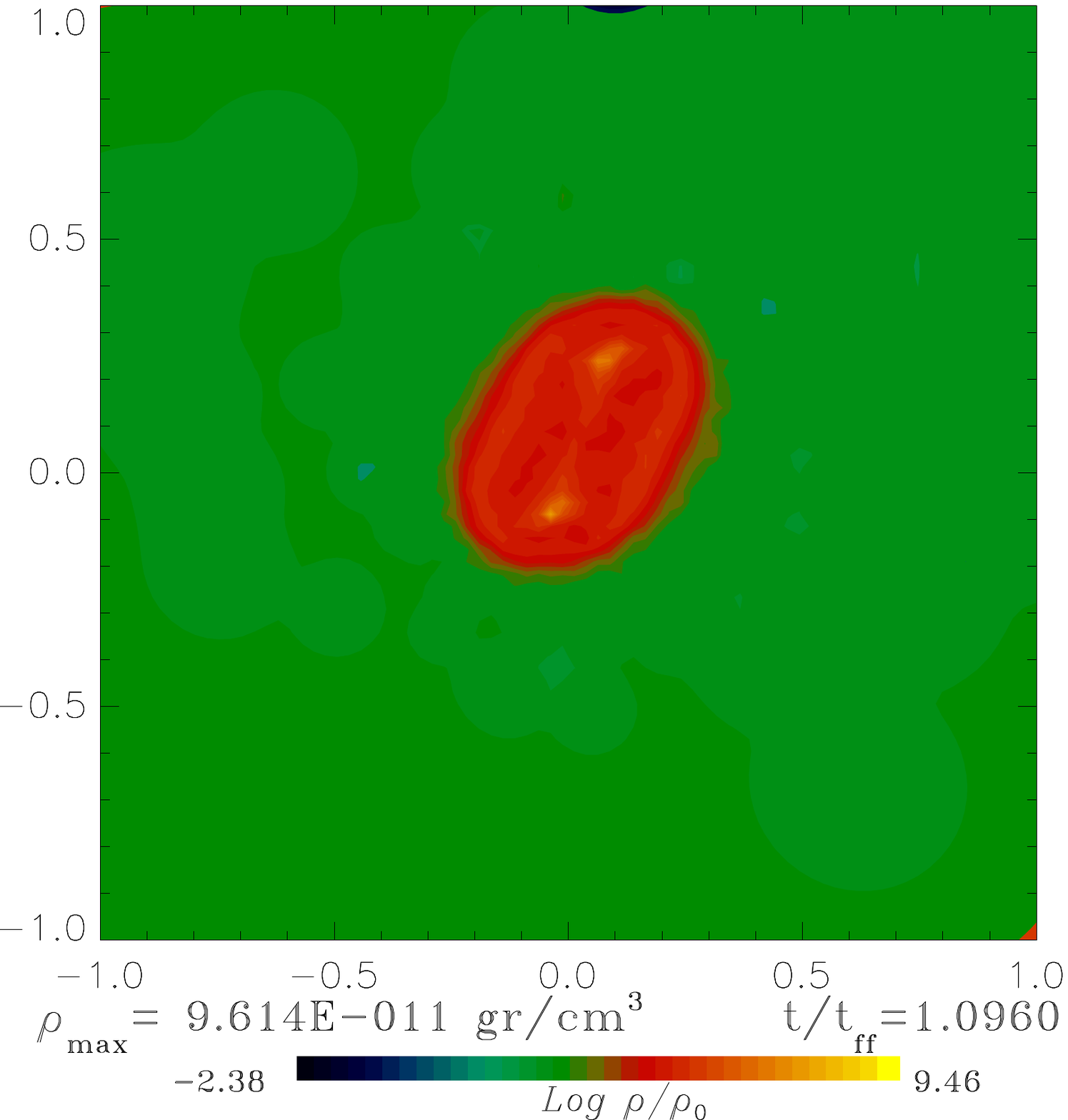} & \includegraphics[width=2.2in]{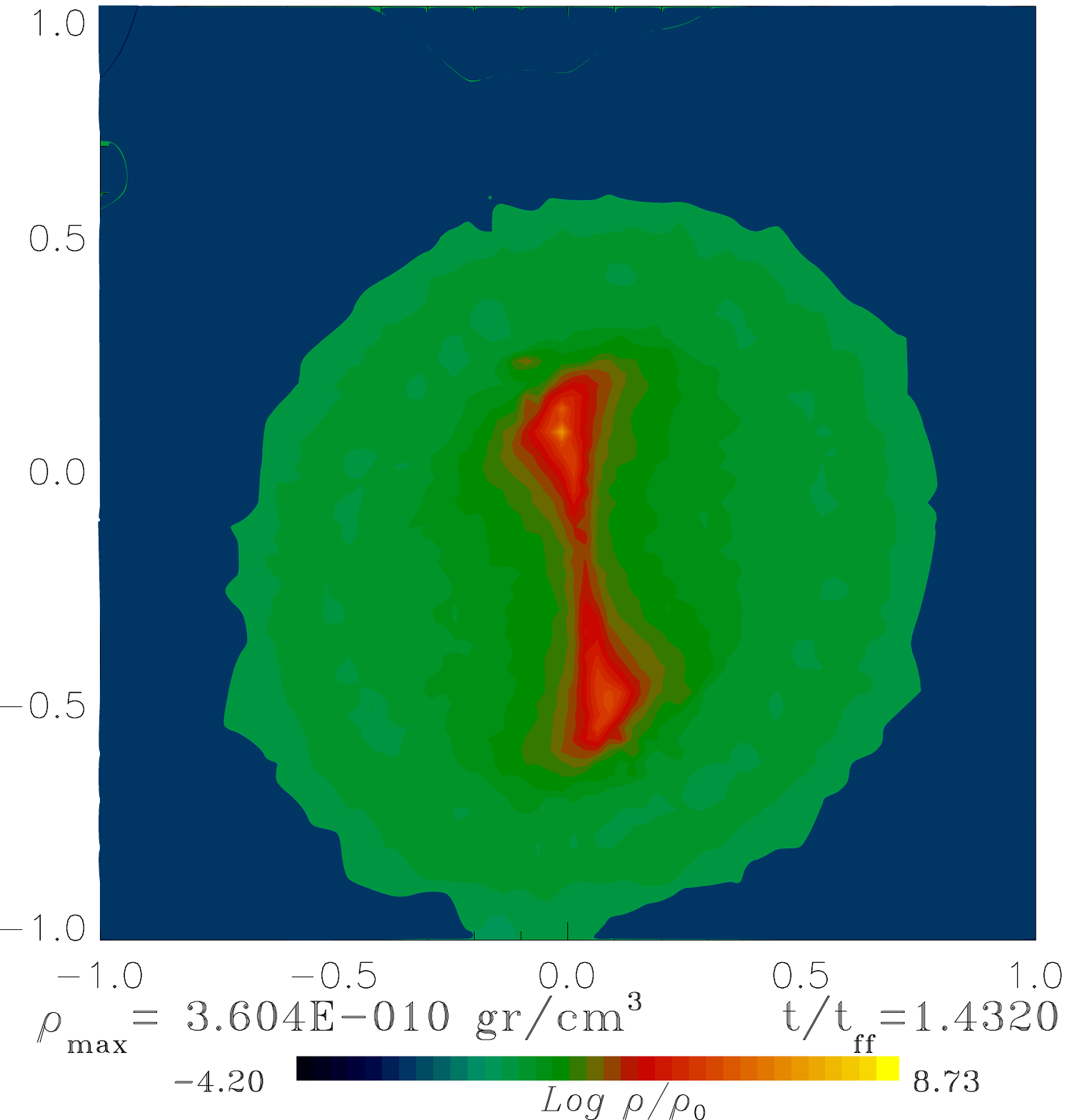} &
\includegraphics[width=2.2in]{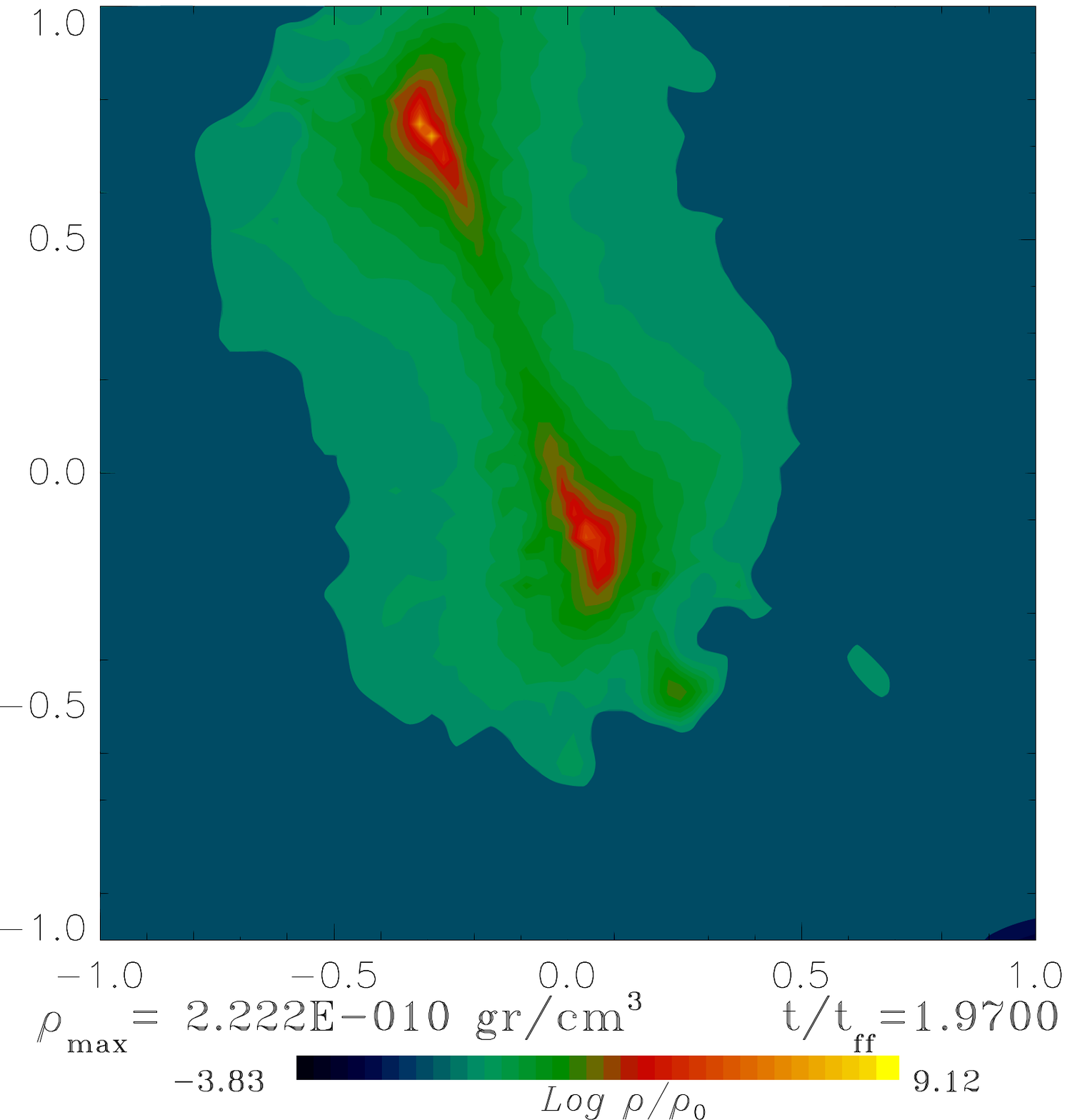} \\
\end{tabular}
\caption{\label{CPrueba30p1} Core models with M$_T$=5 M$_{\odot}$, $a=$0.1
and $\alpha=$0.1; the corresponding $\beta$ are
(left) 0.1
(middle) 0.3
(right) 0.48.}
\end{center}
\end{figure}

\begin{figure}
\begin{center}
\begin{tabular}{cc}
\includegraphics[width=2.2in]{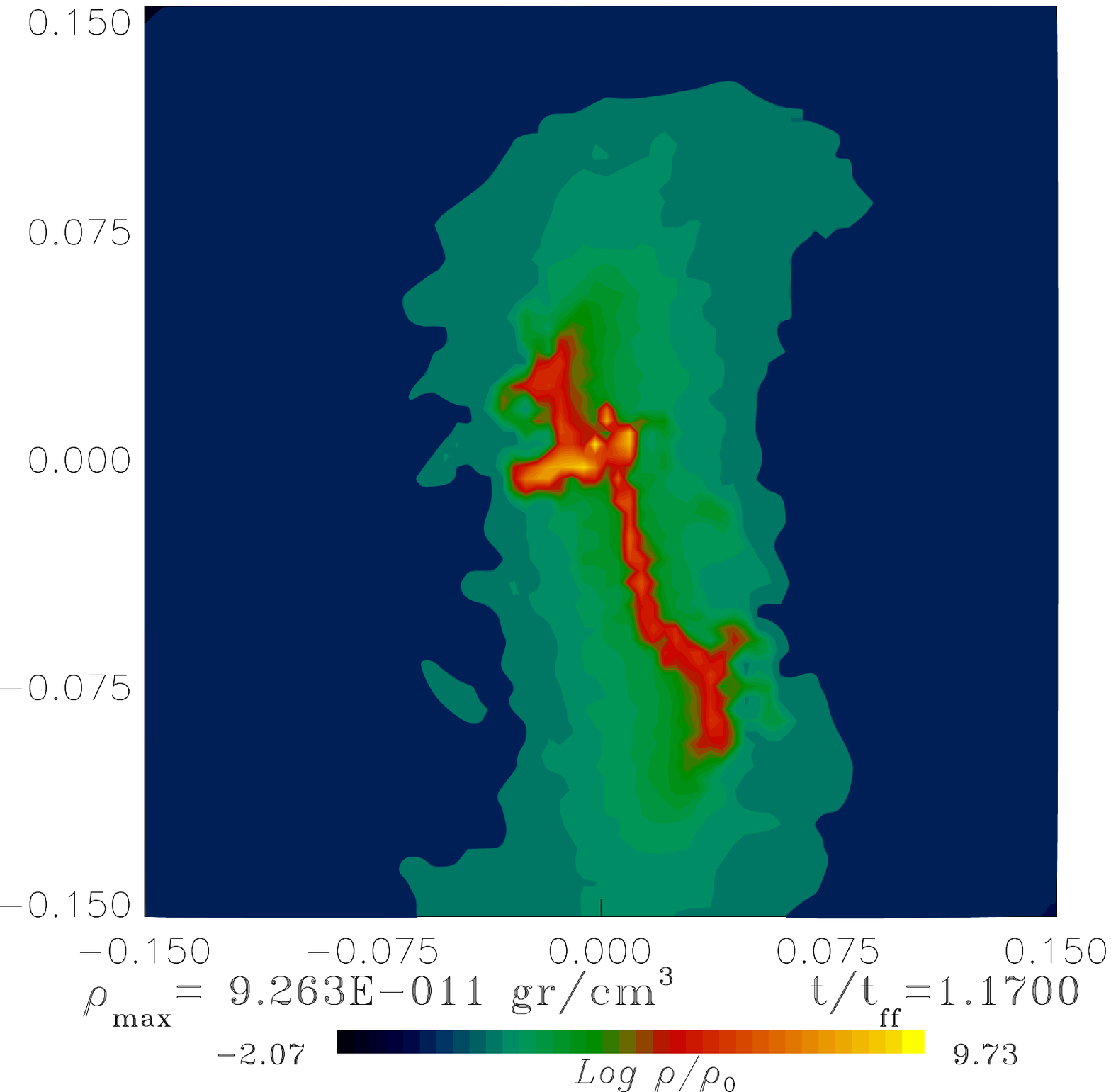} & \includegraphics[width=2.2in]{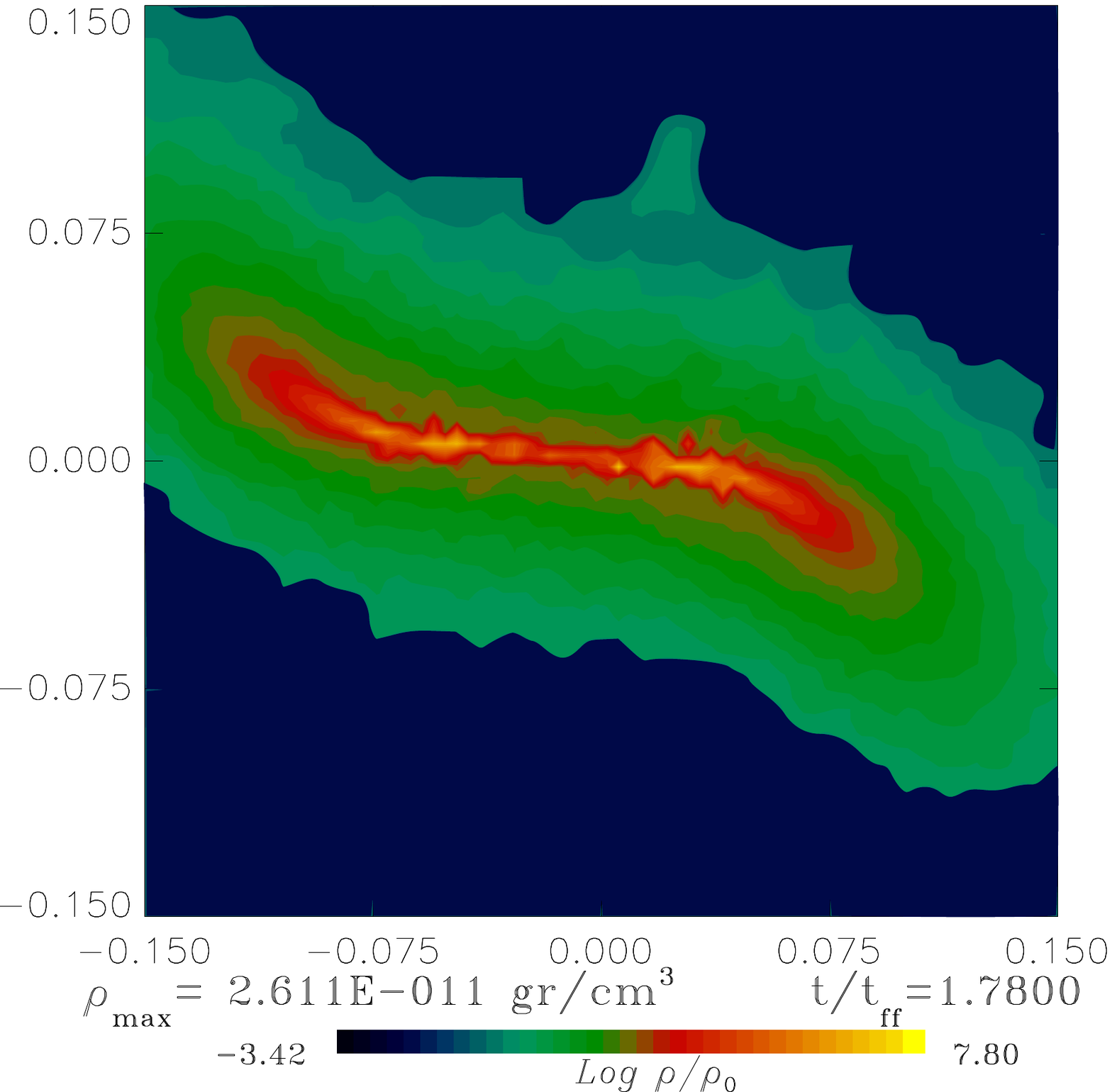}
\end{tabular}
\caption{\label{CPrueba30p3} Core models with M$_T$= 5 M$_{\odot}$, $a=$0.1 and $\alpha=$0.3;
the corresponding $\beta$ are
(left) 0.1
(right) 0.3.}
\end{center}
\end{figure}

For all the values of $\beta$, we see that a pair of mass
condensations connected by a very smooth filament are formed, as can
be seen in Fig.\ref{CPrueba30p1}. When $\alpha$ is increased
from 0.1 to 0.2, the mass condensations are very weak or they even
disappear, while the filaments are thinner and more pronounced; see
Fig. \ref{CPrueba30p2}. For the highest value of $\alpha=$0.3 considered
here, we see in Fig.\ref{CPrueba30p3} that the corresponding
filaments become thinner and narrower and again a rotationally
supported central primary surrounded by long spiral arms is seen as
the last collapsing configuration for the highest value of $\beta$.

\begin{figure}
\begin{center}
\begin{tabular}{ccc}
\includegraphics[width=2.2in]{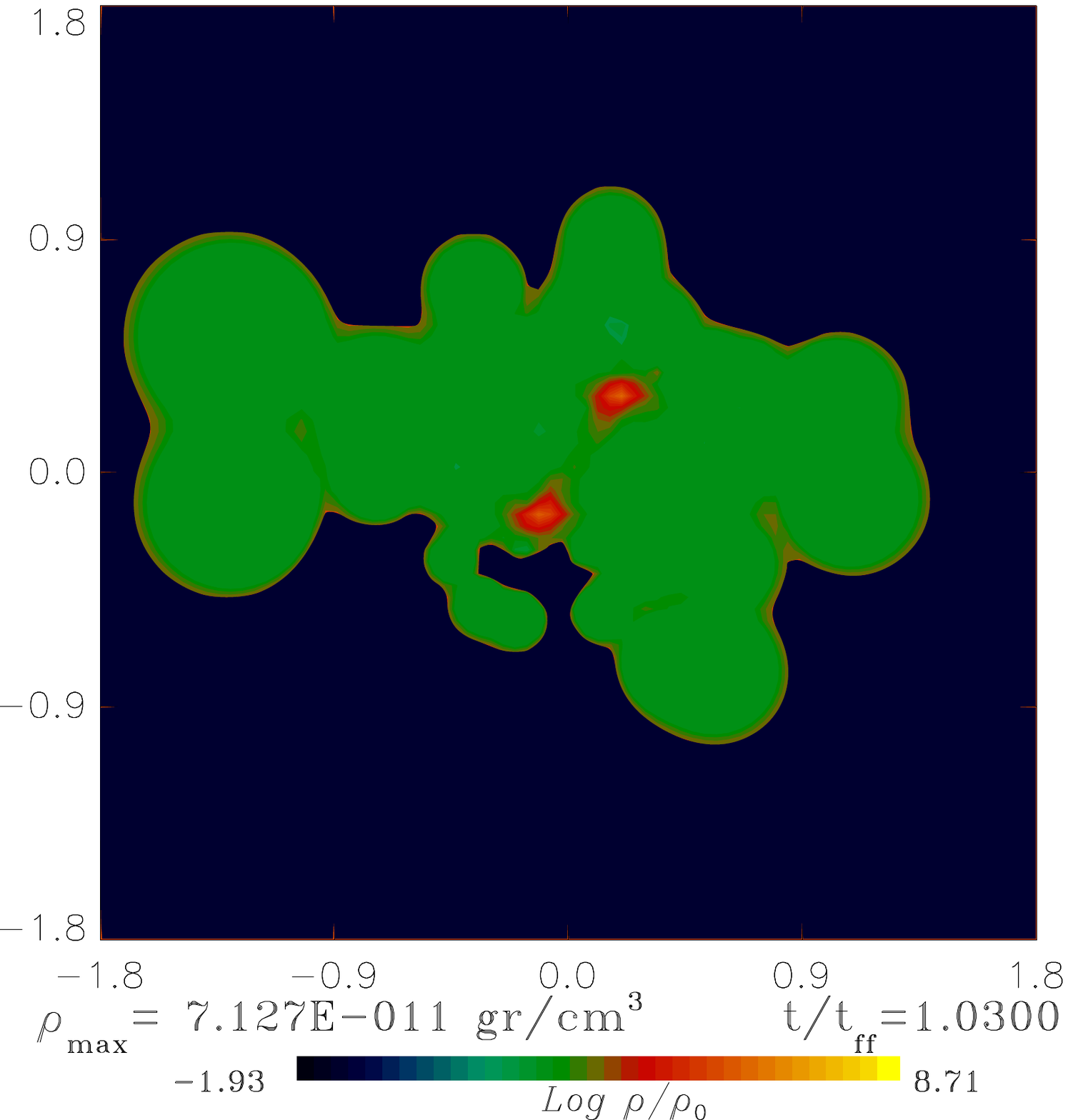} & \includegraphics[width=2.2in]{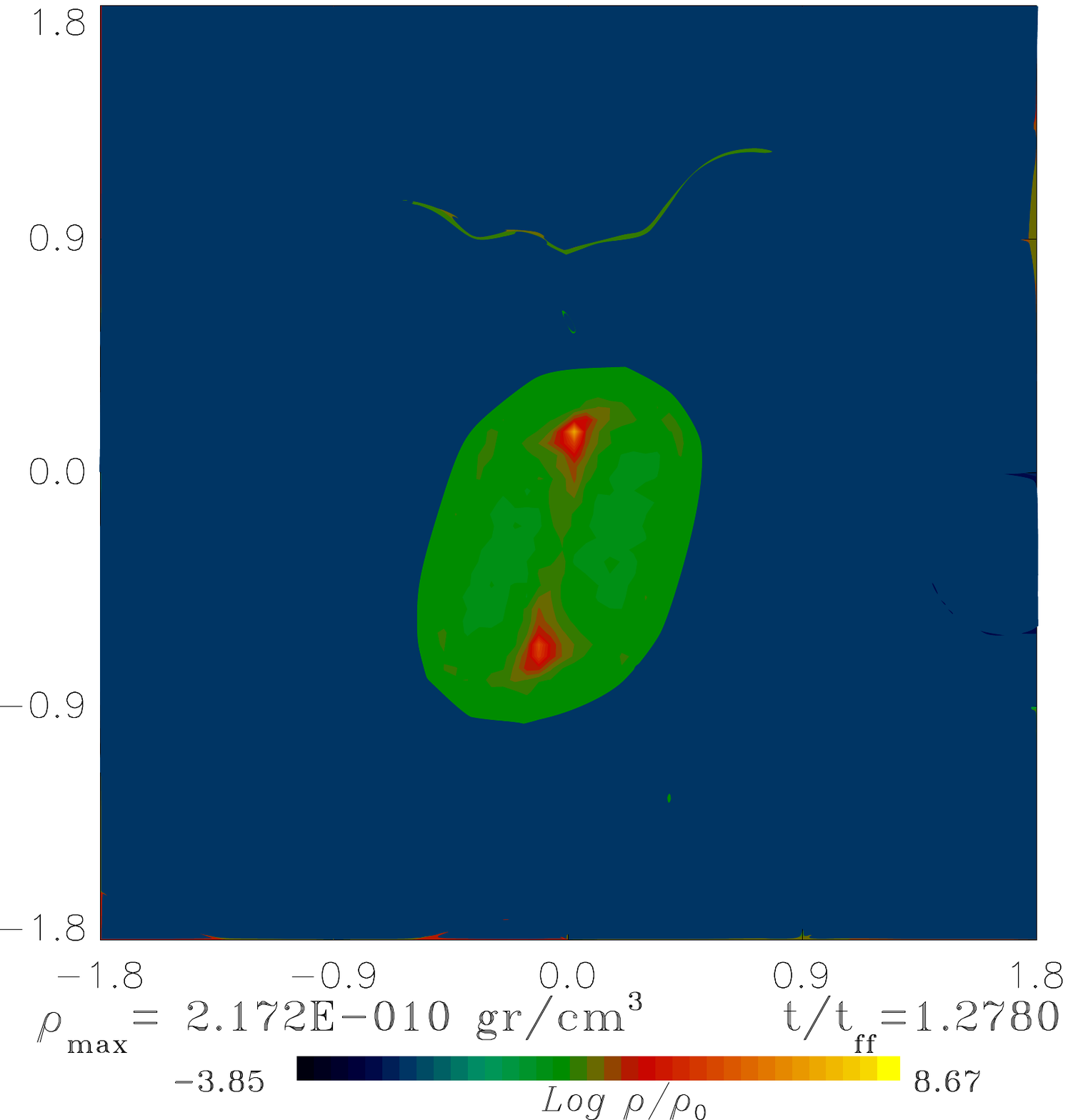}&
\includegraphics[width=2.2in]{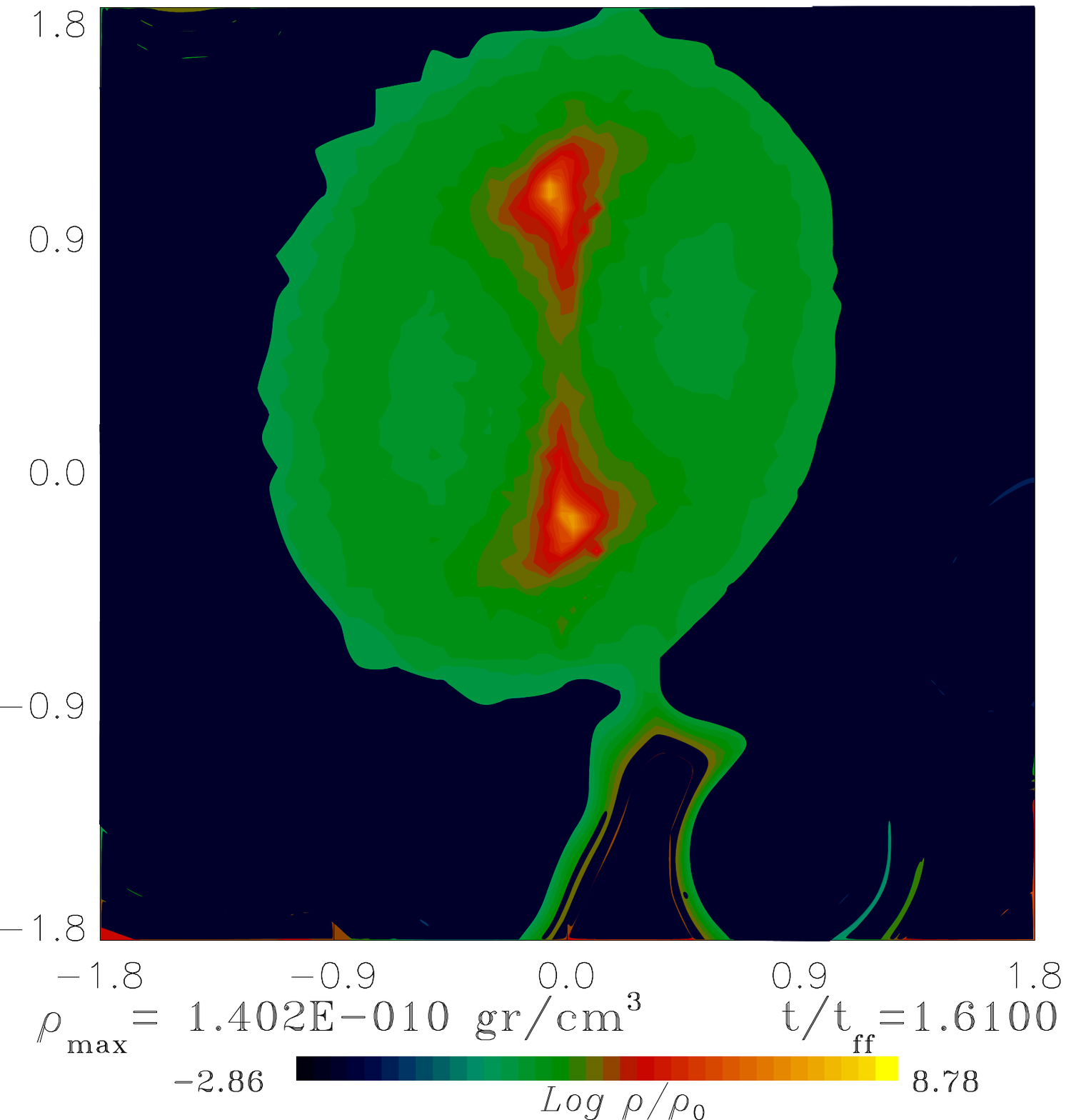} \\
\end{tabular}
\caption{\label{CPrueba40p1} Core models with M$_T$=5 M$_{\odot}$, $a=$0.25
and $\alpha=$0.1; the corresponding $\beta$ are
(left) 0.1
(middle) 0.3
(right) 0.48.}
\end{center}
\end{figure}

When the perturbation scale takes the new value of $a=0.25$, the
filaments almost disappear and instead a pair of mass
condensations are seen located face-to-face and more
pronounced than before, as one can see in
Fig.\ref{CPrueba40p1}. Later, when $\alpha$ increases to 0.2, the mass condensations
are weaker but still face-to-face, while no dense filaments are seen in
Fig.\ref{CPrueba40p2}. We observe again that the binary separation is
significantly increased in these runs, shown in Fig.\ref{CPrueba40p1} and
Fig.\ref{CPrueba40p2}, as compared with those of Fig.\ref{CPrueba30p1} and
Fig.\ref{CPrueba30p2}.

\begin{figure}
\begin{center}
\begin{tabular}{cc}
\includegraphics[width=2.2in]{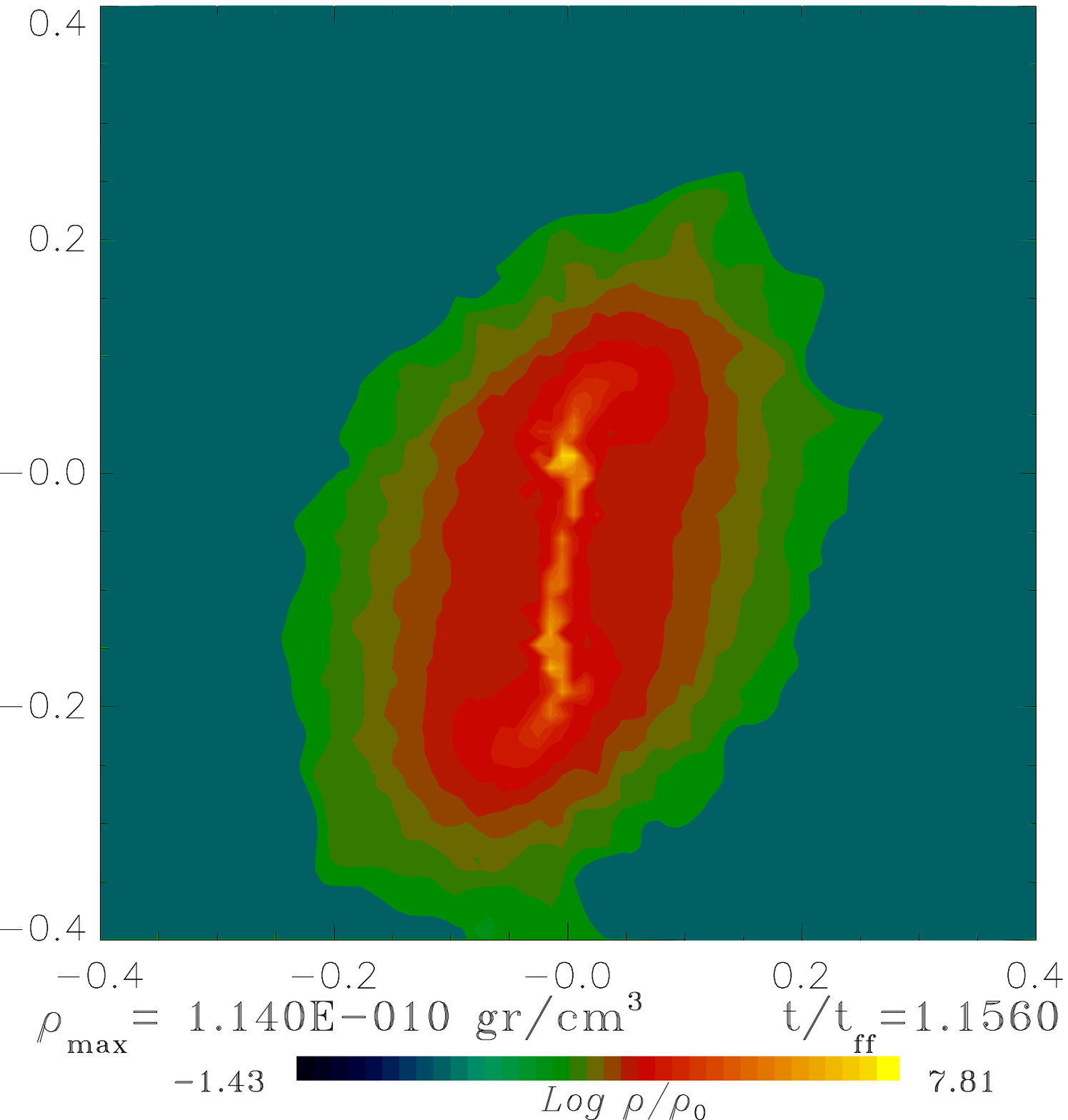} & \includegraphics[width=2.2in]{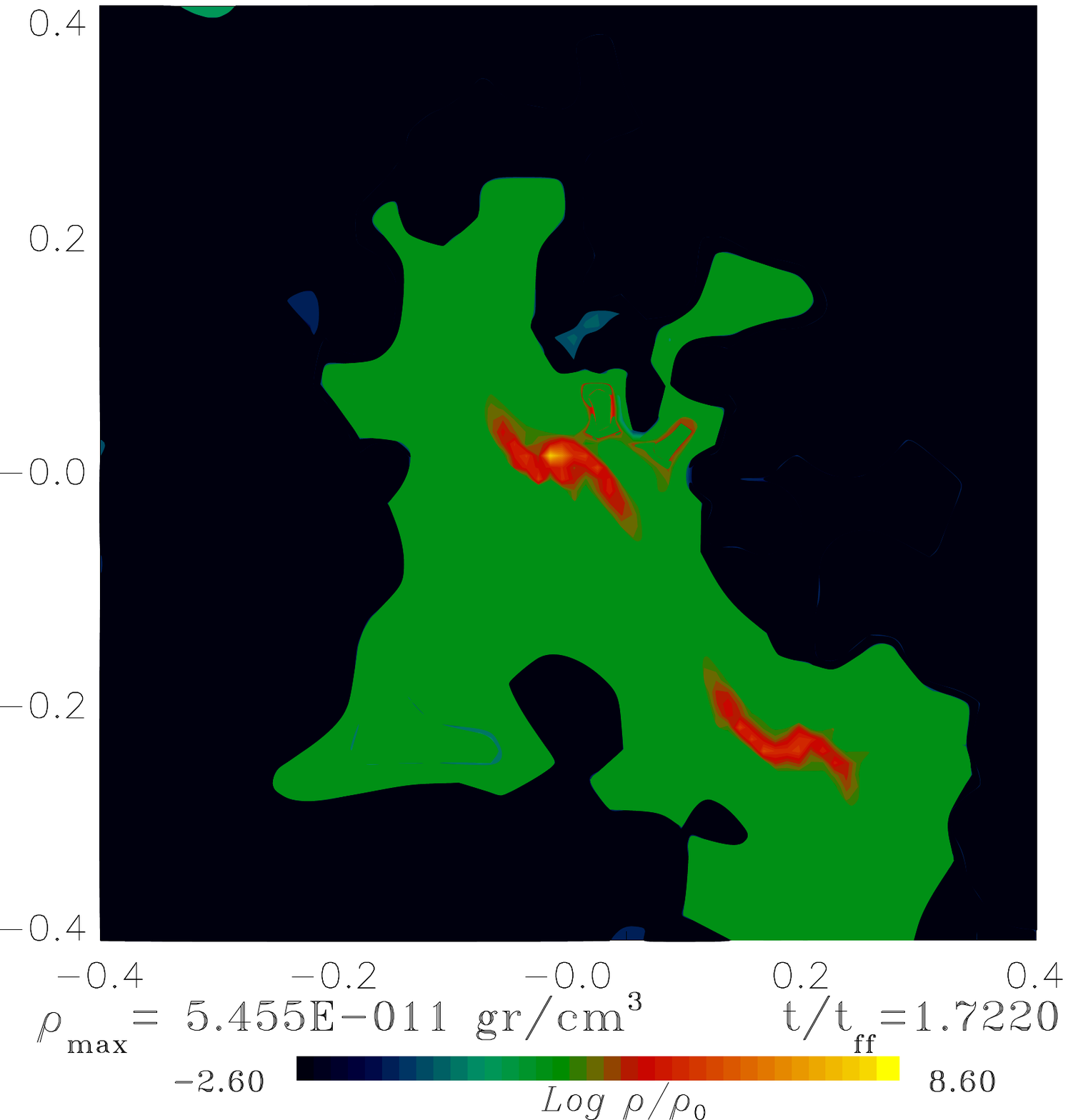}
\end{tabular}
\caption{\label{CPrueba40p3} Core models with M$_T$= 5 M$_{\odot}$, $a=$0.25 and $\alpha=$0.3;
the corresponding $\beta$ are
(top left) 0.1
(top right) 0.3.}
\end{center}
\end{figure}

For the highest $\alpha=$0.3 value considered here, we see
in Fig.\ref{CPrueba40p3} the appearance of weaker filaments and
again the transition to a central primary.

In any run of the models with M$_T$= 5 M$_{\odot}$ we have not directly observed the
formation of an orbiting binary system.
%%%%%%%%%%%%%%%%%%%%%%%%%%%%%%%%%%%%%%%
%%%%%%%%%%%%%%%%%%%%%%%%%%%%%%%%%%%%%%
\subsection{The clump models}
\label{subsec:resclump}

\begin{figure}
\begin{center}
\begin{tabular}{ccc}
\includegraphics[width=2.2in]{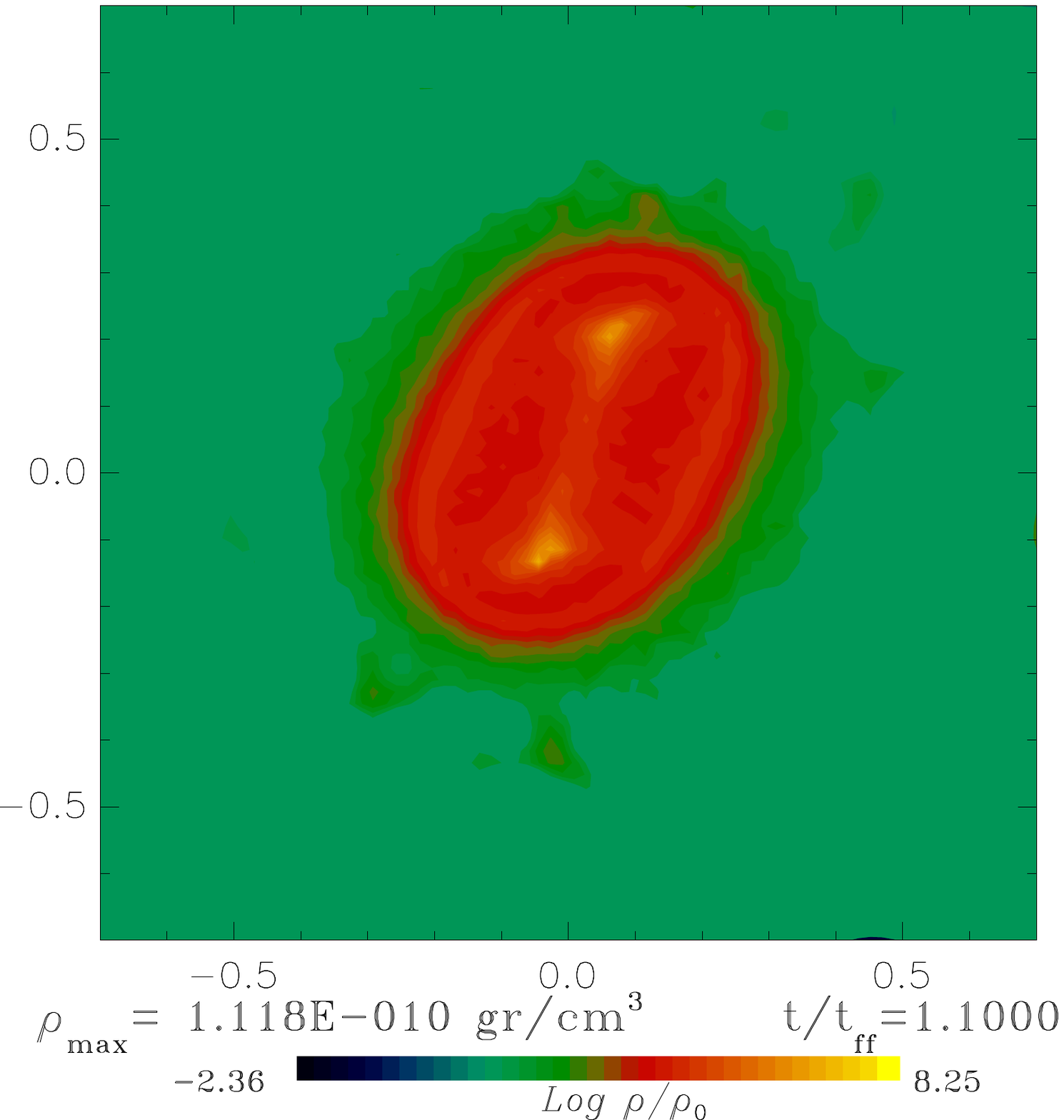} & \includegraphics[width=2.2in]{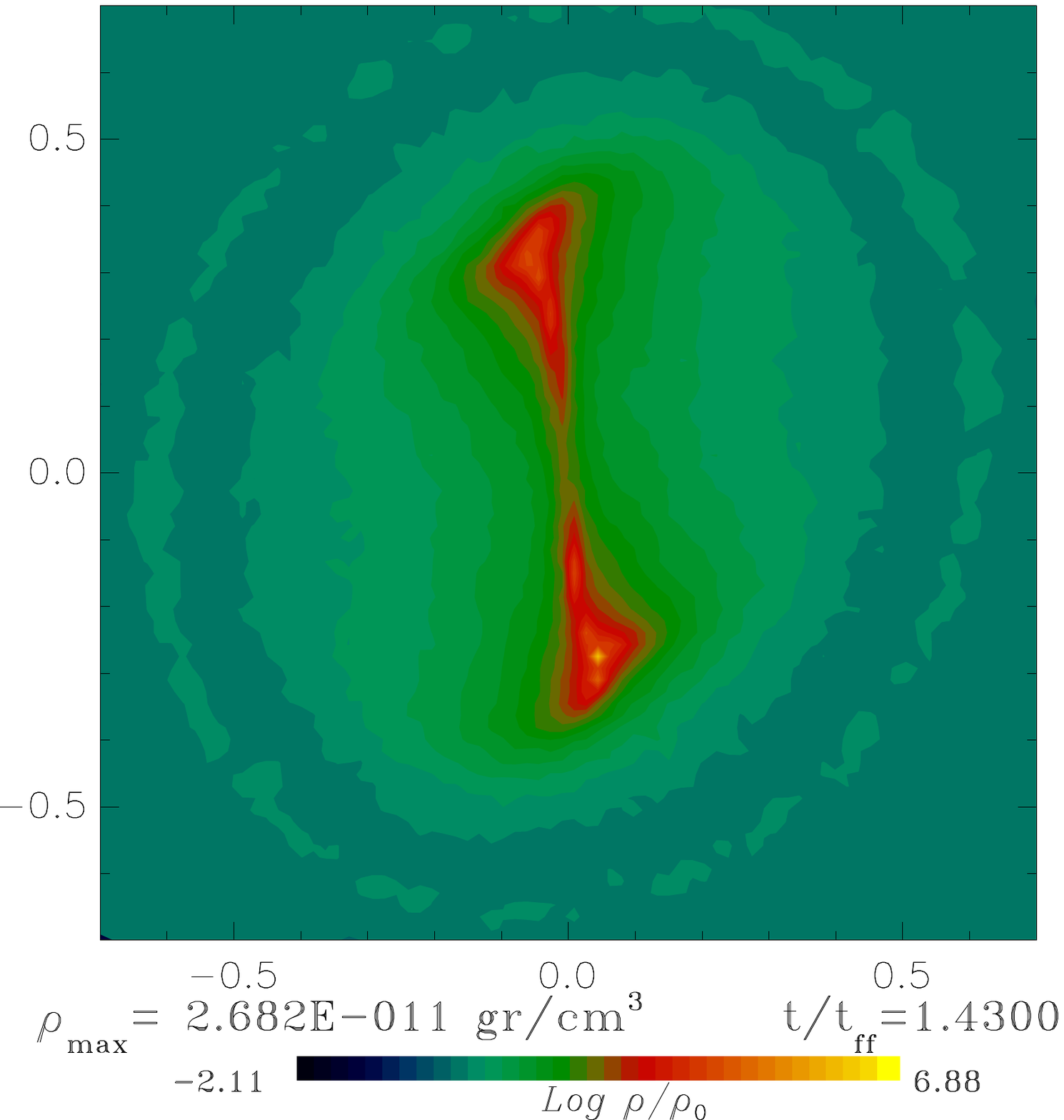} &
\includegraphics[width=2.2in]{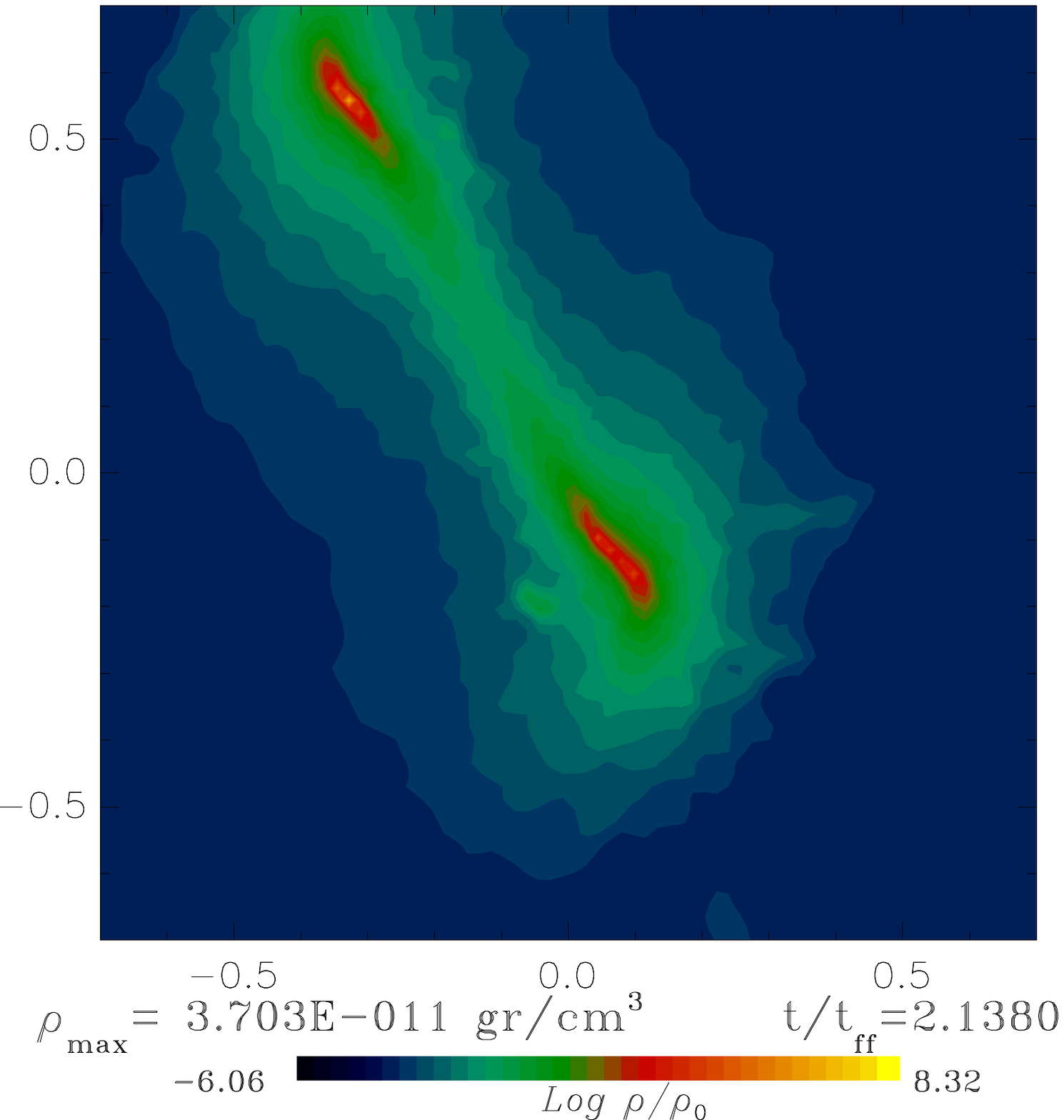} \\
\end{tabular}
\caption{\label{CPrueba80p1} Core models with M$_T$=50 M$_{\odot}$, $a=$0.1
and $\alpha=$0.1; the corresponding $\beta$ are
(left) 0.1
(middle) 0.3
(right) 0.48.}
\end{center}
\end{figure}

\begin{figure}
\begin{center}
\begin{tabular}{ccc}
\includegraphics[width=2.2in]{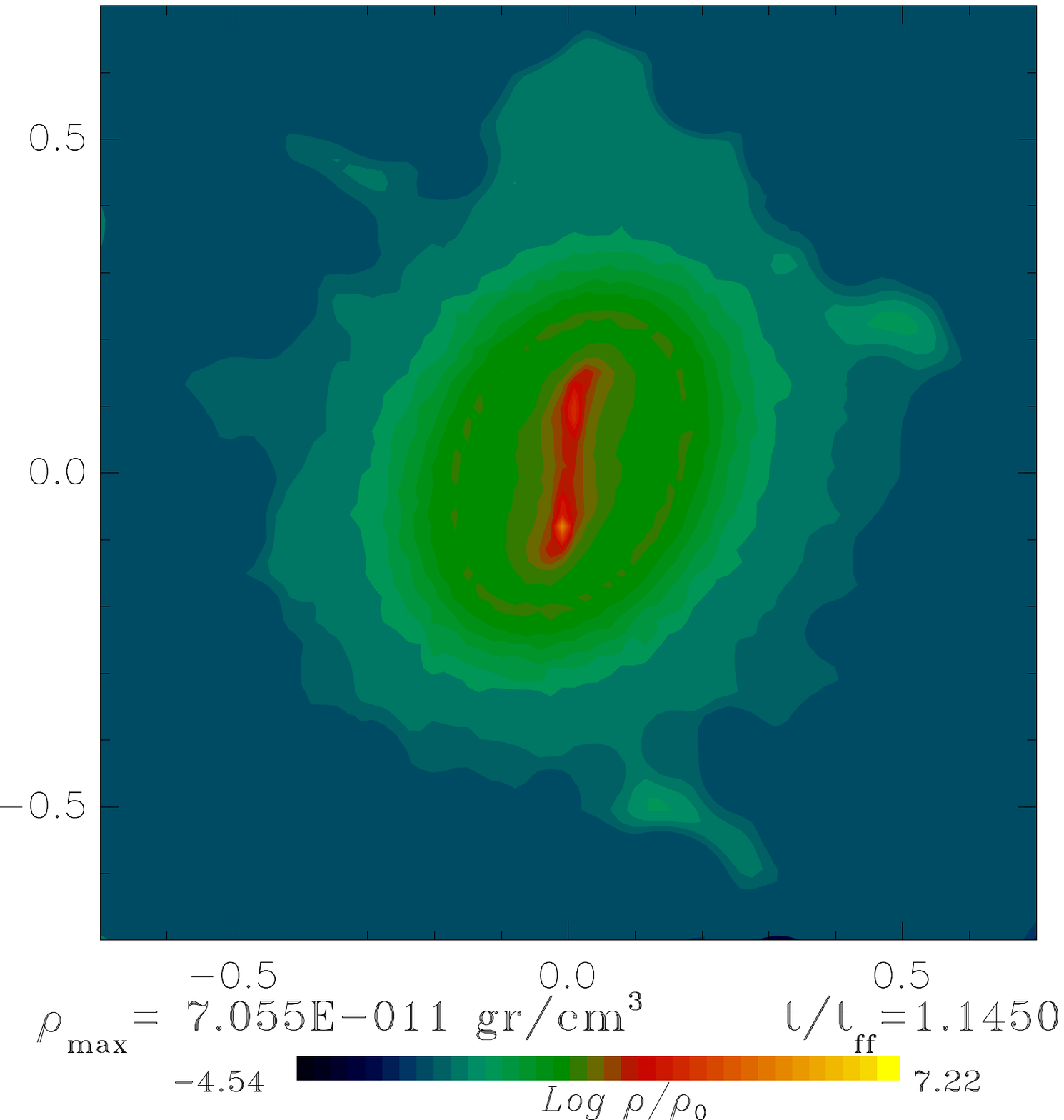} & \includegraphics[width=2.2in]{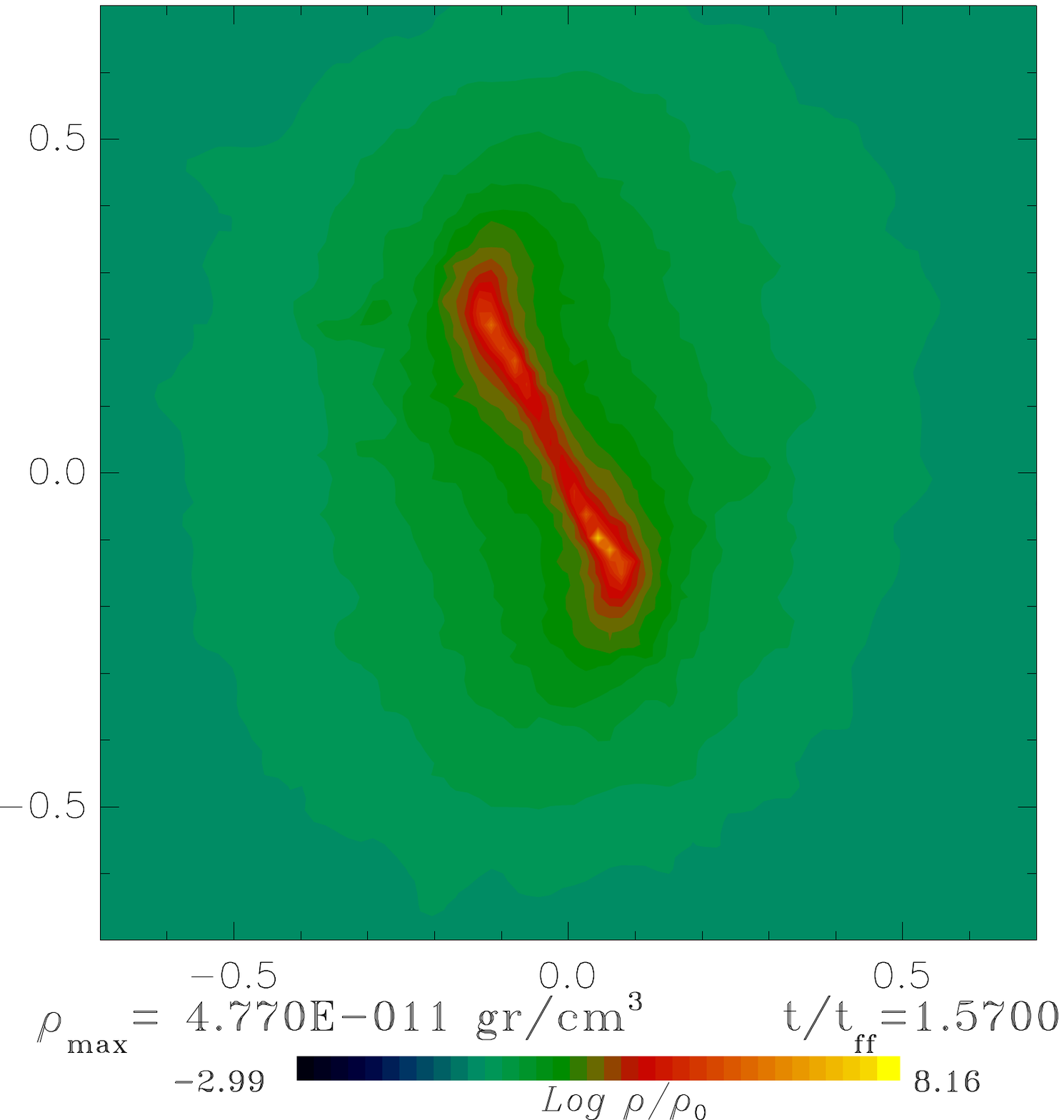} &
\includegraphics[width=2.2in]{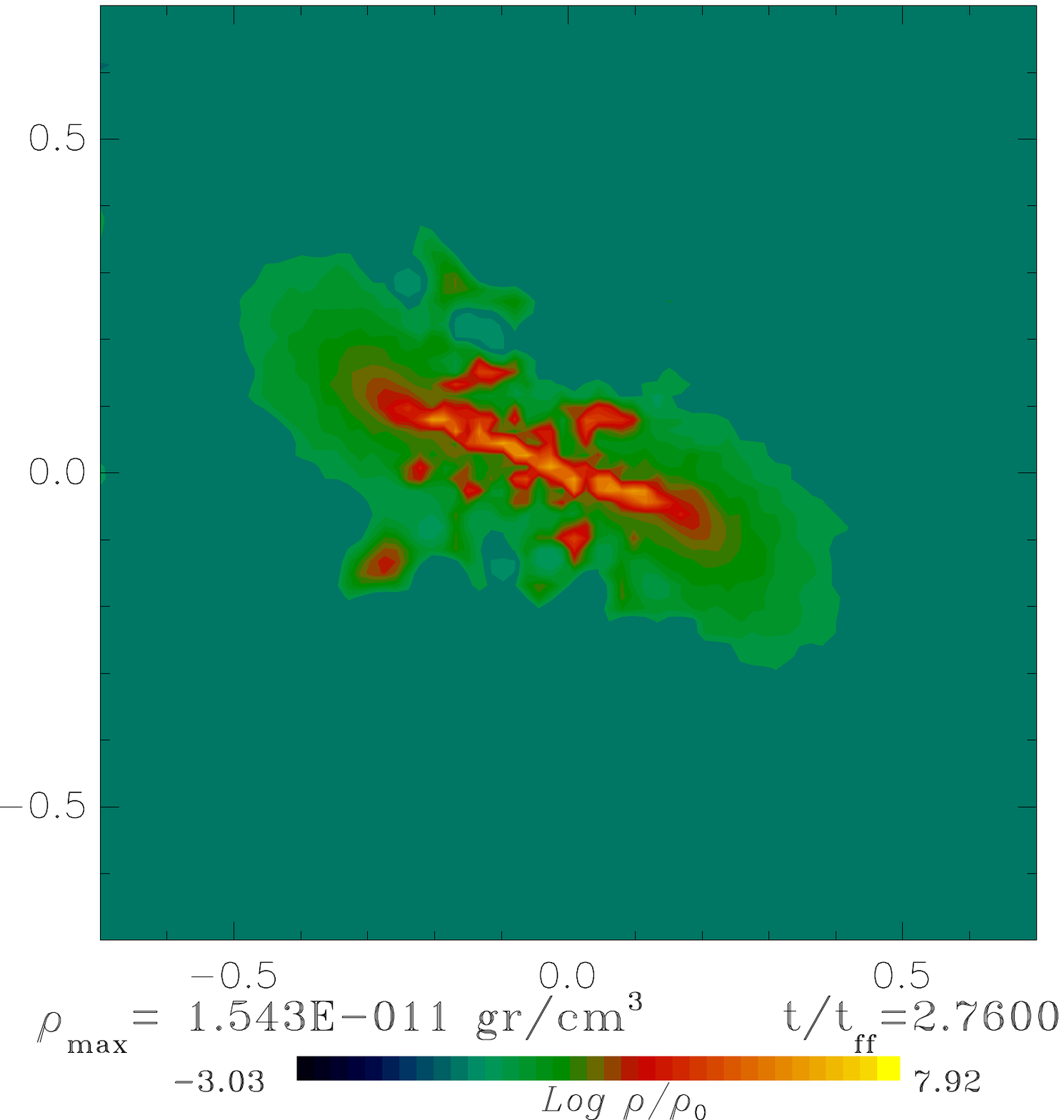} \\
\end{tabular}
\caption{\label{CPrueba80p2} Core models with M$_T$=50 M$_{\odot}$, $a=$0.1
and $\alpha=$0.2; the corresponding $\beta$ are
(left) 0.1
(middle) 0.3
(right) 0.48.}
\end{center}
\end{figure}

\begin{figure}
\begin{center}
\begin{tabular}{cc}
\includegraphics[width=2.2in]{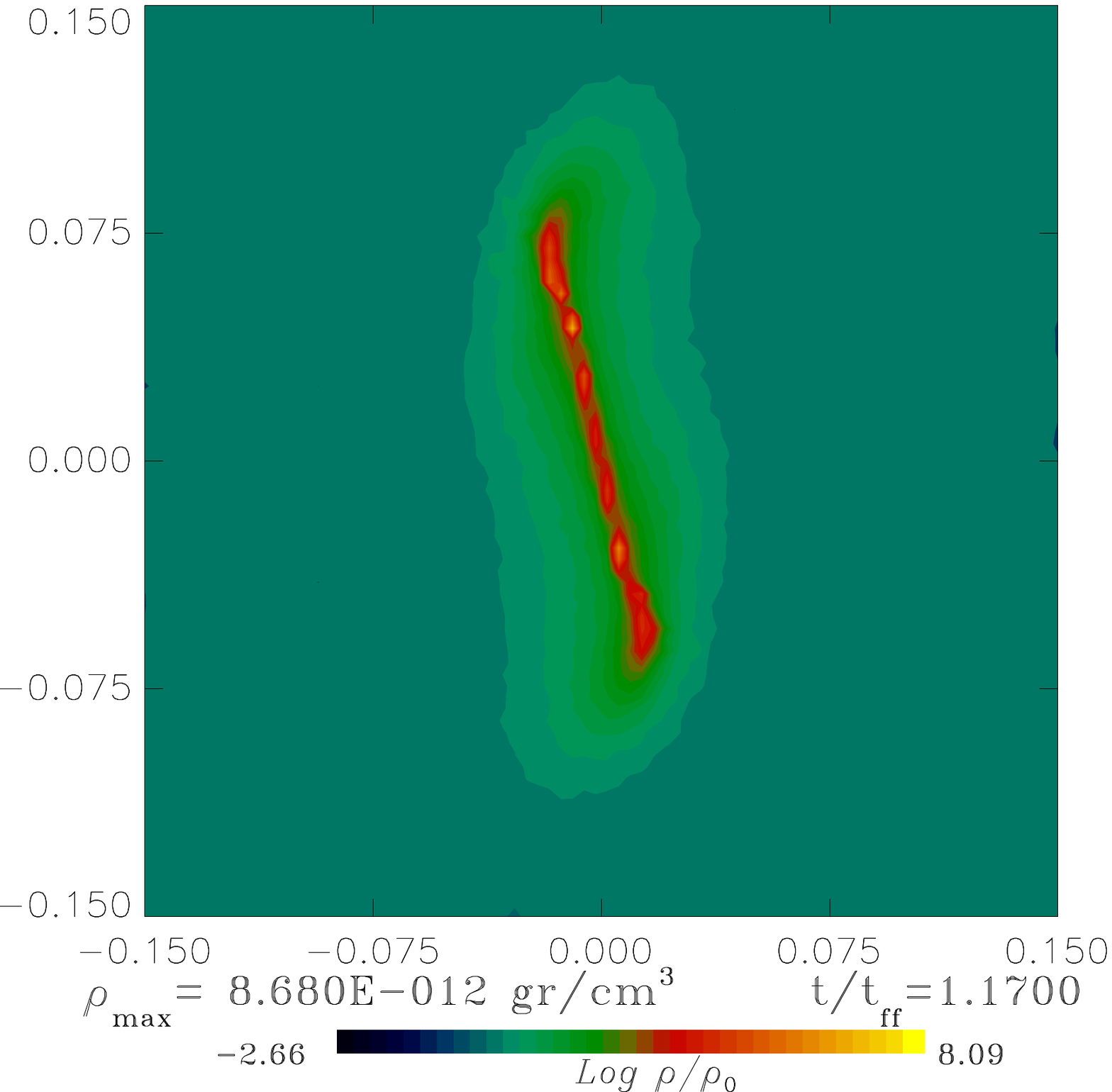} & \includegraphics[width=2.2in]{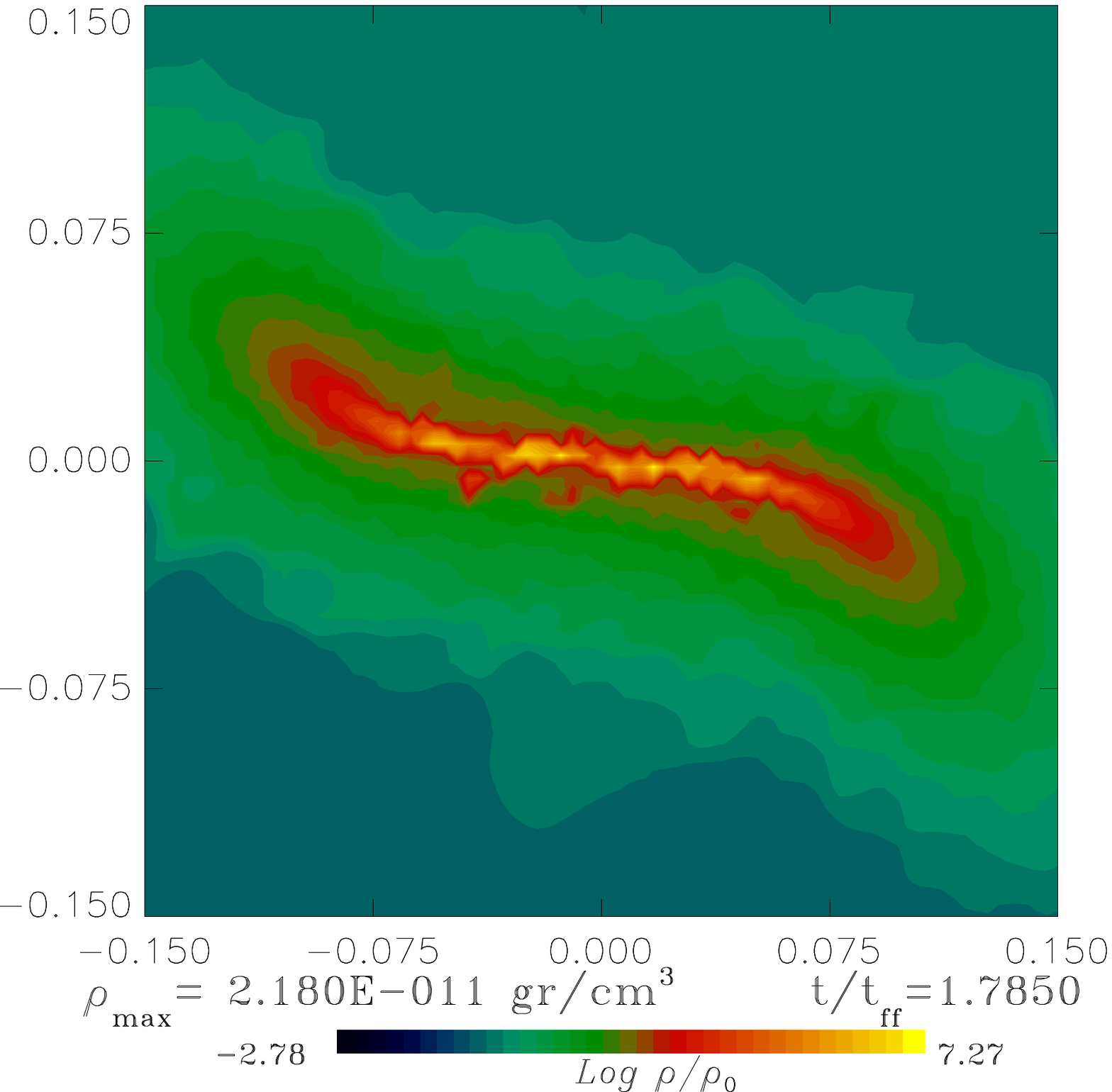} \\
\end{tabular}
\caption{\label{CPrueba80p3} Core models with M$_T$=50 M$_{\odot}$, $a=$0.1
and $\alpha=$0.3; the corresponding $\beta$ are
(left) 0.1
(middle) 0.3.}
\end{center}
\end{figure}

\begin{figure}
\begin{center}
\begin{tabular}{ccc}
\includegraphics[width=2.2in]{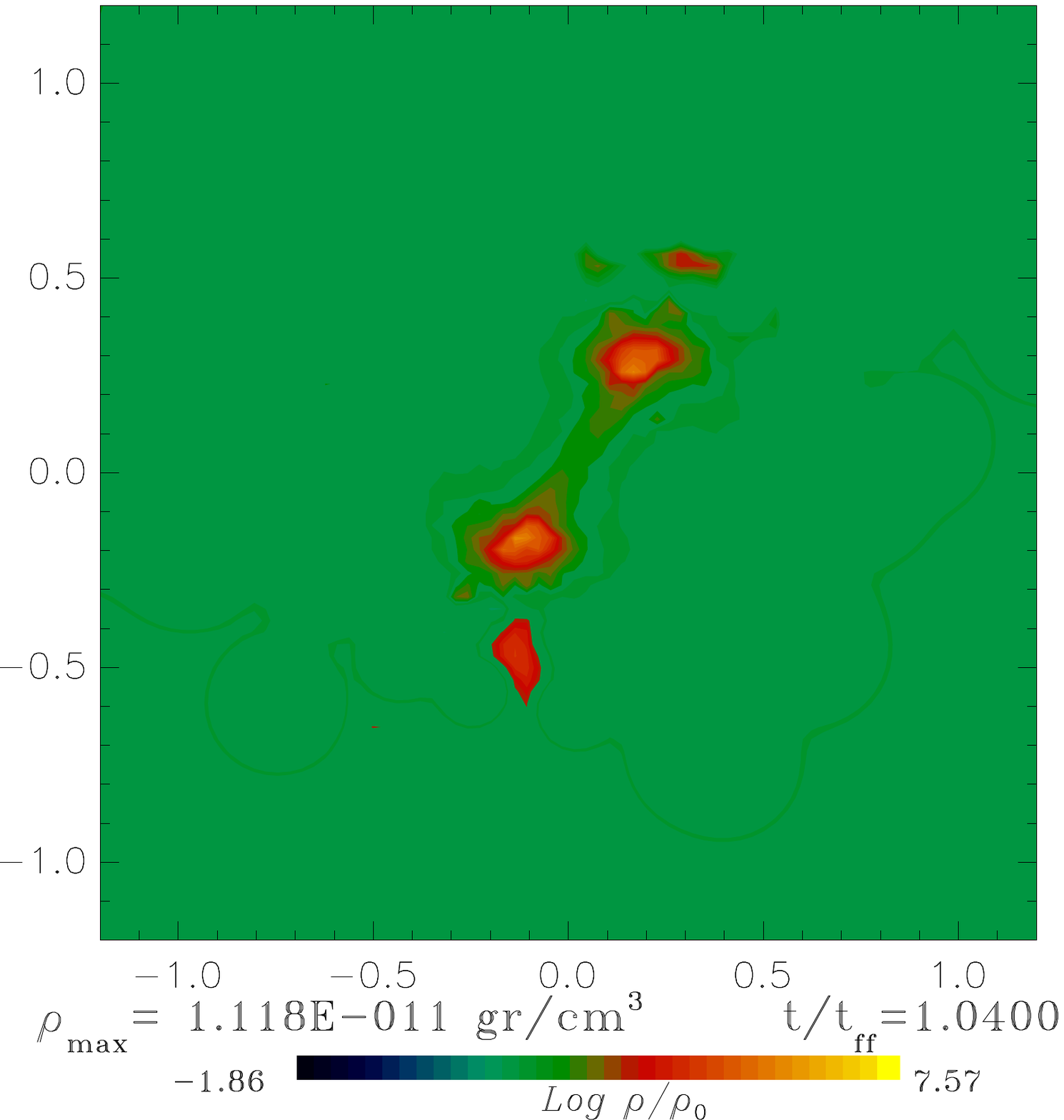} & \includegraphics[width=2.2in]{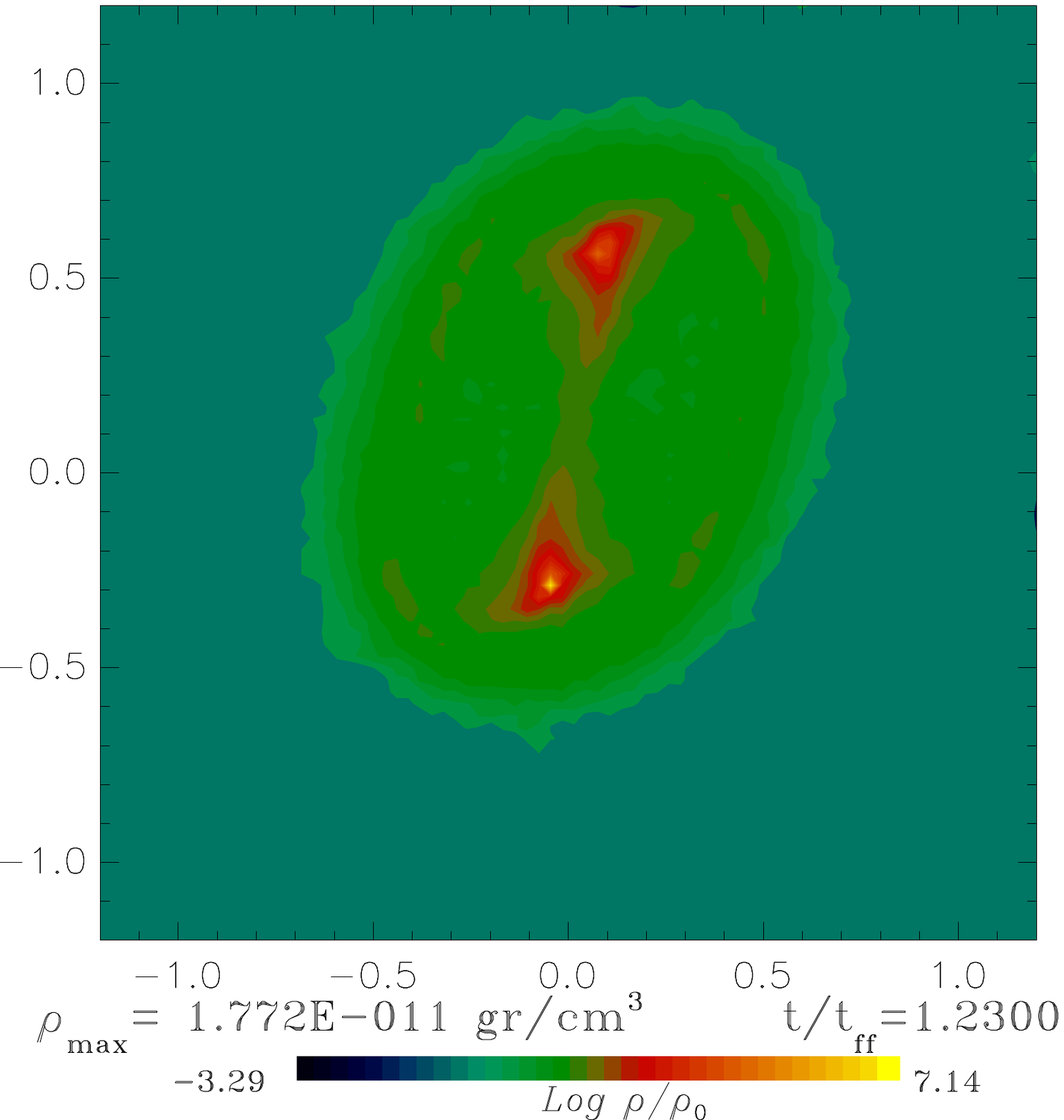} &
\includegraphics[width=2.2in]{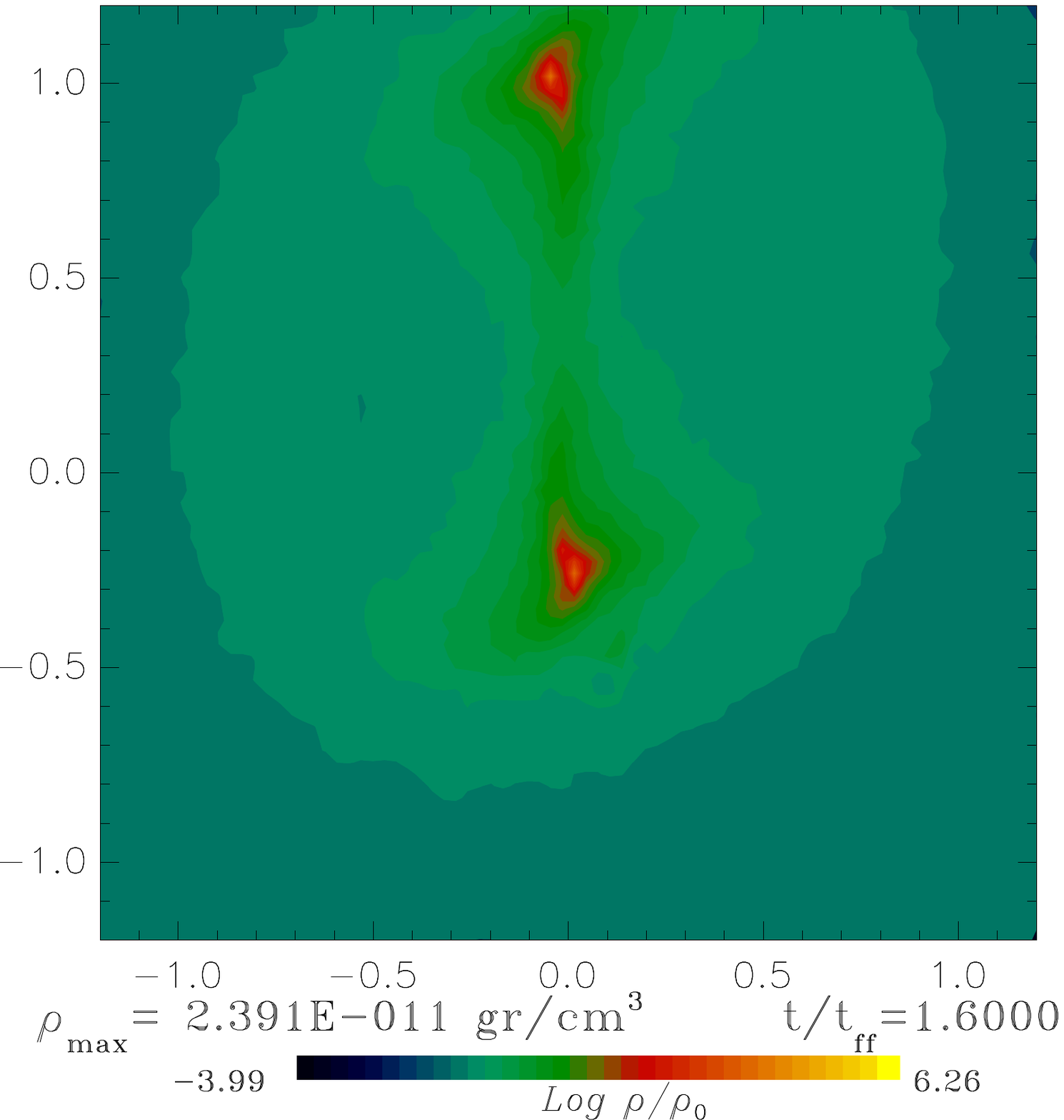} \\
\end{tabular}
\caption{\label{CPrueba90p1} Core models with M$_T$=50 M$_{\odot}$, $a=$0.25
and $\alpha=$0.1; the corresponding $\beta$ are
(left) 0.1
(middle) 0.3
(right) 0.48.}
\end{center}
\end{figure}

\begin{figure}
\begin{center}
\begin{tabular}{ccc}
\includegraphics[width=2.2in]{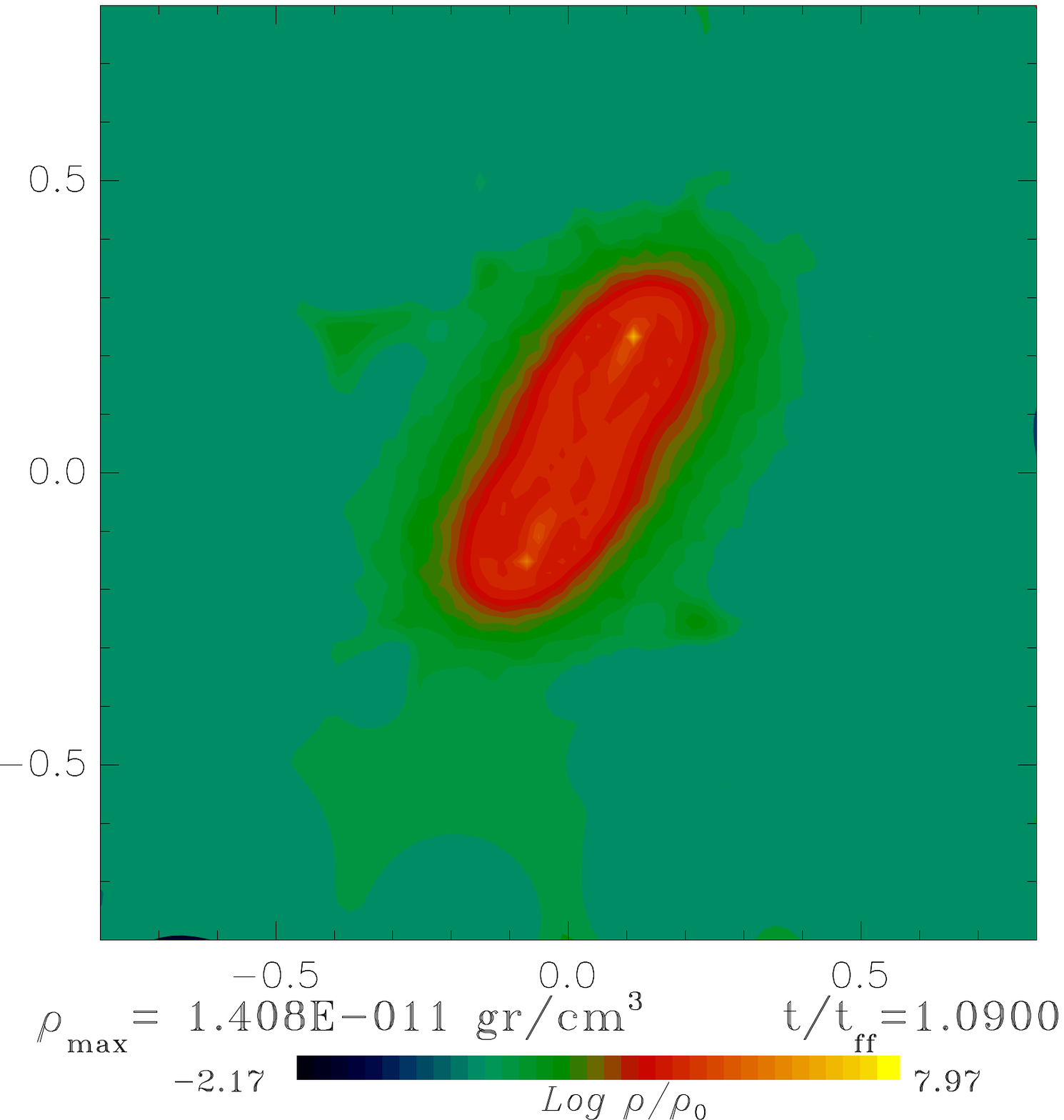} & \includegraphics[width=2.2in]{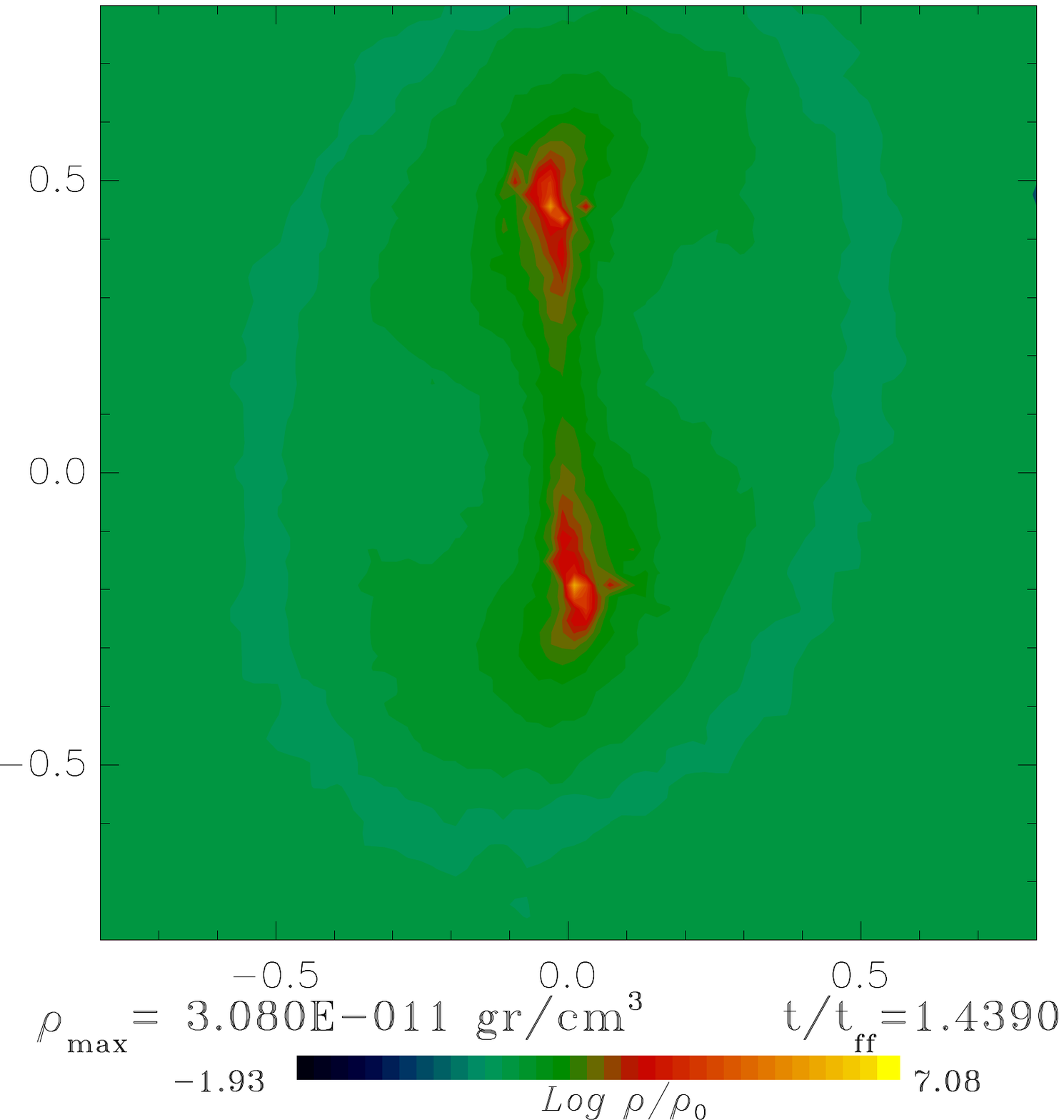} &
\includegraphics[width=2.2in]{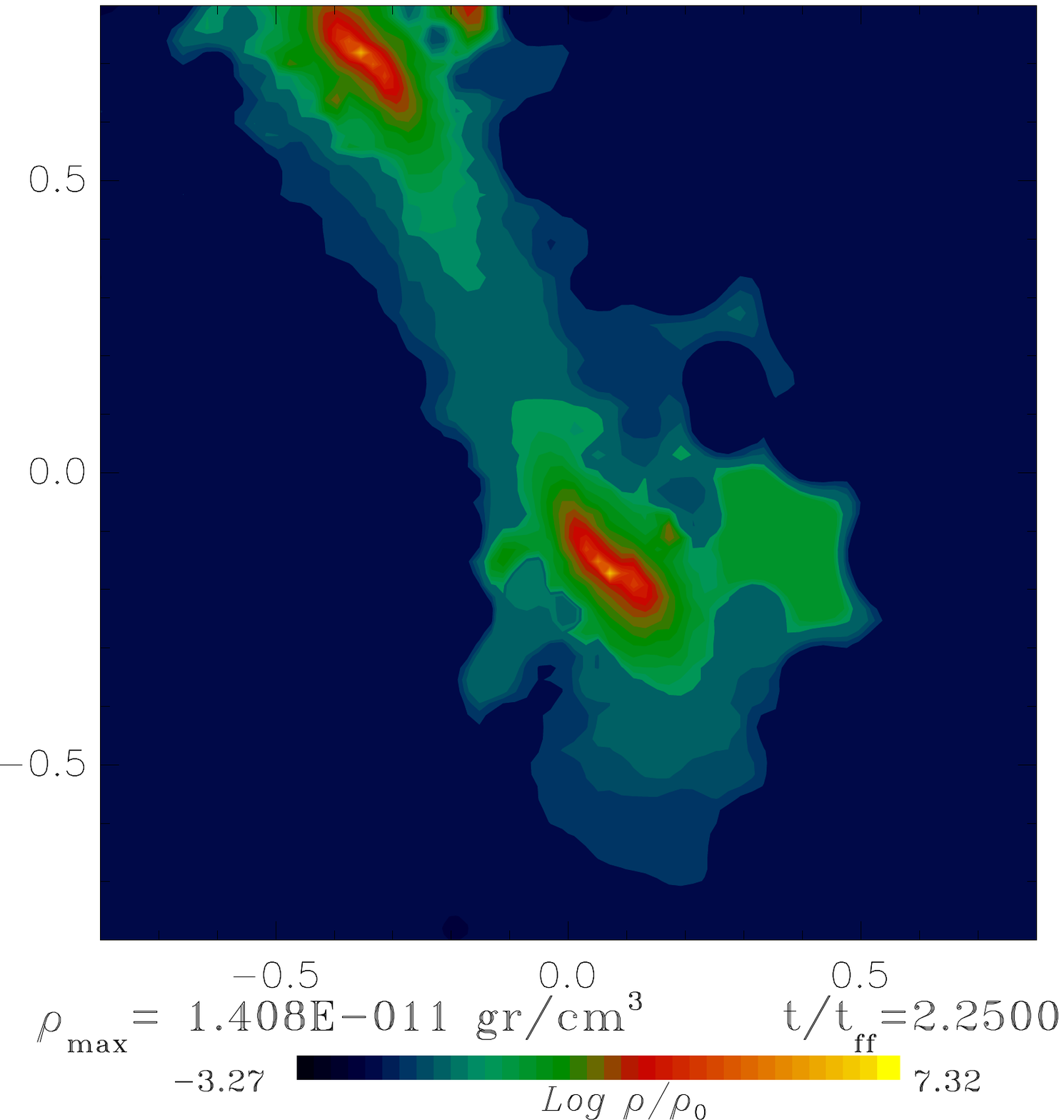} \\
\end{tabular}
\caption{\label{CPrueba90p2} Core models with M$_T$=50 M$_{\odot}$, $a=$0.25
and $\alpha=$0.2; the corresponding $\beta$ are
(left) 0.1
(middle) 0.3
(right) 0.48.}
\end{center}
\end{figure}

\begin{figure}
\begin{center}
\begin{tabular}{cc}
\includegraphics[width=2.2in]{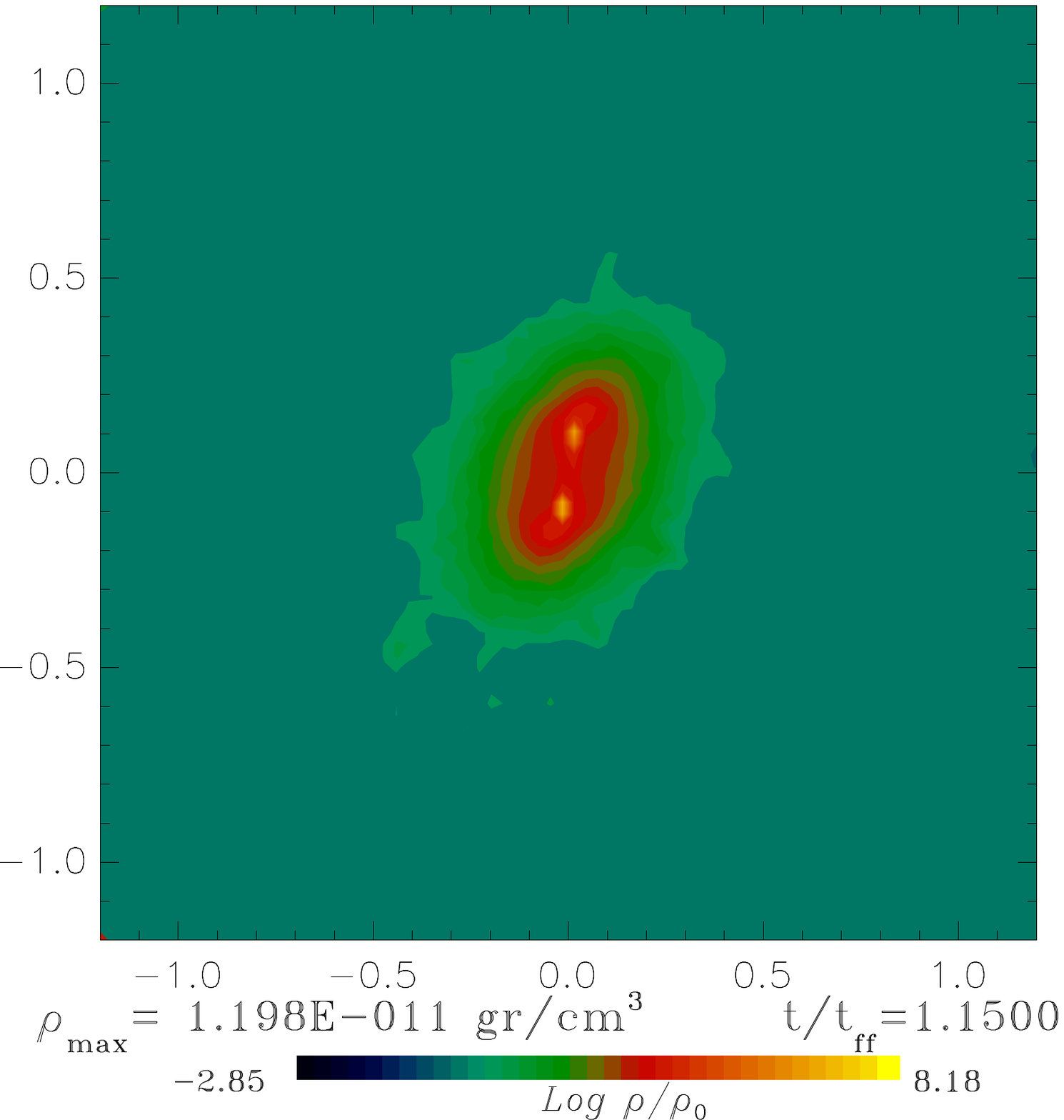} & \includegraphics[width=2.2in]{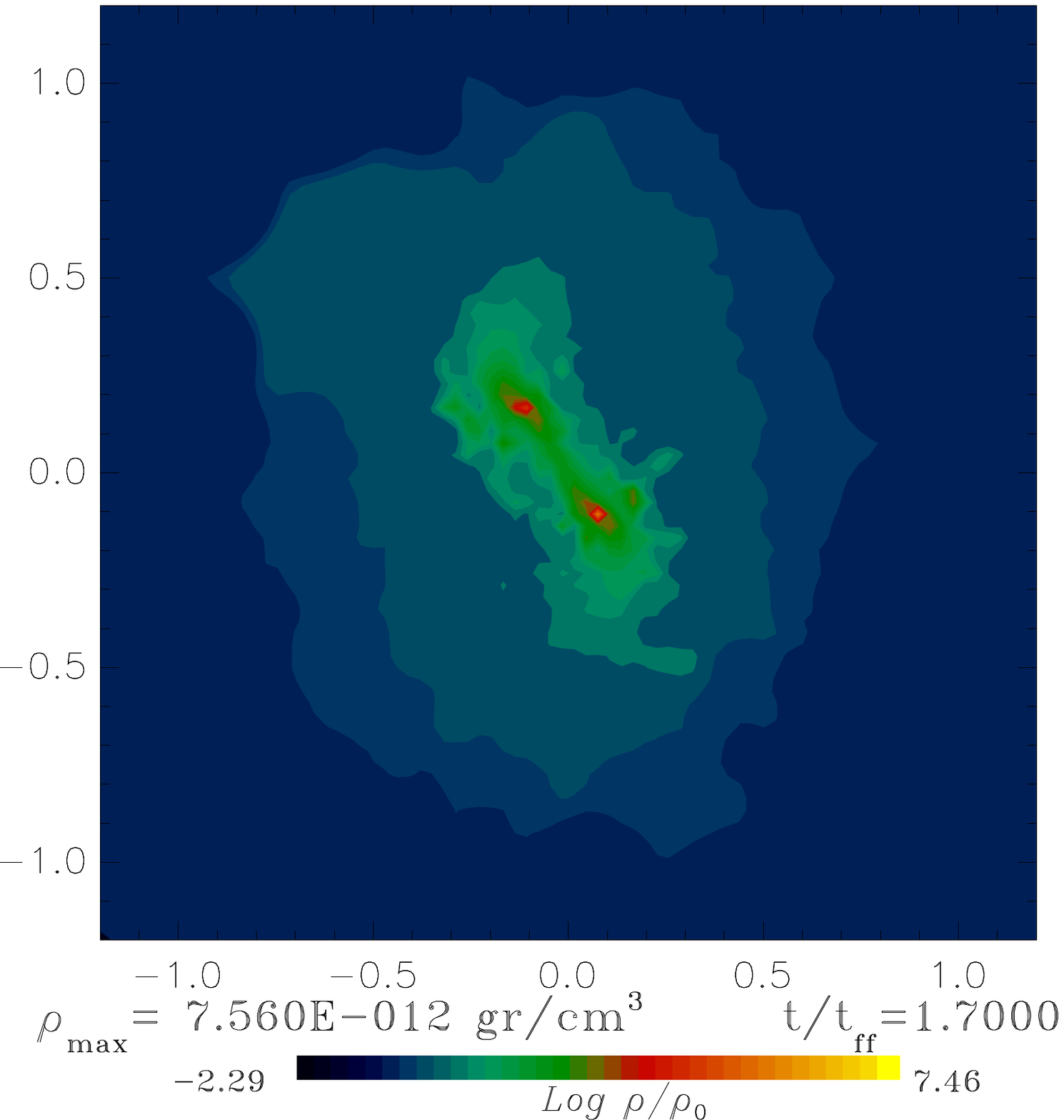} \\
\end{tabular}
\caption{\label{CPrueba90p3} Core models with M$_T$=50 M$_{\odot}$, $a=$0.25
and $\alpha=$0.3; the corresponding $\beta$ are
(left) 0.1
(middle) 0.3.}
\end{center}
\end{figure}

We first notice that there is almost no difference between the configurations obtained
for the clump of M$_T$= 50 M$_{\odot}$ with those obtained for the M$_T$= 400 M$_{\odot}$, as can
be seen by comparing Figs.\ref{CPrueba80p1},\ref{CPrueba80p2},\ref{CPrueba80p3} and
\ref{CPrueba90p1},\ref{CPrueba90p2}, \ref{CPrueba90p3} with the corresponding figures of
the largest mass clump. We take advantage of this similarity by using only the last collapsing
configurations of the M$_T$= 400 M$_{\odot}$ clump to compare with the lower mass models of
previous sections.

In Fig.~\ref{CPrueba80p1} one can see that configurations of the face-to-face type
connected by a weak filament are mainly formed for the two lower values of $\beta$. These
filaments are fainter than those observed in the M$_T$= 1 M$_{\odot}$ model
shown in Fig.\ref{CPrueba60p1} while they are very similar to those
obtained for the M$_T$= 5 M$_{\odot}$ model shown in Fig.\ref{CPrueba30p1}.

\begin{figure}
\begin{center}
\begin{tabular}{ccc}
\includegraphics[width=2.2in,height=2.2in]{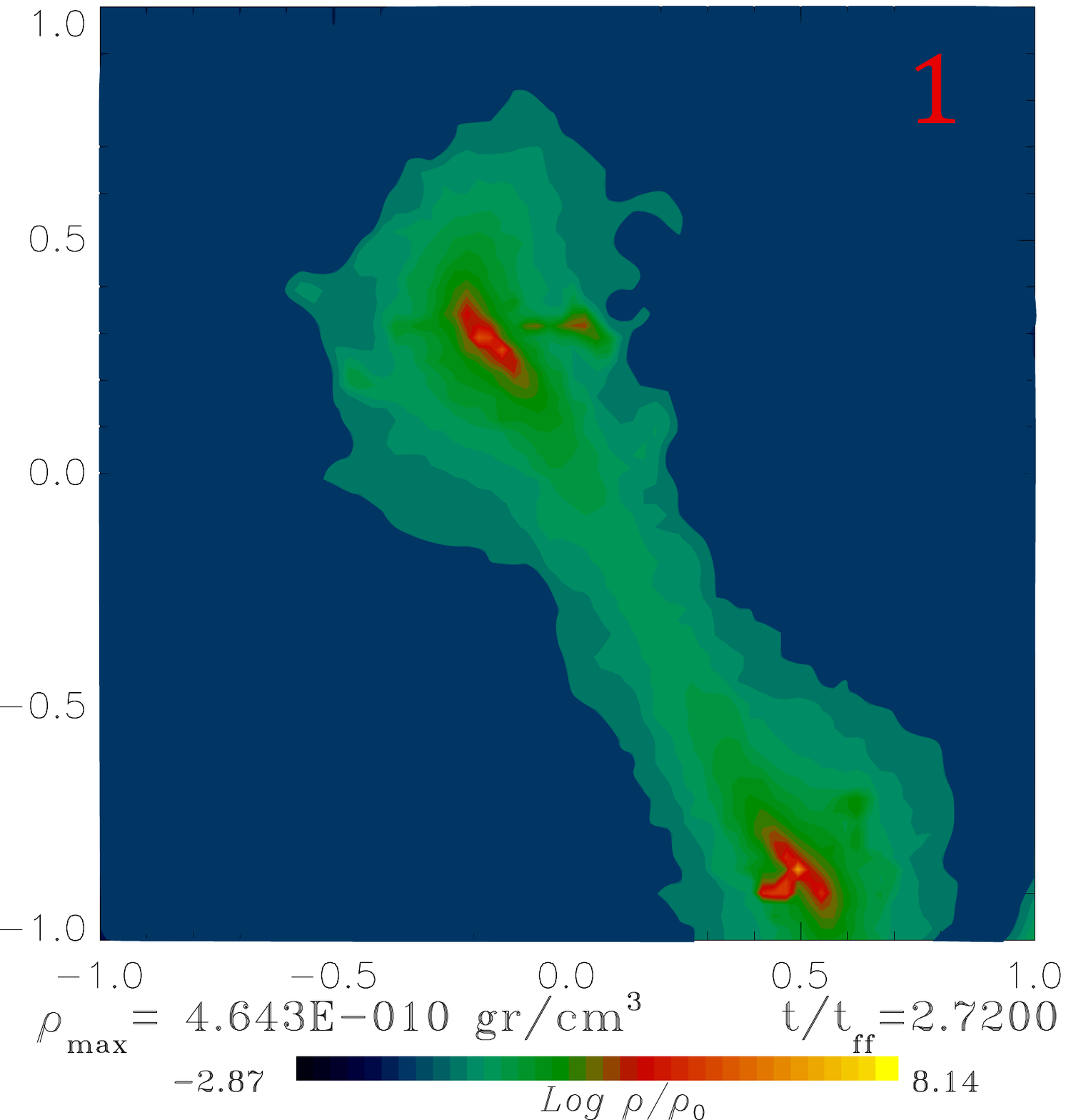} &
&
\includegraphics[width=2.2in,height=2.2in]{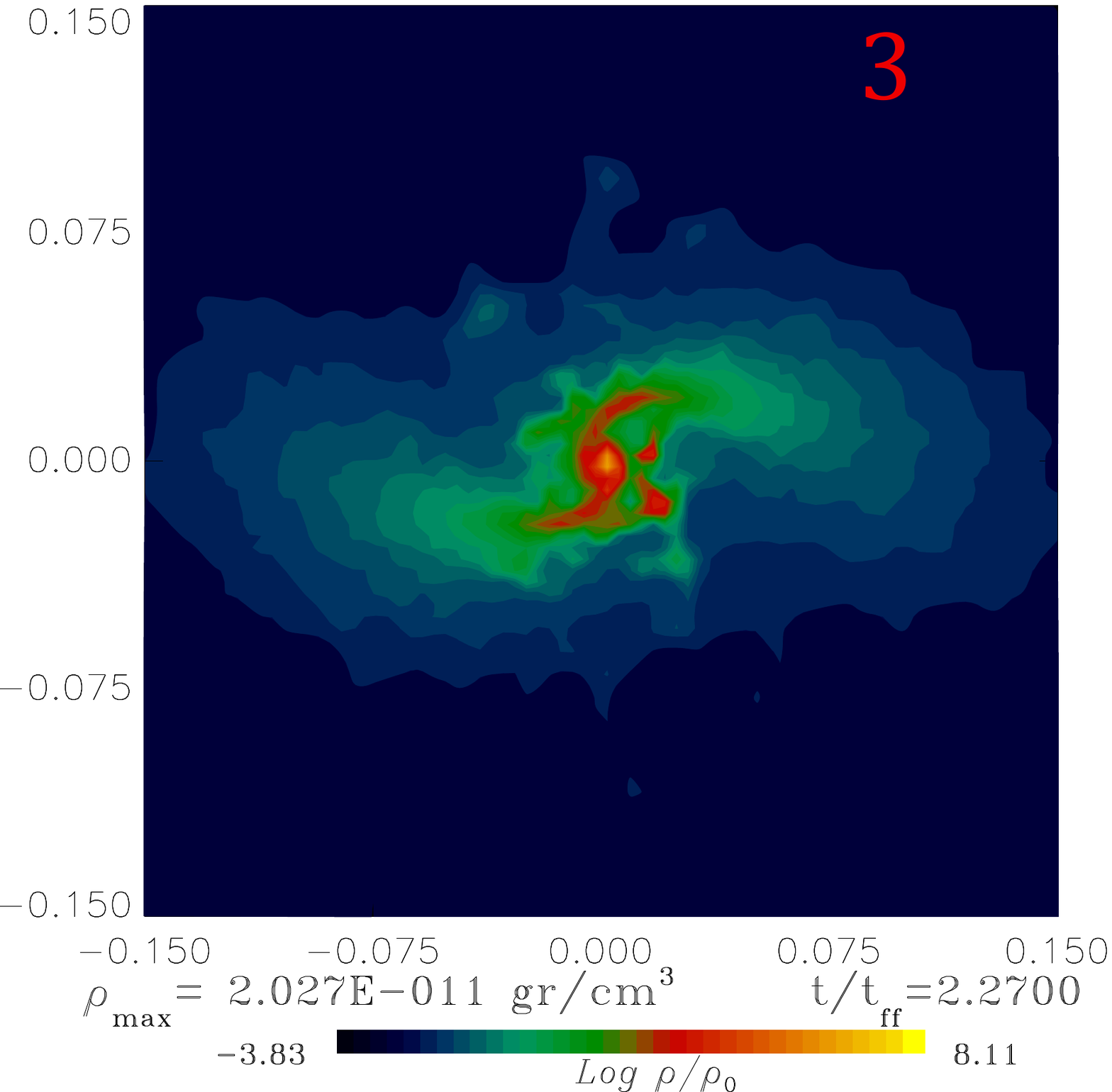} \\
\includegraphics[width=2.2in,height=2.2in]{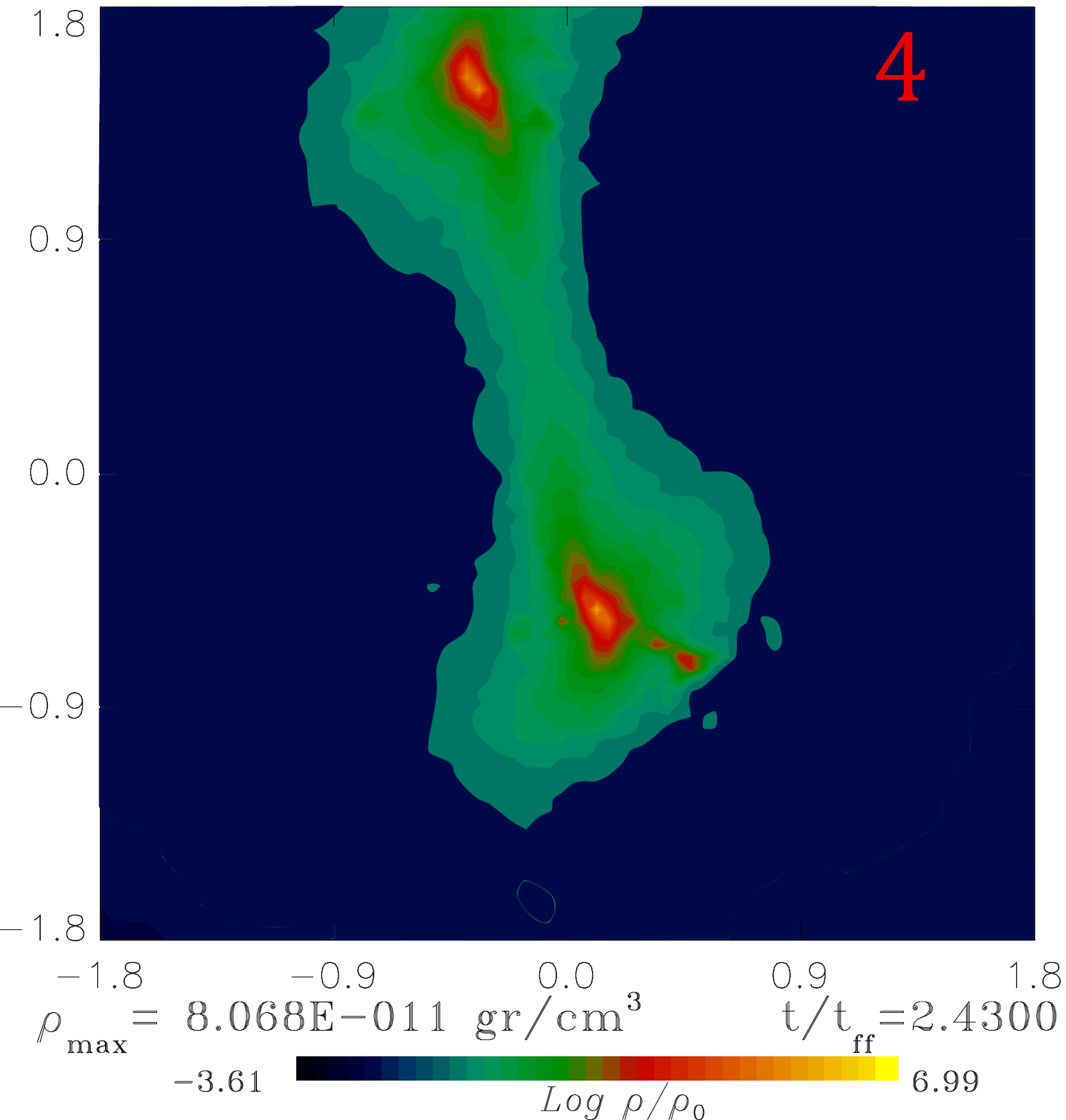} &
\includegraphics[width=2.2in,height=2.2in]{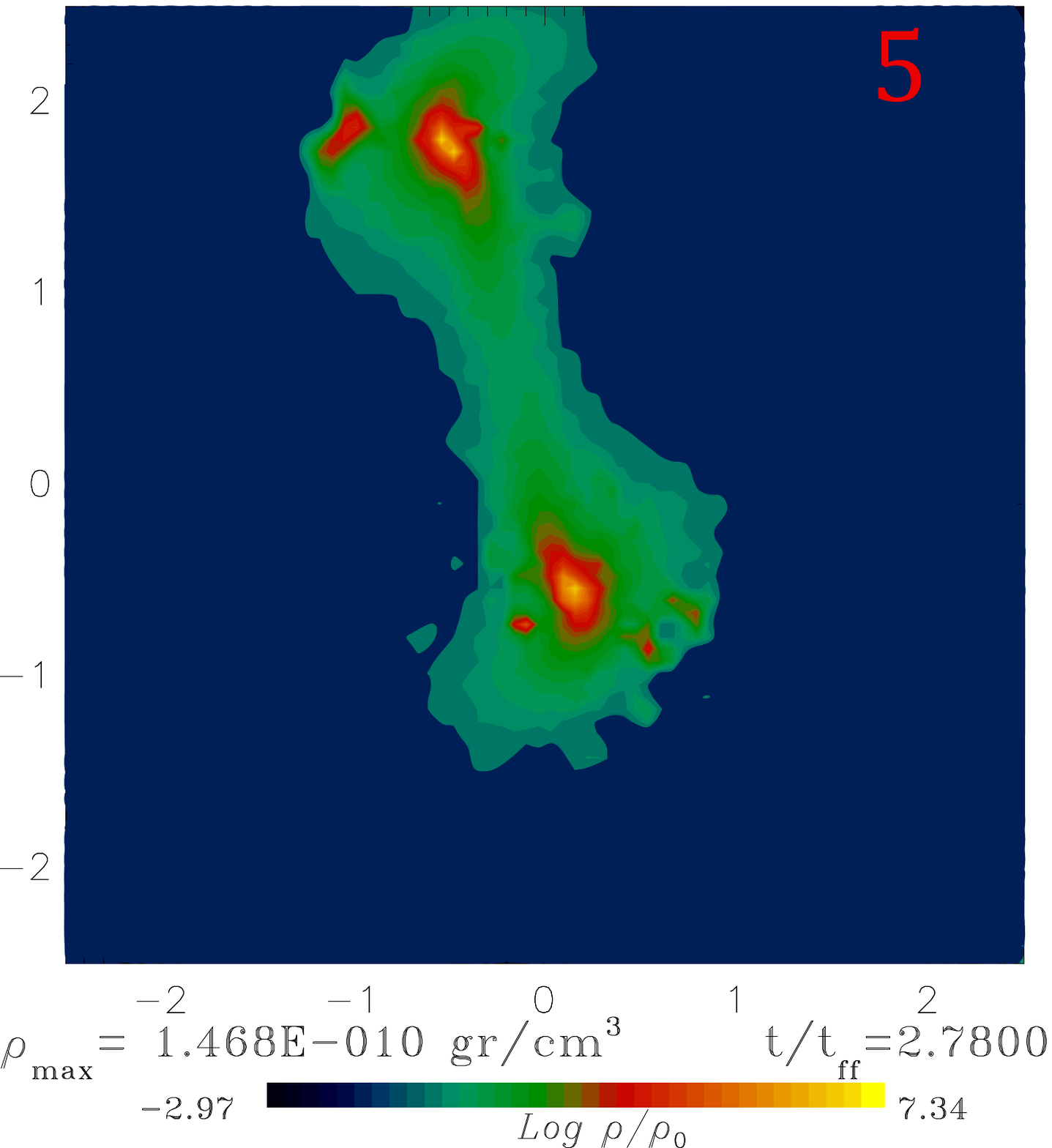} &
\includegraphics[width=2.2in,height=2.2in]{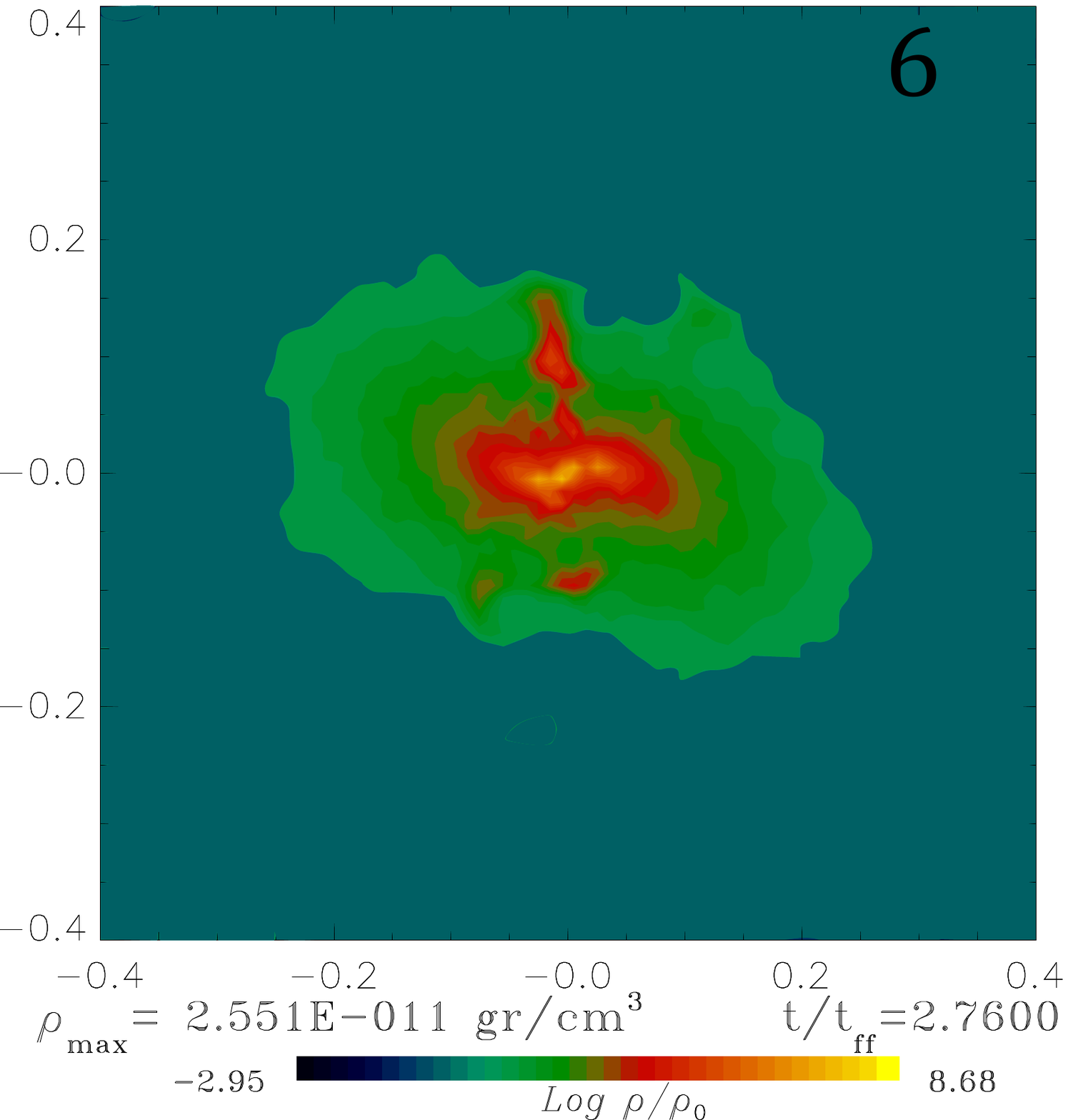} \\
\end{tabular}
\caption{\label{lastcolconfM5}
(1) M$_T$=5 M$_{\odot}$, $a=$0.1, $\alpha=$0.1, $\beta=$0.6118;
(3) M$_T$=5 M$_{\odot}$, $a=$0.1, $\alpha=$0.3, $\beta=$0.3594;
(4) M$_T$=5 M$_{\odot}$, $a=$0.25,$\alpha=$0.1, $\beta=$0.6989;
(5) M$_T$=5 M$_{\odot}$, $a=$0.25,$\alpha=$0.2, $\beta=$0.7467;
(6) M$_T$=5 M$_{\odot}$, $a=$0.25,$\alpha=$0.3, $\beta=$0.3918;
}
\end{center}
\end{figure}

\begin{figure}
\begin{center}
\begin{tabular}{ccc}
\includegraphics[width=2.2in,height=2.2in]{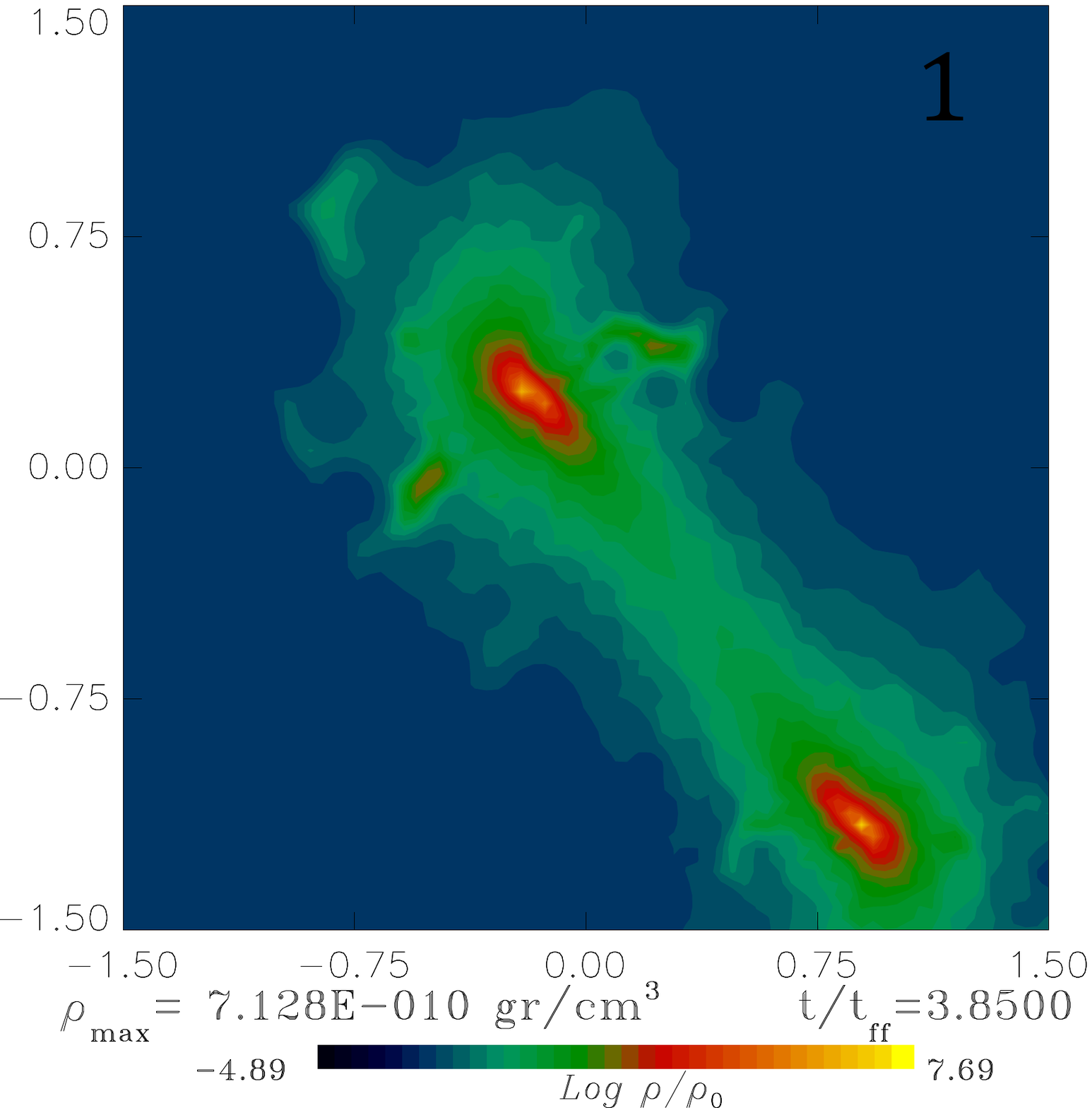} &
\includegraphics[width=2.2in,height=2.2in]{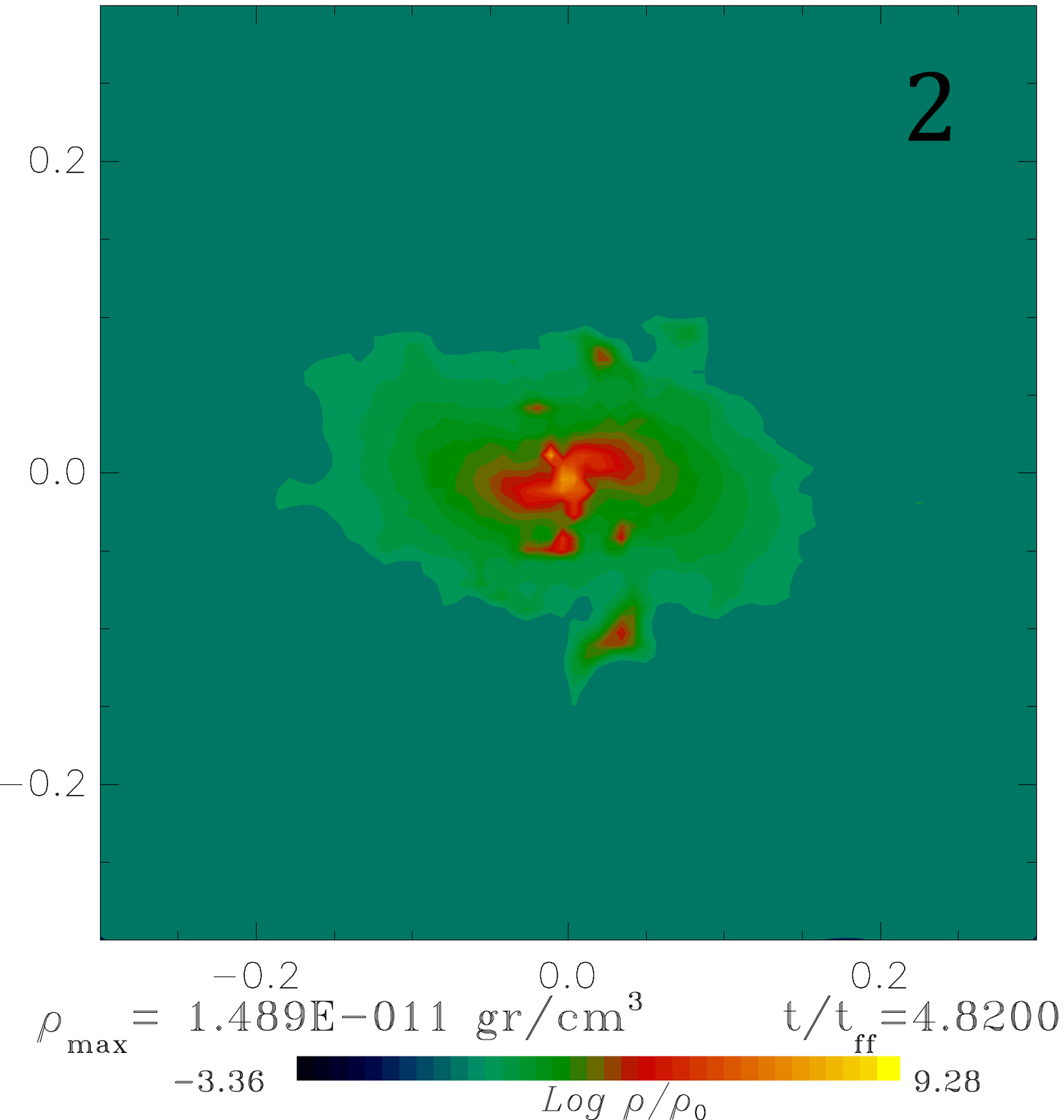} &
\includegraphics[width=2.2in,height=2.2in]{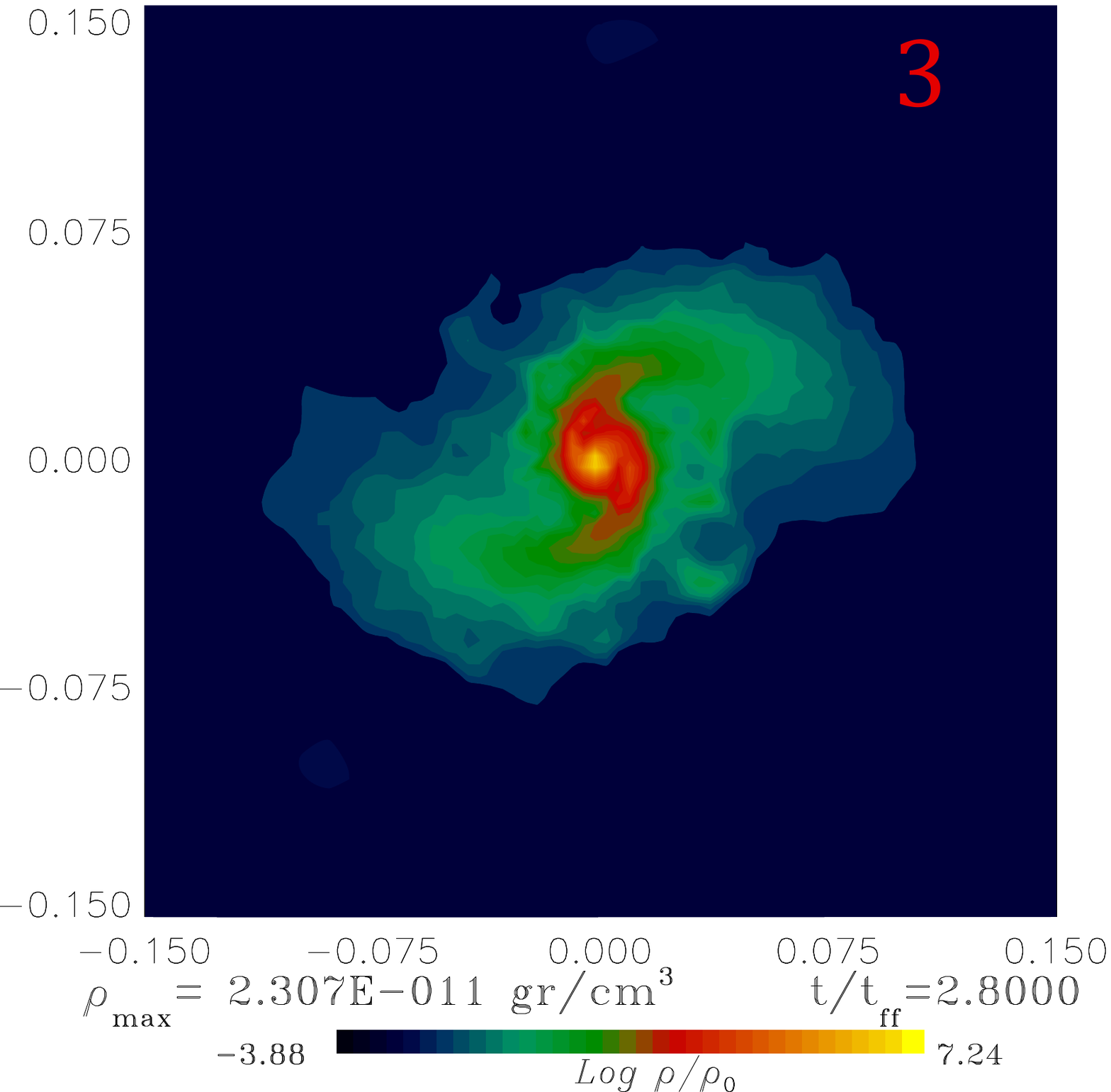}\\
\includegraphics[width=2.2in,height=2.2in]{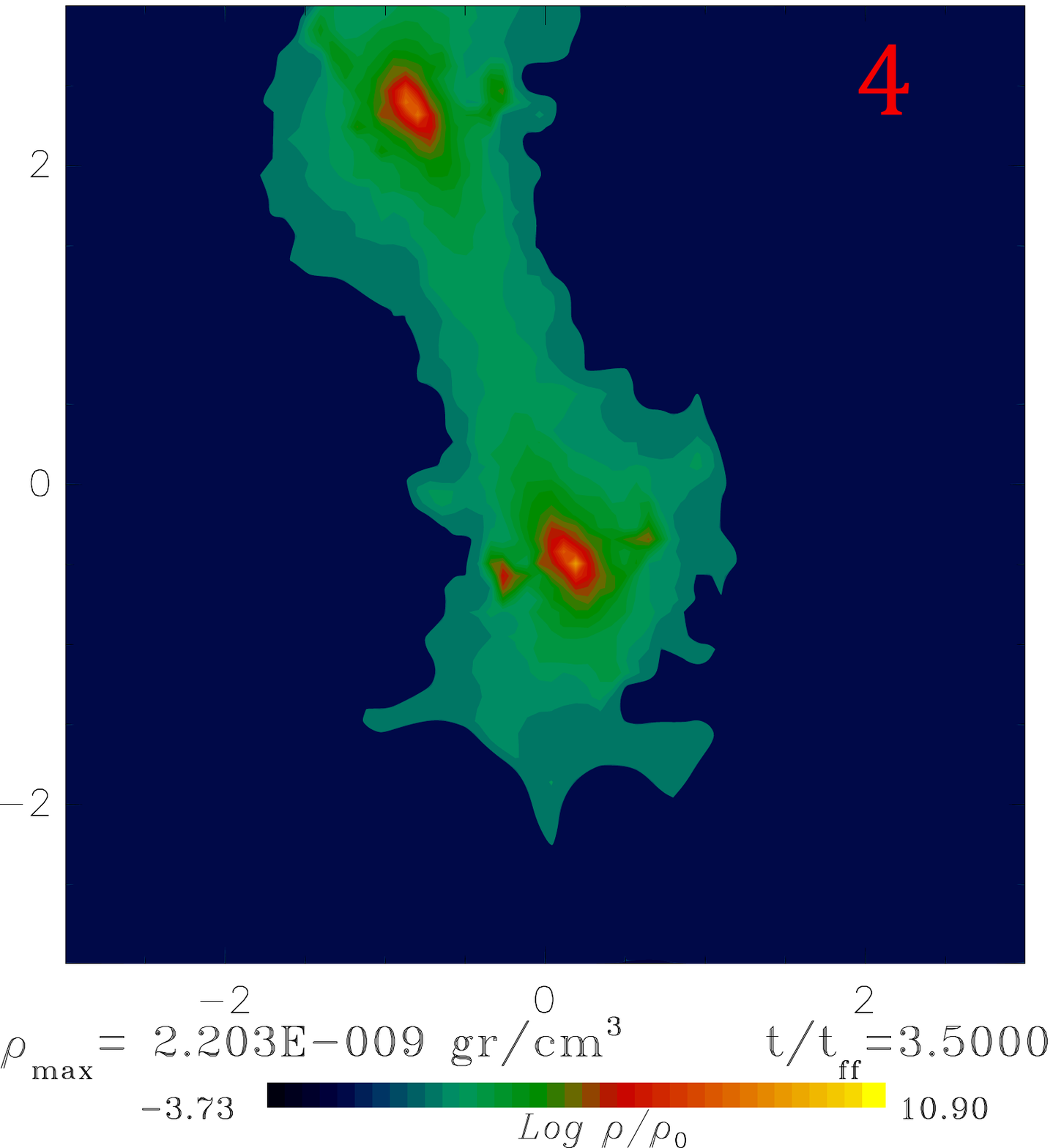} &
\includegraphics[width=2.2in,height=2.2in]{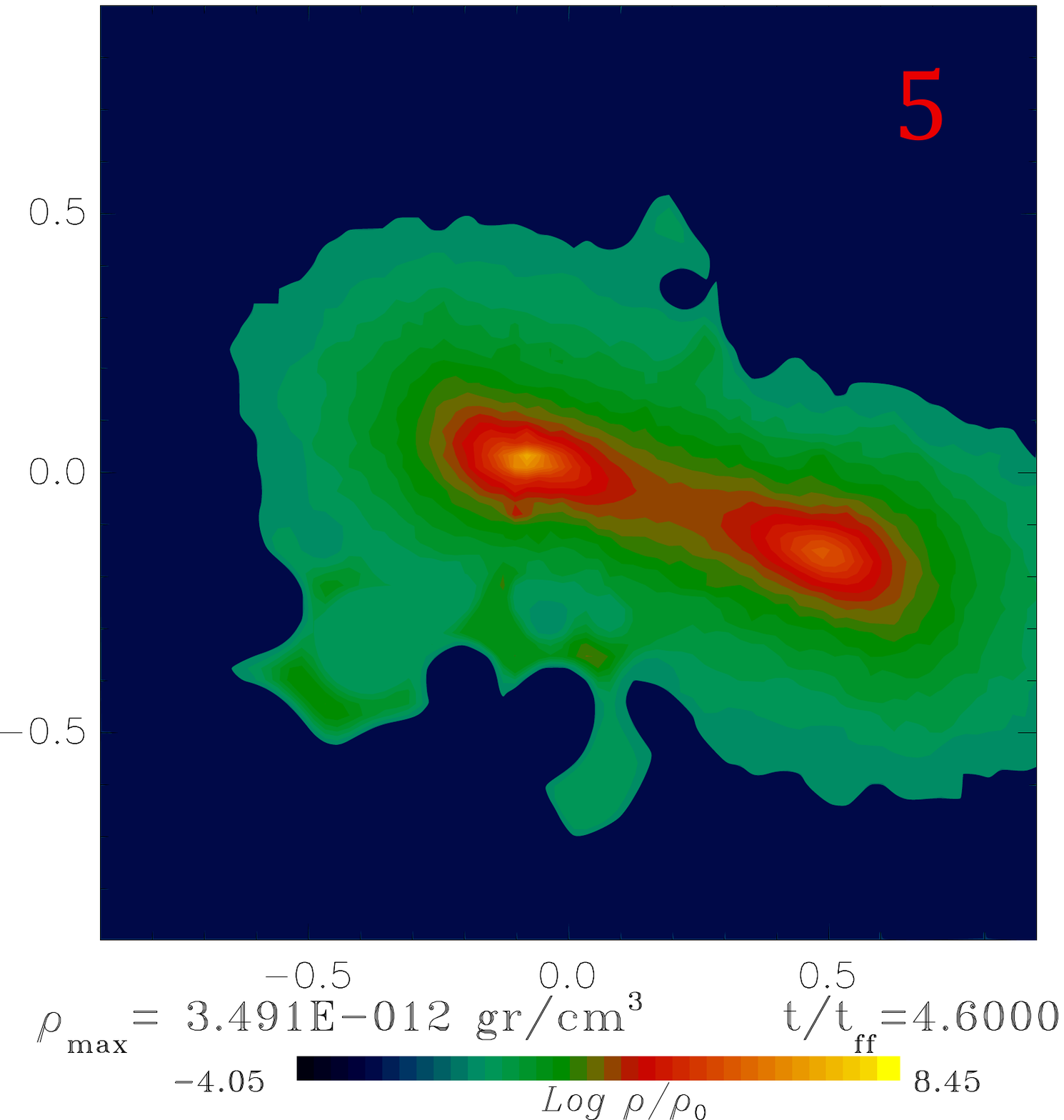} &
\includegraphics[width=2.2in,height=2.2in]{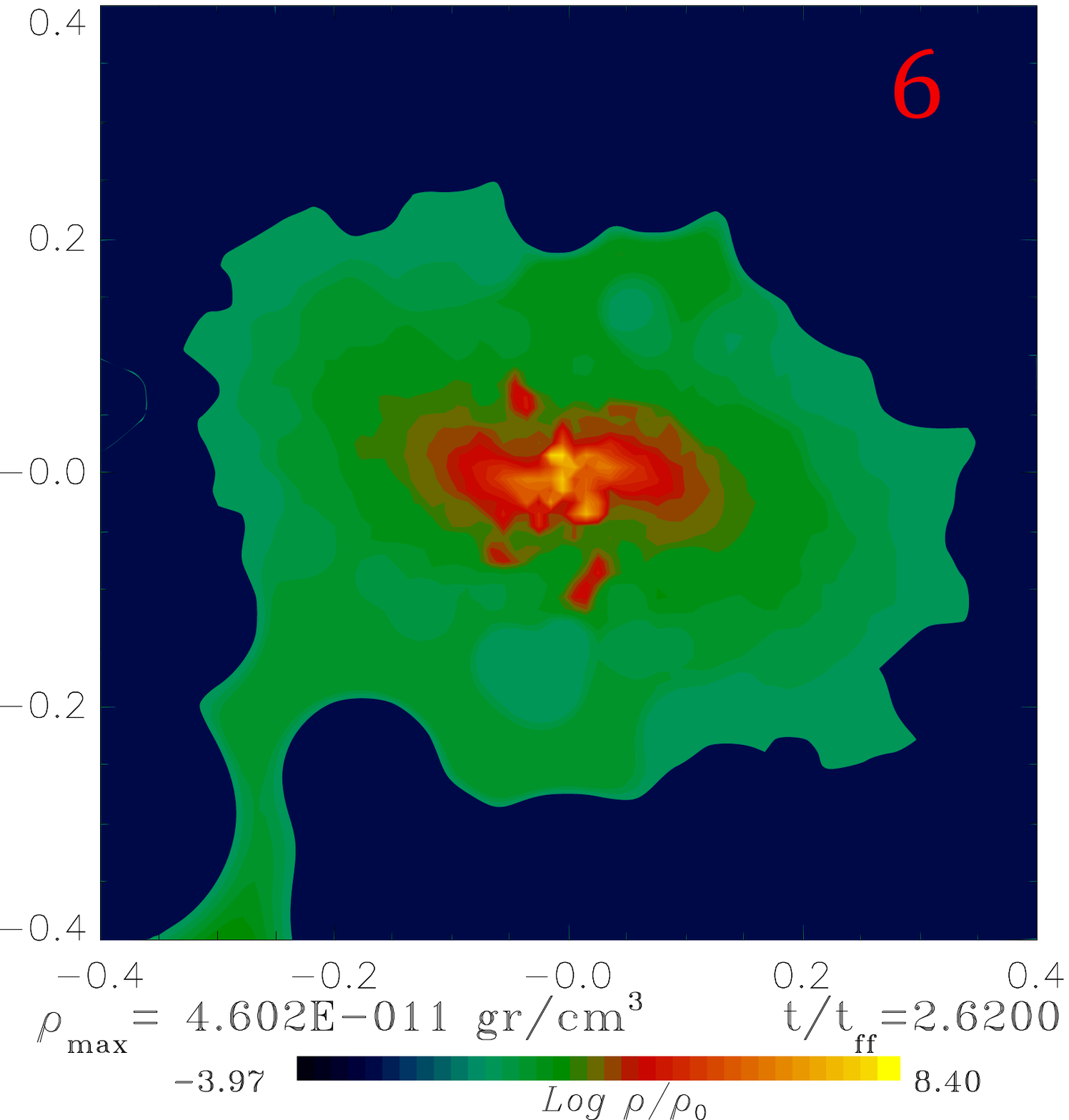}
\end{tabular}
\caption{\label{lastcolconfM400}
(1) M$_T$=400 M$_{\odot}$,$a=$0.1,$\alpha=$0.1, $\beta=$0.7;
(2) M$_T$=400 M$_{\odot}$,$a=$0.1,$\alpha=$0.2, $\beta=$0.55;
(3) M$_T$=400 M$_{\odot}$,$a=$0.1,$\alpha=$0.3, $\beta=$0.38;
(4) M$_T$=400 M$_{\odot}$,$a=$0.25,$\alpha=$0.1,$\beta=$0.81;
(5) M$_T$=400 M$_{\odot}$,$a=$0.25,$\alpha=$0.2,$\beta=$0.55;
(6) M$_T$=400 M$_{\odot}$,$a=$0.25,$\alpha=$0.3,$\beta=$0.385;
}
\end{center}
\end{figure}

In the right panel of Fig.~\ref{CPrueba0p1} and panel 1 of
Fig.\ref{lastcolconfM400}, which corresponds
to a $\beta=$0.3 and $\beta=$0.7,
respectively, still show the face-to-face
configuration, as was the case for the
two models of smaller total mass considered; see the panels 1 of
Fig.\ref{lastcolconfM1} and Fig.\ref{lastcolconfM5}.

When $\alpha$ takes the value of 0.2, the filament formed for a
large $\beta$ is fragmenting
as can be seen in the right panel of Fig.~\ref{CPrueba80p2}, while the opposite case
was seen for the M$_T$= 5 M$_{\odot}$ model in Fig.\ref{CPrueba30p2}; that
is, the filament becomes denser and thinner while in the
M$_T$= 5 M$_{\odot}$ model shown in Fig.\ref{CPrueba60p2} there is no
filament for comparison.

\begin{figure}
\begin{center}
\begin{tabular}{cc}
\includegraphics[width=2.1in,height=1.75in]{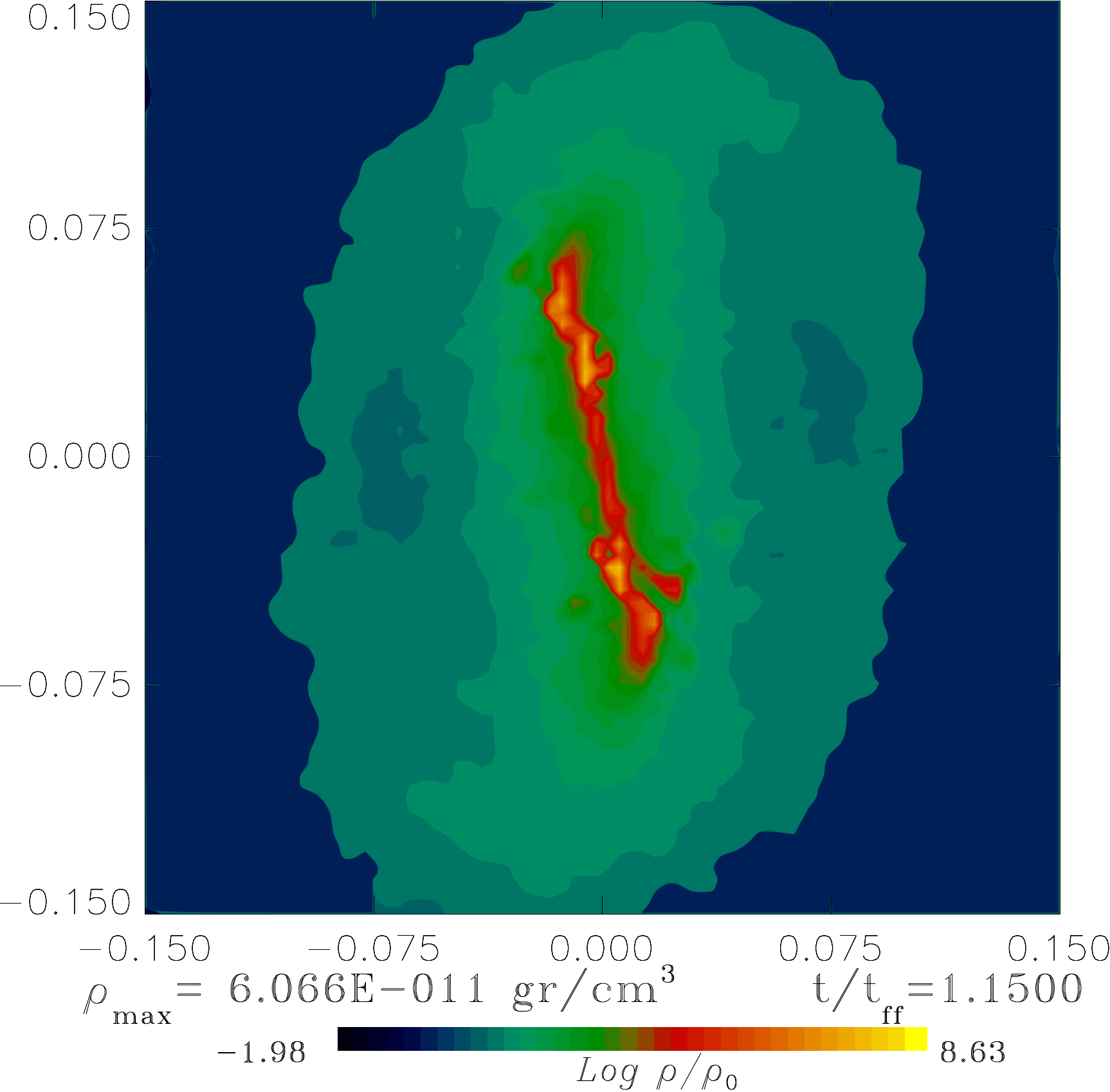} & \includegraphics[width=2.2in]{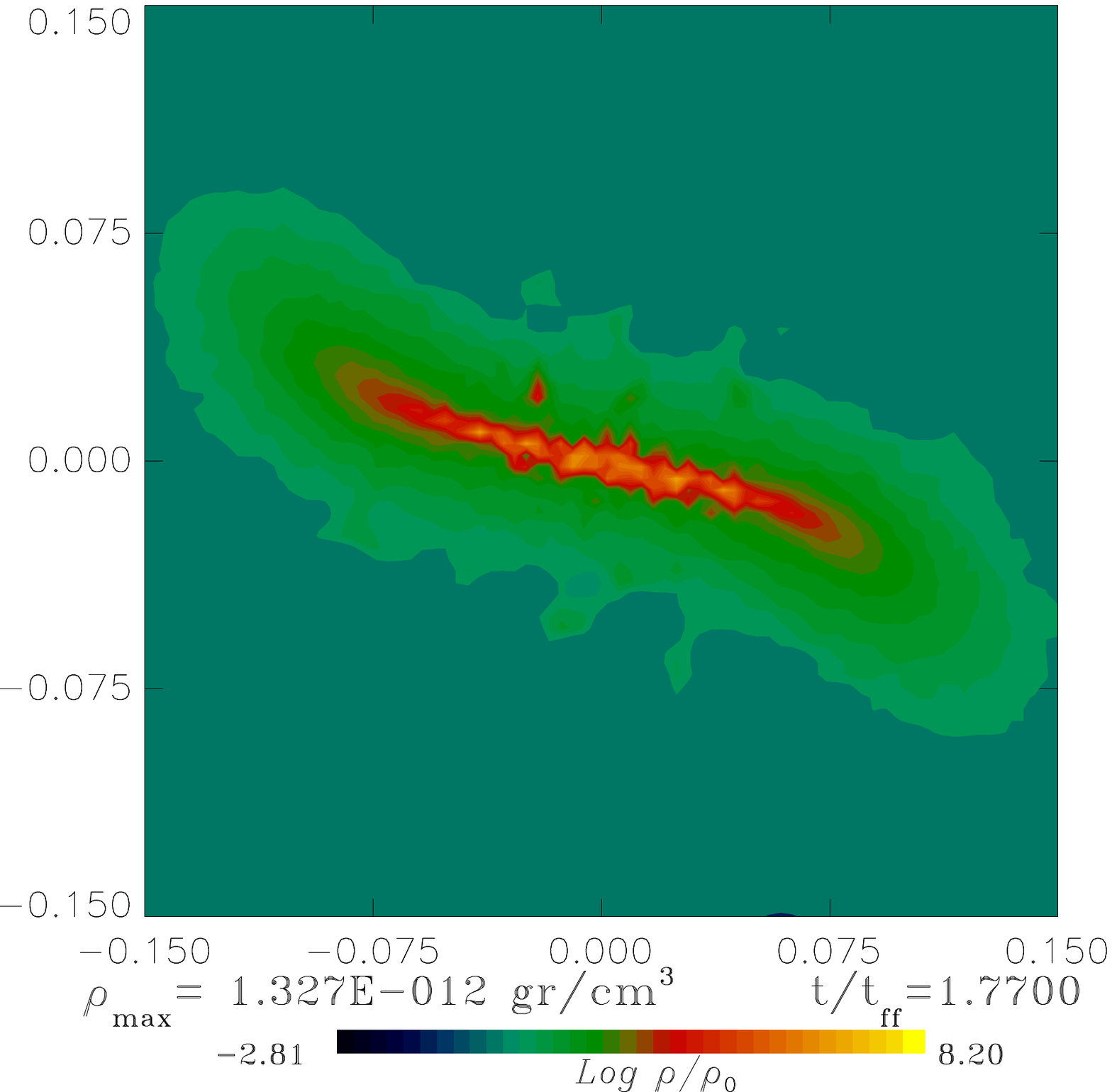}
\end{tabular}
\caption{\label{CPrueba0p3} Clump models with M$_T$= 400 M$_{\odot}$, $a=$0.1 and $\alpha=$0.3;
the corresponding $\beta$ are
(left) 0.1
(right) 0.3. }
\end{center}
\end{figure}

When $\beta$ takes its highest possible value, still in the collapsing
regime, the filament of the M$_T$= 400 M$_{\odot}$ model is
replaced by a central primary configuration surrounded by spiral
arms; see the panel 2 and 3 of Fig.~\ref{lastcolconfM400} and the
bottom left panel of Fig.~\ref{CPrueba0p3}. This behavior was also
observed in the M$_T$= 5 M$_{\odot}$ models in
Fig.\ref{CPrueba30p2} and Fig.\ref{CPrueba30p3} but was not observed
in the M$_T$= 1 M$_{\odot}$ models with $\alpha=$0.2 in
Fig.\ref{CPrueba60p2} nor with $\alpha=$0.1 in Fig.\ref{CPrueba60p1}, although
it occurred for this model with $\alpha=$0.3, as can be seen in
Fig.\ref{CPrueba60p3}.

\begin{figure}
\begin{center}
\begin{tabular}{ccc}
\includegraphics[width=2.2in]{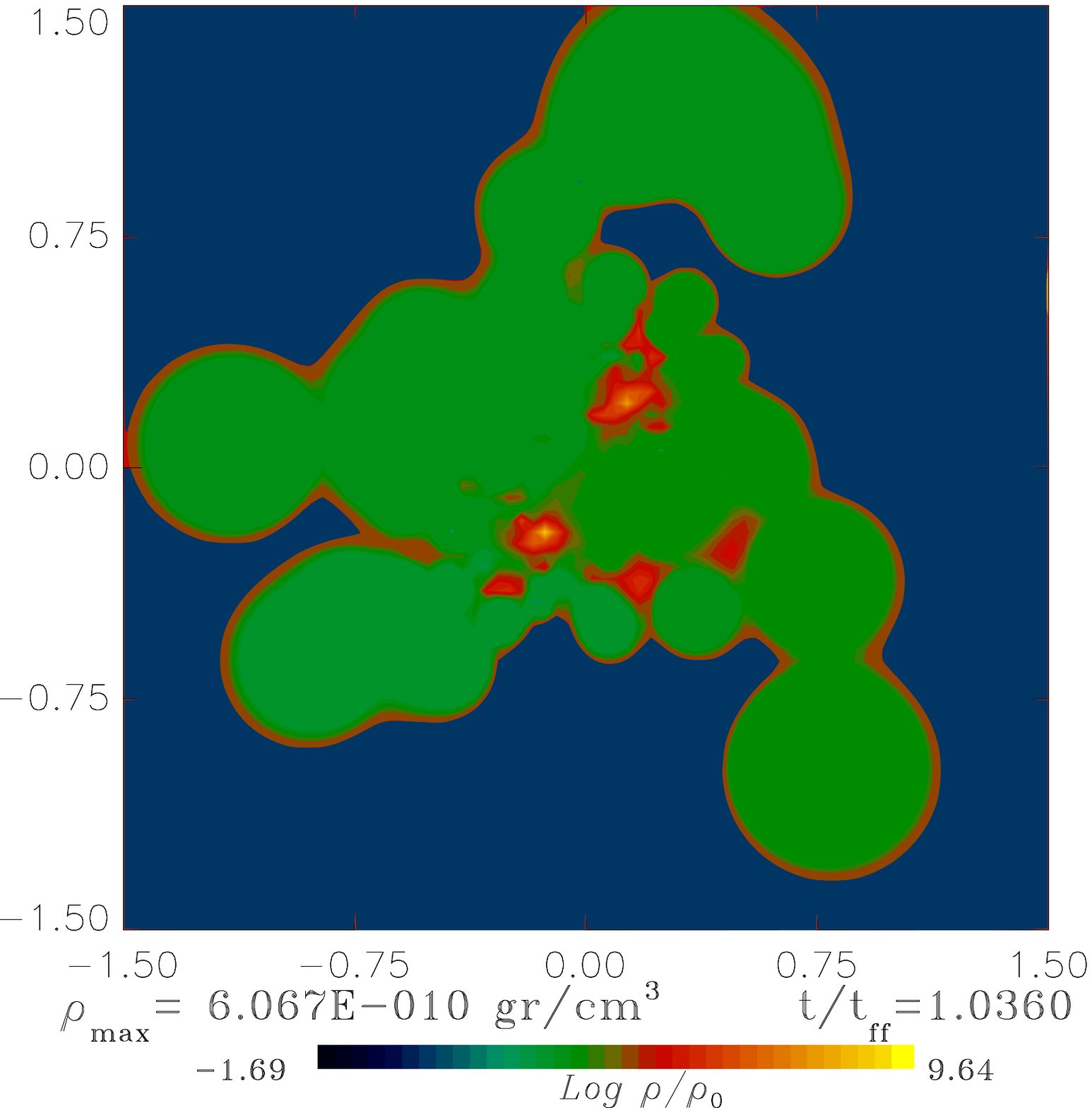} &\includegraphics[width=2.2in]{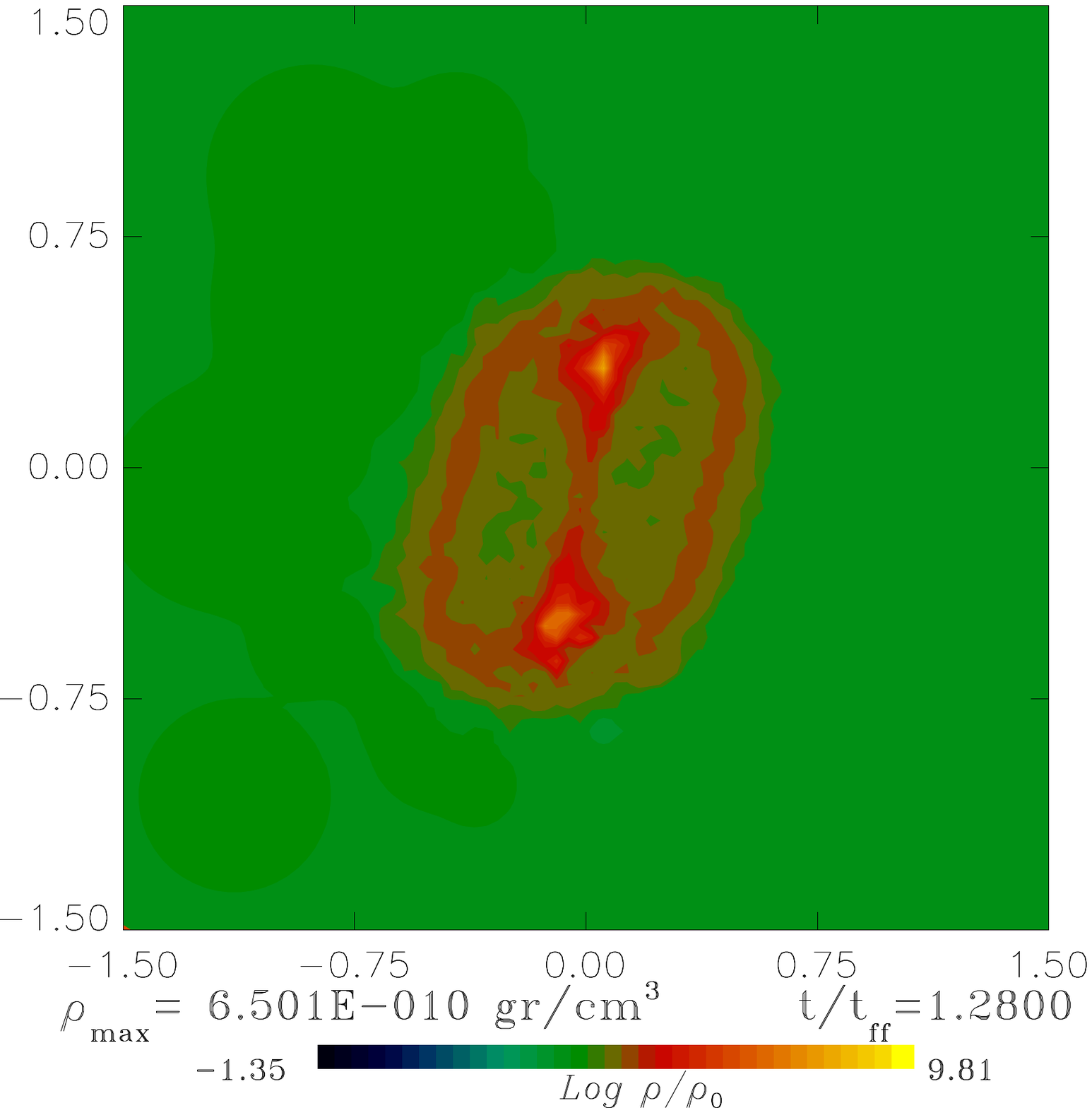} &
\includegraphics[width=2.2in]{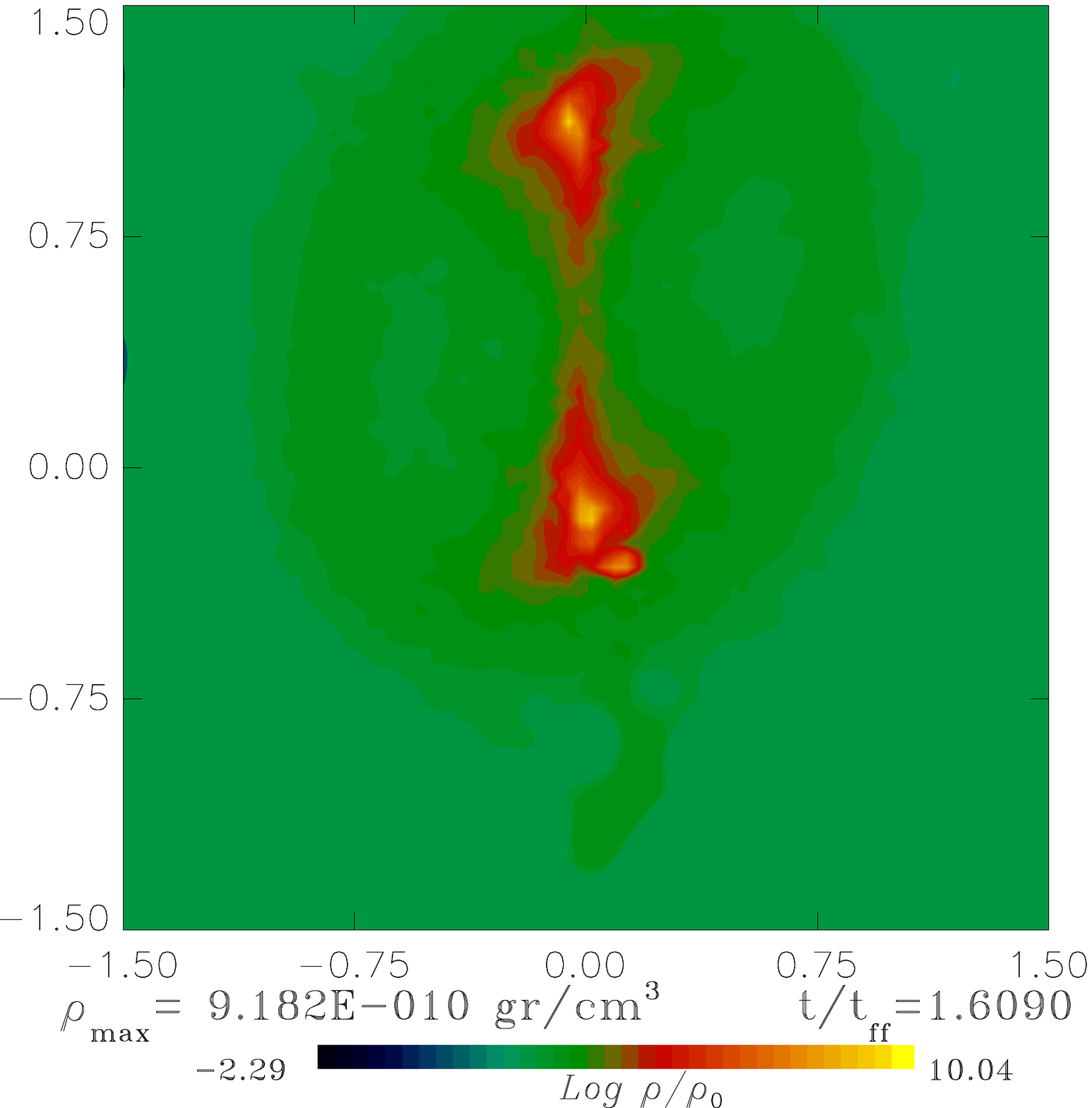} \\
\end{tabular}
\caption{\label{CPrueba50p1} Clump models with M$_T$= 400 M$_{\odot}$, $a=$0.25 and $\alpha=$0.1;
the corresponding $\beta$ are
(left) 0.1
(middle) 0.3
(right) 0.48.}
\end{center}
\end{figure}

When the mass perturbation is increased to $a=$0.25, more
configurations of the face-to-face type appear connected by a
tenuous bridge of gas; see Fig.~\ref{CPrueba50p1}. This behavior was
also observed in the M$_T$= 5 M$_{\odot}$ model shown in
Fig.\ref{CPrueba40p1}, but it must be mentioned that the binary
separation is much larger in the former model than in the latter.
Something similar was also observed for the M$_T$= 1
M$_{\odot}$ model, but there the filaments are more pronounced, as can be
seen in Fig.\ref{CPrueba70p1}.

The runs with $\alpha=$0.3 are similar for the M$_T$= 50 M$_{\odot}$ models and
M$_T$= 5 M$_{\odot}$ shown in Fig.~\ref{CPrueba50p2} and
Fig.\ref{CPrueba40p1}, respectively, in which we see configurations of
the face-to-face type. It should be noted that in the
M$_T$=1 M$_{\odot}$ model, the configurations formed are not at all
similar to the ones seen for the corresponding two models with larger parent
mass, as an orbiting binary configuration is only seen in
Fig.\ref{CPrueba70p2}.

\begin{figure}
\begin{center}
\begin{tabular}{cc}
\includegraphics[width=2.2in]{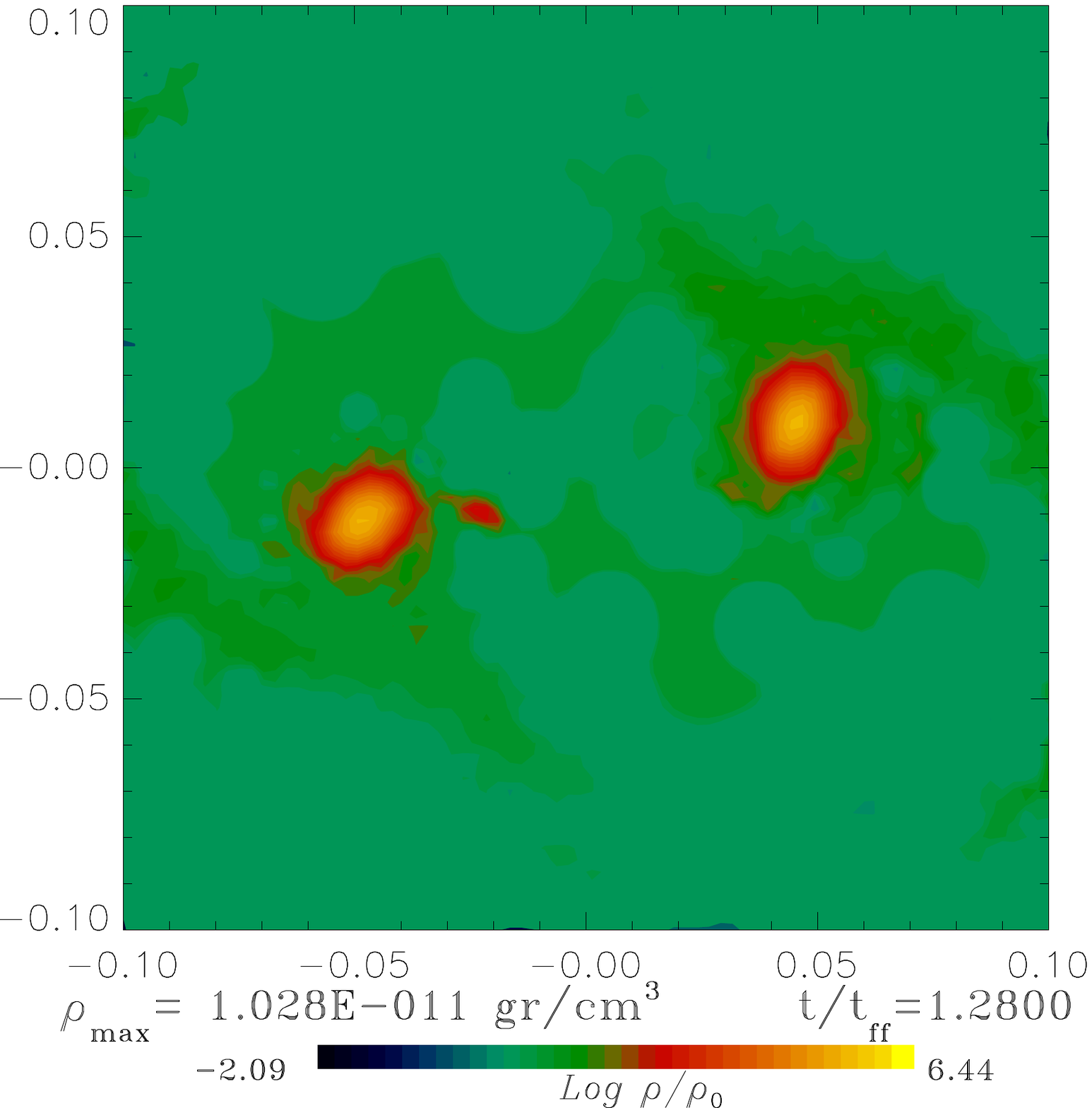} & \includegraphics[width=2.2in]{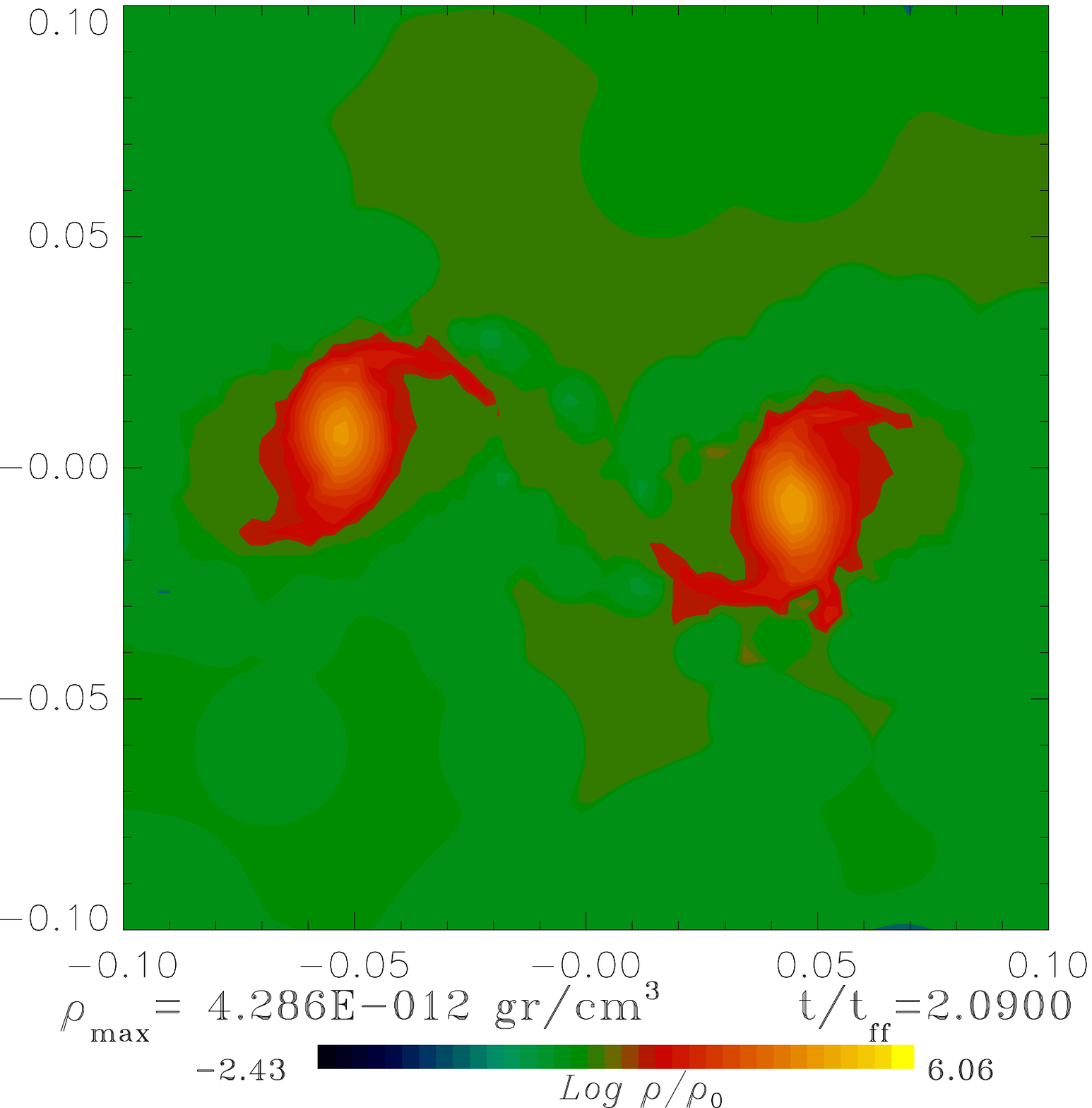} \\
\end{tabular}
\caption{\label{CPrueba70p3} Core models with M$_T$=1 M$_{\odot}$, $a=$0.25 and $\alpha=$0.3;
the corresponding $\beta$ are
(left) 0.1
(right) 0.3.}
\end{center}
\end{figure}

\begin{figure}
\begin{center}
\begin{tabular}{cc}
\includegraphics[width=2.2in]{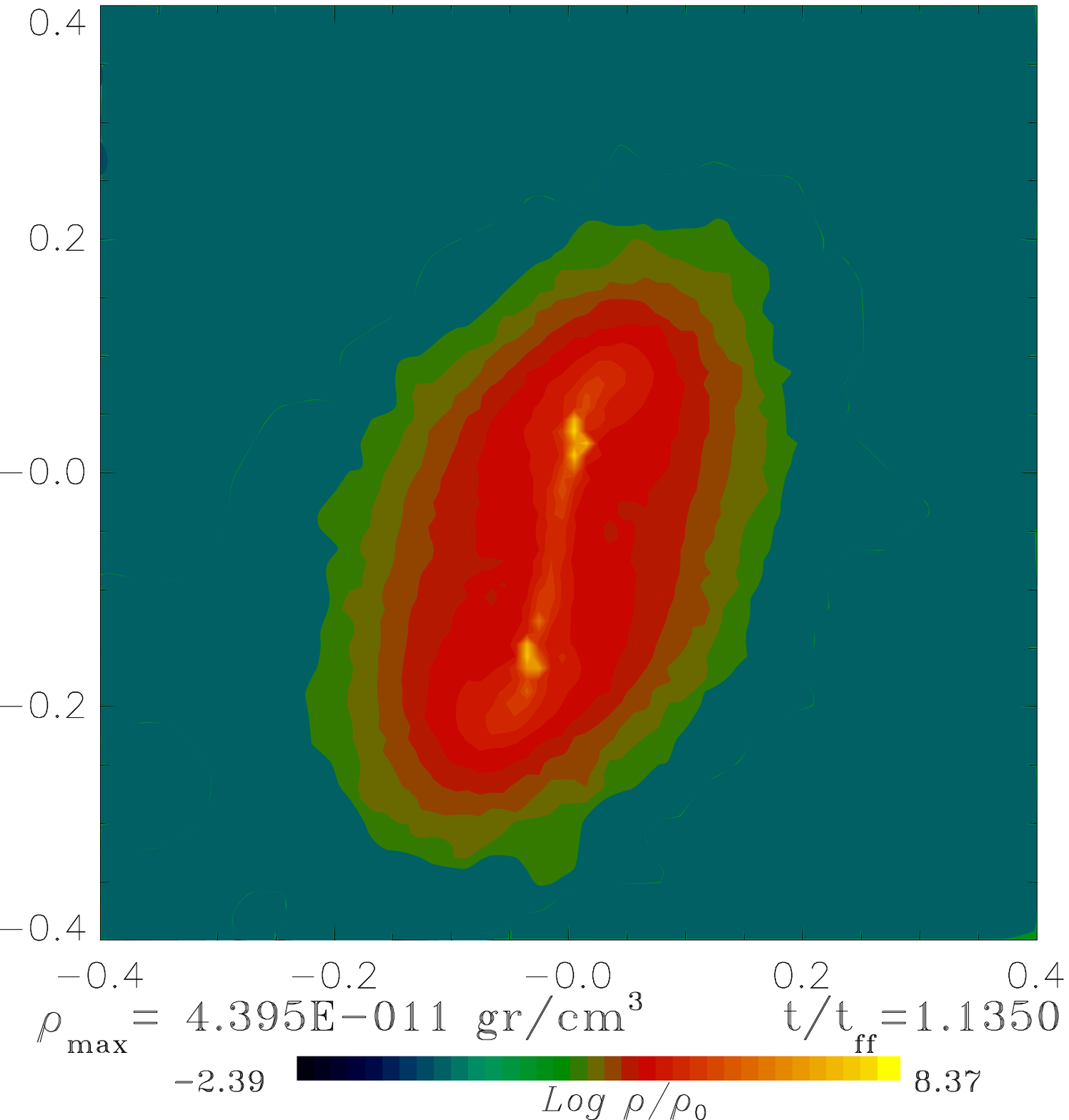} & \includegraphics[width=2.2in]{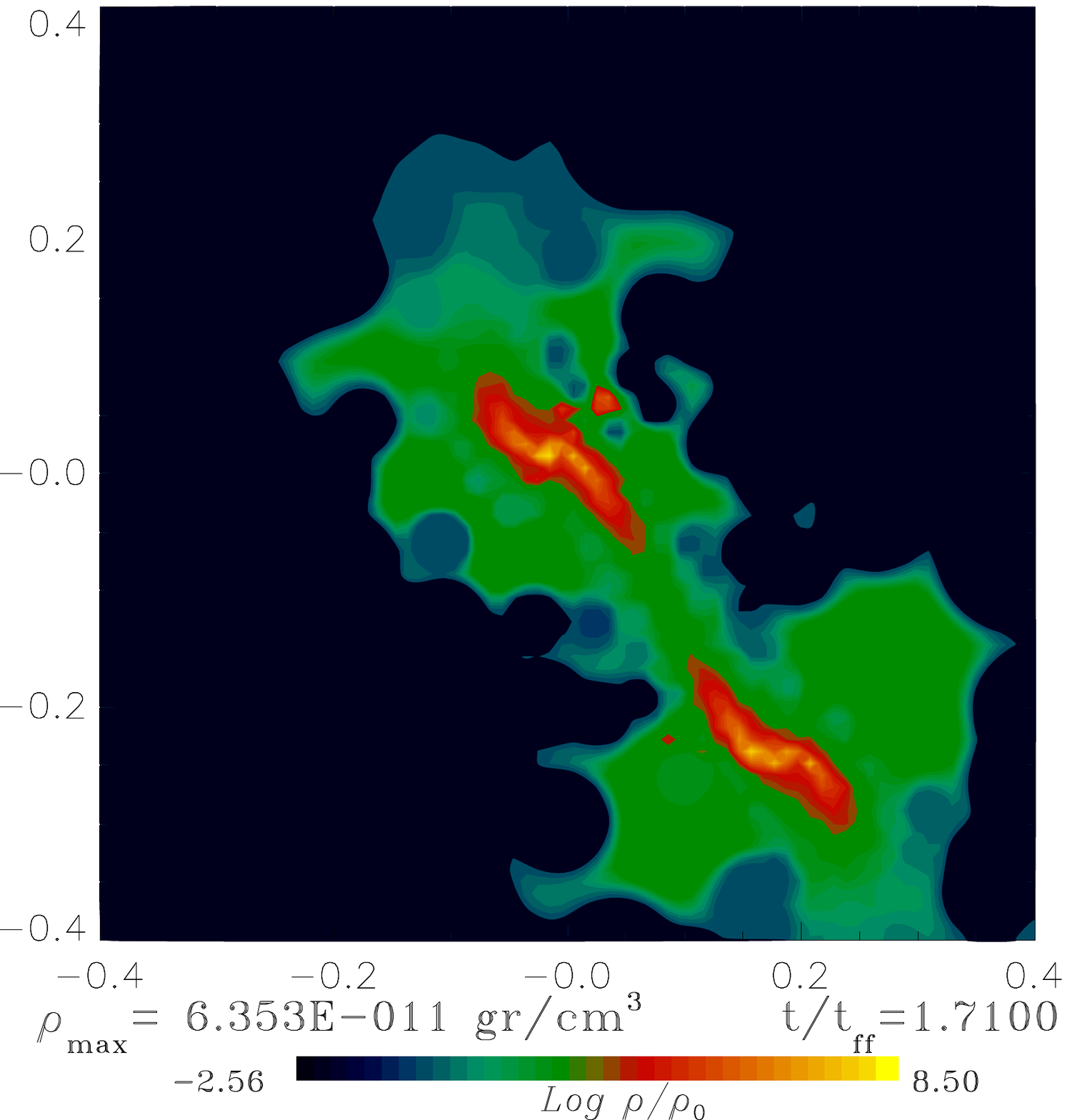}
\end{tabular}
\caption{\label{CPrueba50p3} Clump models with M$_T$= 400 M$_{\odot}$, $a=$0.25 and $\alpha=$0.3;
the corresponding $\beta$ are
(left) 0.1
(right) 0.3.}
\end{center}
\end{figure}

For the $M_T=$ 400 $M_{\odot}$ models with $\alpha=$0.3, we see a
well-defined filament that is transformed to a central primary
configuration when $\beta$ takes its highest value; see
Fig.~\ref{CPrueba50p3}; this was also observed in Fig.\ref{CPrueba40p3}
for the M$_T$= 5 M$_{\odot}$ model. But again, there is no possible comparison with the
configuration obtained for the M$_T$= 1 M$_{\odot}$ model shown in
Fig.\ref{CPrueba70p3}, in which orbiting binaries were indeed observed even for this
higher value of $\alpha$.

For all the models of M$_T$= 50 M$_{\odot}$, we did not observe
the formation of any orbiting binary system.

%%%%%%%%%%%%%%%%%%%%%%%%%%%%%%%%%%%%%%%%%%%%%%%%%%%%%
%%%%%%%%%%%%%%%%%%%%%%%%%%%%%%%%%%%%%%%%%%%%%%%%%%%%%
\subsection{The last collapsing configuration}
\label{subsec:lastcoll}

We measure the uppermost $\beta$ values for all
models except for the clump of M$_T$= 50 M$_{\odot}$,
so that the parent structure
is still within the collapsing regime to
the left of these curves, while to the right of each curve, there is no
collapse. The configurations found are shown in
Figs.~\ref{lastcolconfM1}, \ref{lastcolconfM5} and \ref{lastcolconfM400}.

\begin{figure}
\begin{center}
\includegraphics[width=3.0in,height=2.5in]{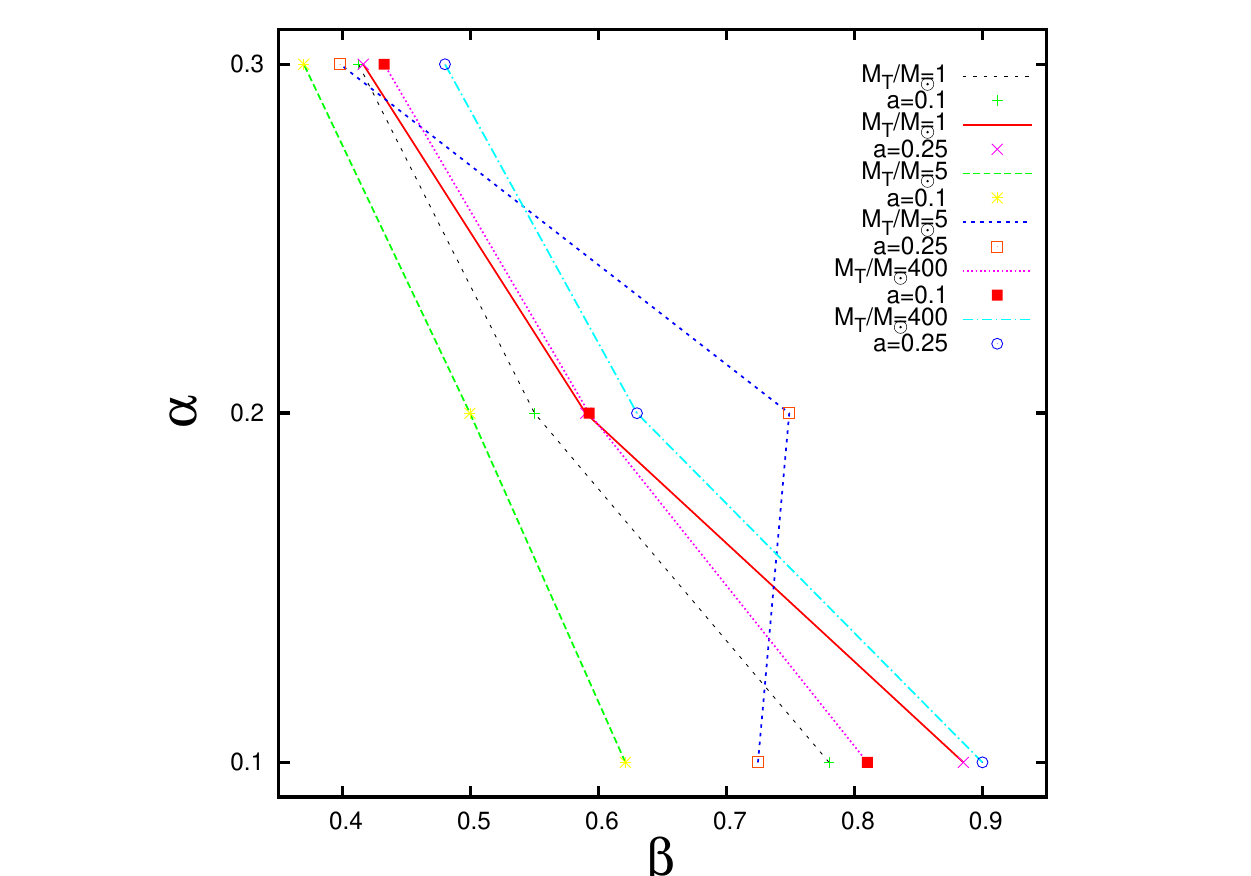}
\caption{\label{lcurve} Limiting curves for all the models, so that in the
region to the right of each curve, the corresponding parent gas structure
does not collapse anymore.}
\end{center}
\end{figure}

In Fig.~\ref{lcurve} we observe that the curve for the models with
a=0.25 are always located to the right of the corresponding curve
for a=0.1; this fact is easy to
understand, as the mass perturbation weighs more, then the resulting
fragments are more massive, so they need more
rotational energy to reach the level in which the equilibrium between
the centrifugal force and the gravitational
force is to be overcome. When this happens, the fragments are separated
indefinitely and the gas structure does not collapse.
%%%%%%%%%%%%%%%%%%%%%%%%%%%%%%%%%%%%%%%%%%%%%%%%%%%%%%%%%%%%%%%%%%%%%%%%%%%%%
\subsection{A quantitative comparison of the resulting configurations}
\label{subsec:qc}

We measure the most important integral properties of
the resulting collapse configurations: for binaries, we consider the fragment
masses and the binary separations; for the central primaries, we consider the central mass and radius;
for all the resulting fragments or primaries, we also calculate the values of the energy ratios
$\alpha_f$ and $\beta_f$.

In order to calculate these properties, we proceeded
as follows: we locate the highest density particle in the fragment's region; this
particle is considered
the center of the fragment. The binary separation is
simply calculated as the distance between the centers associated with
each fragment. We next find all the particles which
have density above or equal to some minimum density value, given
in advance as
$\log_{10} \left( \rho_{min}/\rho_0 \right)=0.0$ for all the models. We finally
check that these particles are also
located within a given maximum radius r$_{max}$ from the fragment's center. This set
of particles allowed us to calculate the
fragment integral properties, as the mass, including the $\alpha_f$ and $\beta_f$;
see \cite{SegRMAA}.

\begin{figure}
\begin{center}
\includegraphics[width=3.0in,height=2.5in]{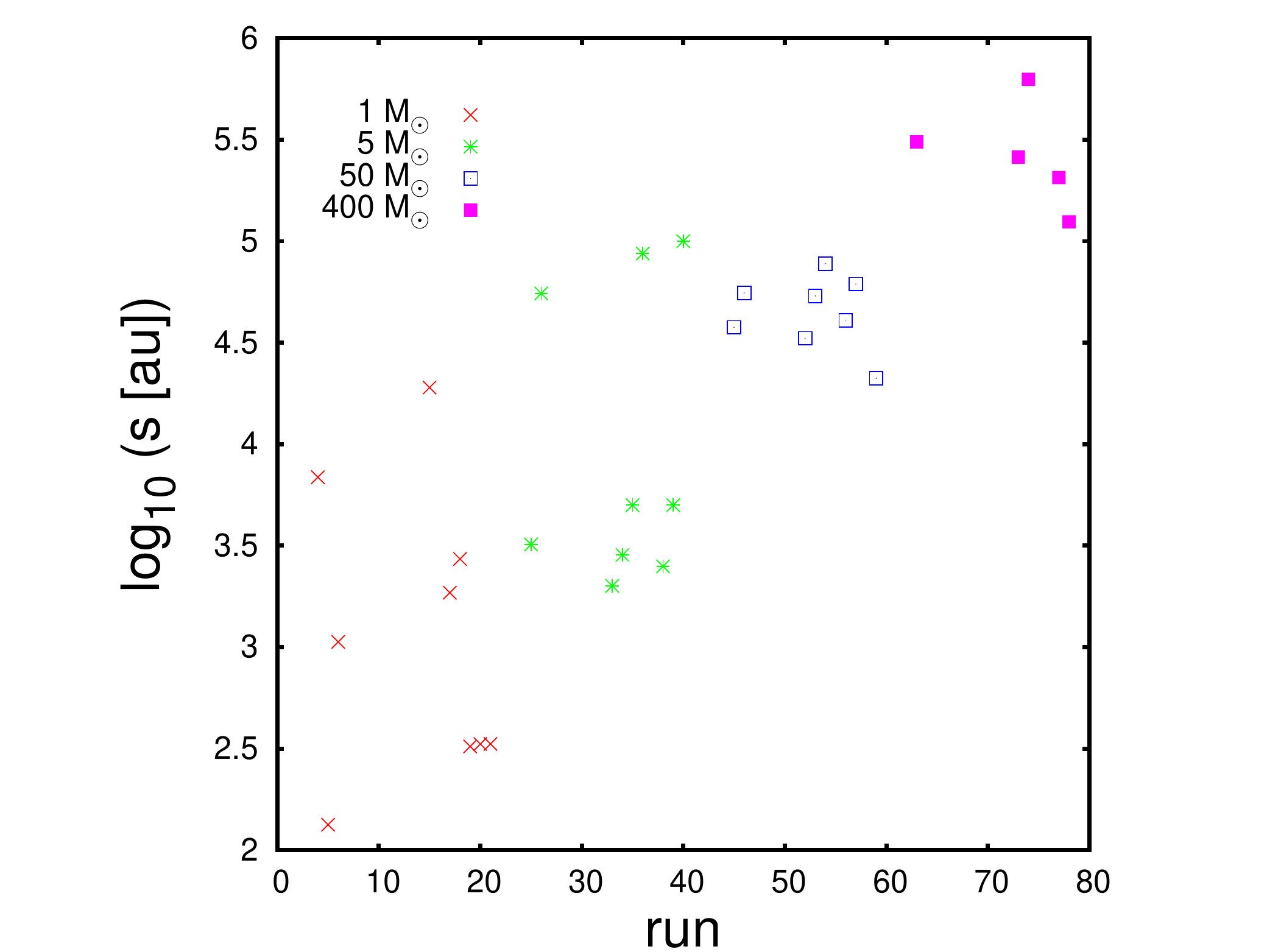}
\caption{\label{SepFrags} Binary separation $s$ in terms of the run number; see Table~\ref{tab:modelos}.}
\end{center}
\end{figure}

\begin{figure}
\begin{center}
\begin{tabular}{cc}
\includegraphics[width=3.0in,height=2.5in]{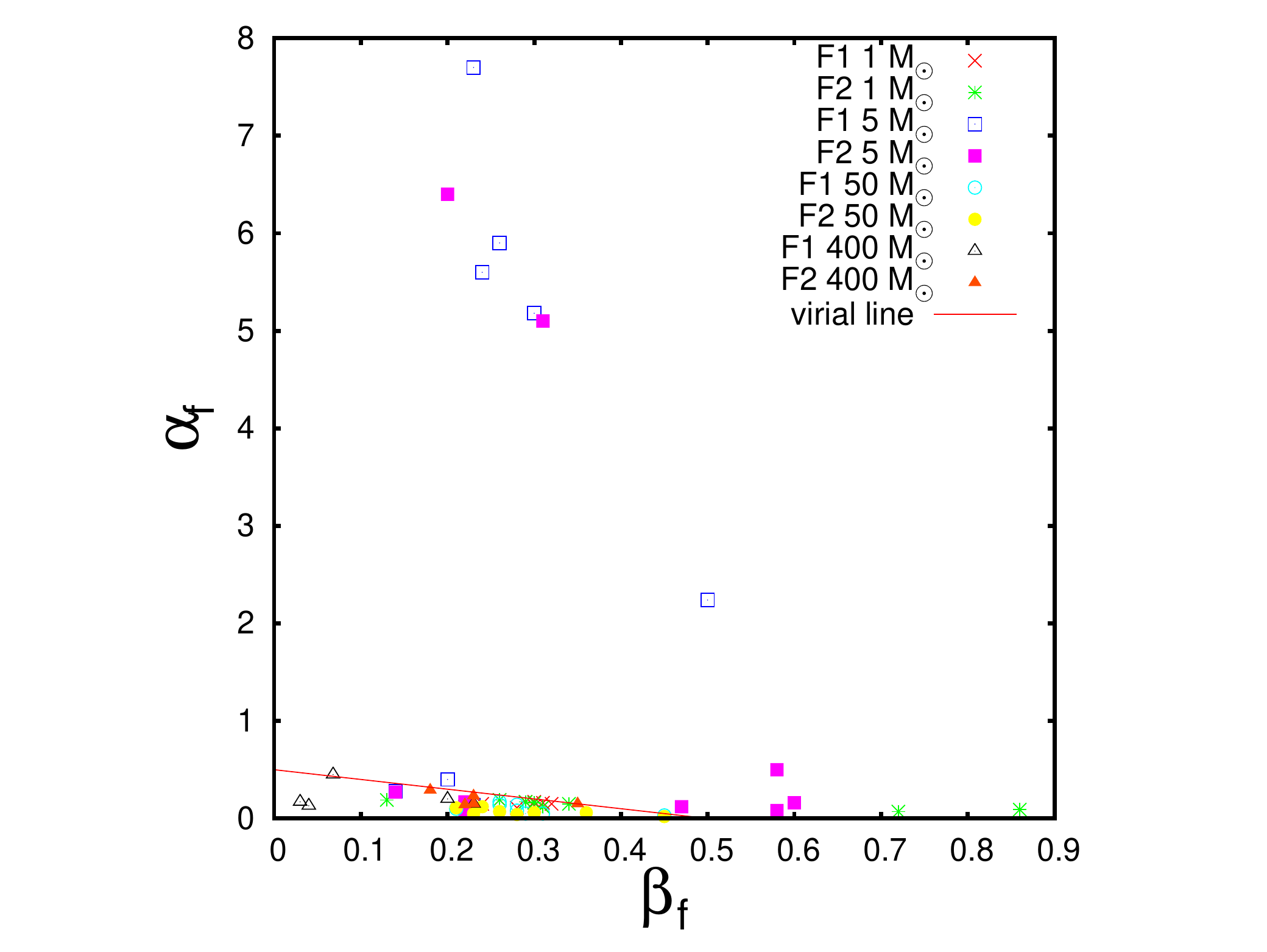} & \includegraphics[width=3.0in,height=2.5in]{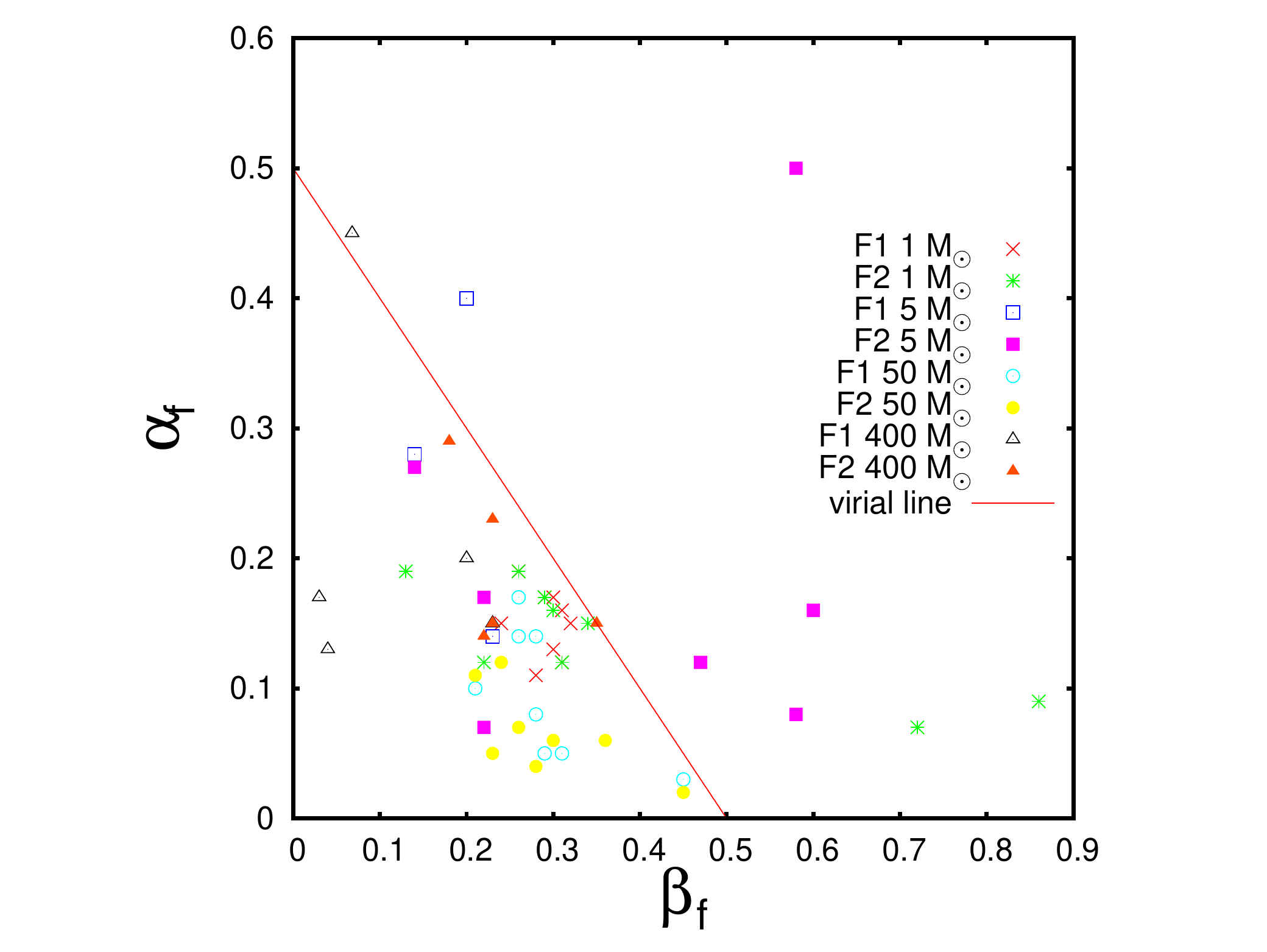}
\end{tabular}
\caption{\label{AlphavsBetaFrags} (left) Integral properties for binary fragments; (right) A zoom in of the left panel.}
\end{center}
\end{figure}

\begin{figure}
\begin{center}
\includegraphics[width=3.0in,height=2.5in]{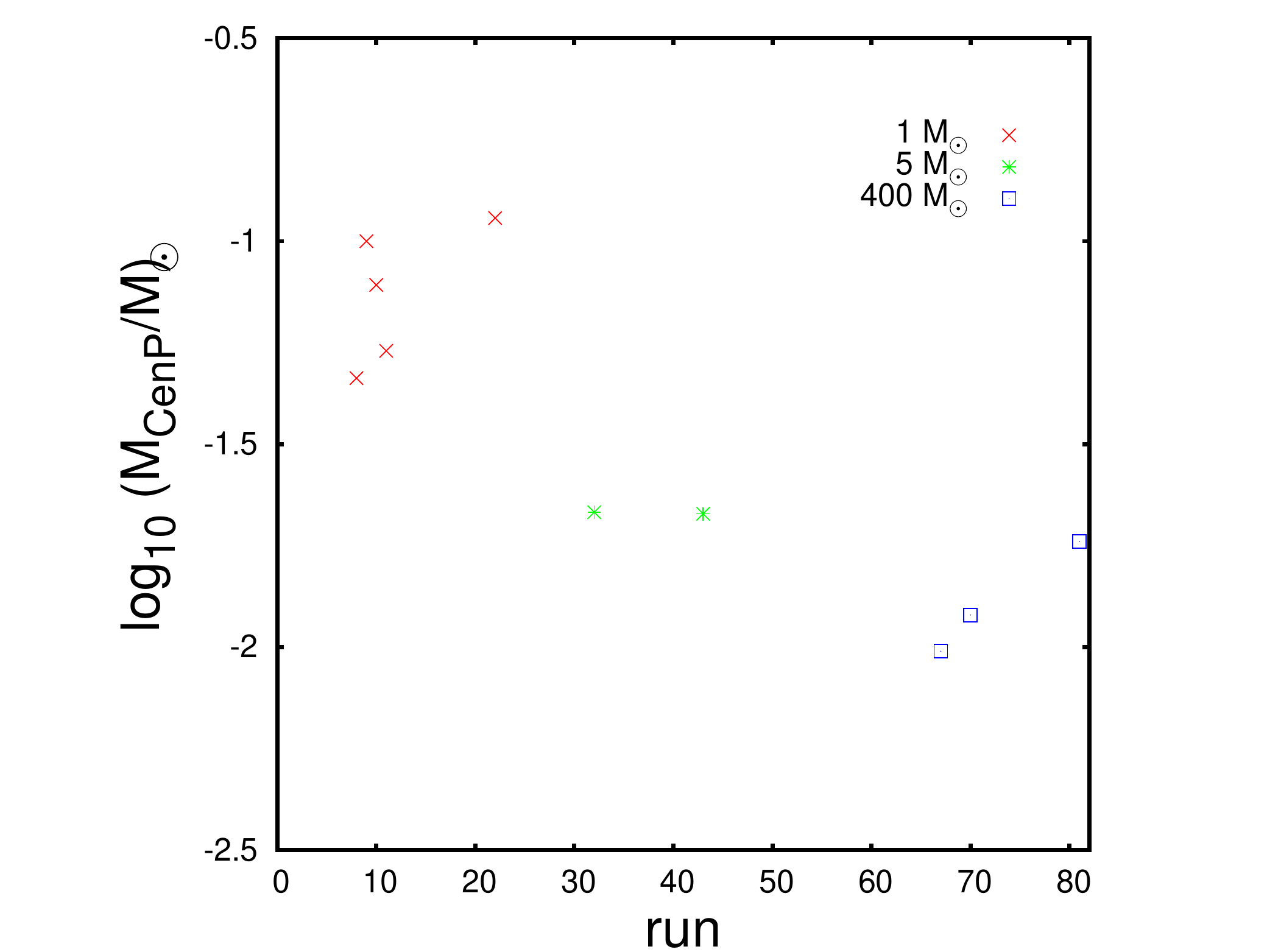}
\caption{\label{MassCenP} Central primary mass in terms of the run number; see Table~\ref{tab:modelos}.}
\end{center}
\end{figure}

\begin{figure}
\begin{center}
\includegraphics[width=3.0in,height=2.5in]{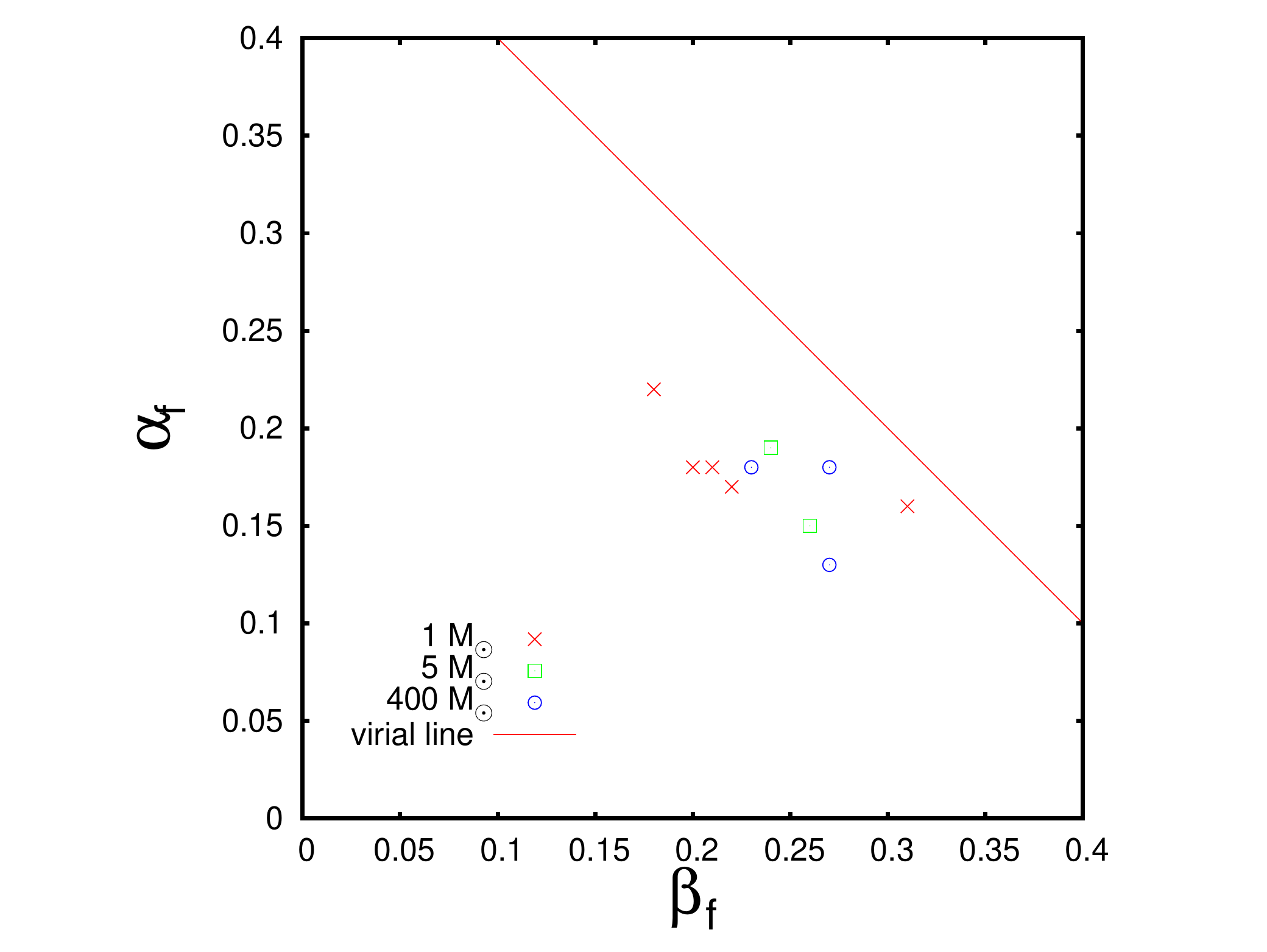}
\caption{\label{AlphavsBetaCenP} Integral properties for central
primaries.}
\end{center}
\end{figure}

The results obtained for the face-to-face configurations are shown in
Figs.\ref{SepFrags}, \ref{MassFrags} and \ref{AlphavsBetaFrags}. For the central primary
configurations, we report the mass of the central condensations results and their integral properties, which
are shown in Figs.\ref{MassCenP} and \ref{AlphavsBetaCenP}.

All the fragment properties of the face-to-face and of central primary configurations
have also been reported in Tables \ref{tab:propfrags} and \ref{tab:propfrags2},
respectively. The entries of these Tables
are as follows. The first column shows the number of the model; the total
parent mass M$_T$ is shown in the second
column; column third shows  r$_{max}$ in terms of the initial sphere
radius R$_0$; columns 4, 5 and 6
of Table \ref{tab:propfrags} show the mass of the fragments given in terms
of M$_T$ and the binary separation
in astronomical units; in the case of Table \ref{tab:propfrags2}, the last column
shows the mass of the central primary region in terms of M$_T$. The last columns give the values
of the $\alpha_f$ and $\beta_f$.

%%%%%%%%%%%%%%%%%%%%%%%%%%%%%%%%%%%%
%%%%%%%%%%%%%%%%%%%%%%%%%%%%%%%%%%%%
%%%%%%%%%%%%%%%%%%%%%%%%%%%%%%%%%%%%
%%%%%%%%%%%%%%%%%%%%%%%%%%%%%%%%%%%%%
\section{Discussion}
\label{sec:dis}

The mass seeds, implemented by means of the mass perturbation of Eq.~\ref{masspert},
accrete mass slowly during the early evolution stage of all the simulations until they
become a well-defined pair of mass condensations that move through a
gas of particles. Shortly thereafter, when the mass condensations are
massive enough, their translational motion can be slowed down or even stopped as a
consequence of the dynamical friction of the surrounding gas. Then, the
mass condensations appear to be face-to-face. At this time, a competition begins
between their gravitational force, which favors their approach, and
the centrifugal force due to the parent structure rotation, which favors
their separation. Also, the mass condensations may or may not be connected by
a filament, and as we have seen, these connecting filaments can be of
very different types depending on the total mass of the parent structure and
on the values of $\alpha$ and $\beta$.

For the M$_T$= 1 M$_{\odot}$ core models, a good combination of these
forces occurs such that the assembled mass condensations are sufficiently massive
yet the centrifugal force is strong enough for their gravitational attraction to
make them approach
each other, swing past each other, and finally separate to form
an orbiting binary system. We have observed that the occurrence of these events
is significantly increased when the mass perturbations weigh more, as the mass
condensations
are naturally more massive to overcome their centrifugal force more easily,
while the $\beta$ must still be high enough to make them swing past one
another.

By contrast, when the mass condensations are too massive, they
approach each other, make contact, and finally merge to form a single central
primary condensation. In this case, the gravitational
force between the mass condensations has too easily overcome the rotational
centrifugal force.

For the M$_T$= 5 M$_{\odot}$ and M$_T$= 50 M$_{\odot}$ models, we observed
that the mass condensations remain face-to-face in most of the runs, such that
if $\beta$ increases further, then the separation of the face-to-face
configuration gets larger up to the point that the parent structure no longer
collapses. However, for higher values of $\beta$, these face-to-face configurations
give rise to a central primary configuration, such that for a slightly higher
$\beta$, the parent structure does not collapse but instead expands outwards
but with spherical symmetry.

\begin{figure}
\begin{center}
\includegraphics[width=3.0in,height=2.5in]{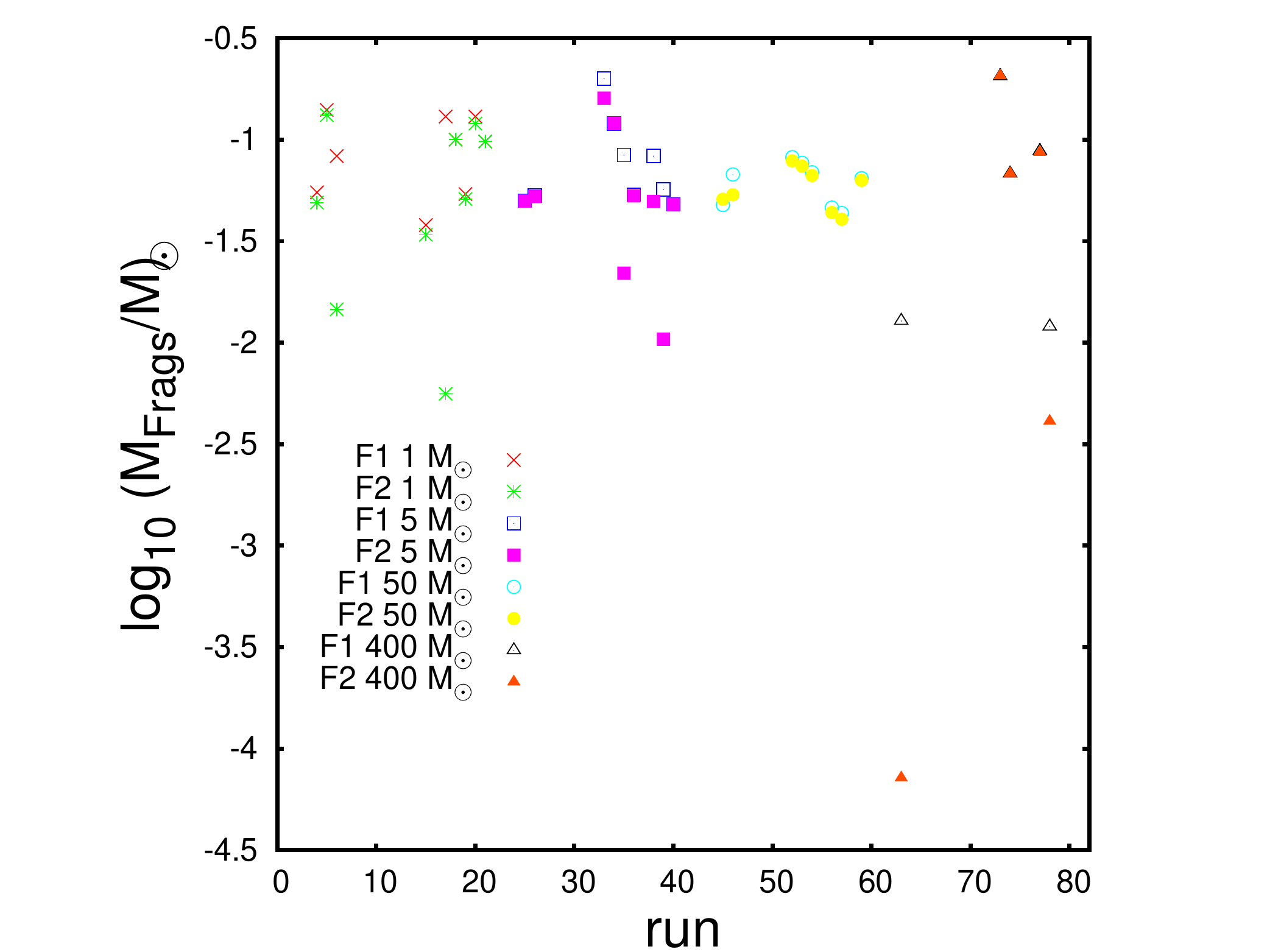}
\caption{\label{MassFrags} Fragment mass in terms of the run number; see Table~\ref{tab:modelos}.}
\end{center}
\end{figure}

We see in Fig.\ref{SepFrags} that the binary separation is
correlated with the mass of the parent gas structure.
It seems that no correlation exists in the case of the
fragment mass, as can be see in Fig.\ref{MassFrags}.
This missing correlation is expected on physical grounds, in fact, for the Taurus
 dark cloud, \citet{myers} reported a correlation between the mass of the newly
formed stars and the mass of the associated dense proto-stellar
cores. More recently, large proto-stellar masses were observed by
\cite{tobin}.

A potentially interesting issue is the fact that some binary systems
have a barycentre that is not at the Cartesian coordinate origin; see for
instance the panels of Fig.\ref{CPrueba30p1}. A
possible explanation for this fact is that the masses of the fragments of some binary
systems are different; see Fig.\ref{MassFrags} and Table\ref{tab:propfrags}. While it
is true that the perturbation mass implemented in Eq.\ref{masspert} has axial symmetry, it
should be remembered that each spherical radial shell was randomly populated by means of a
Monte Carlo method; see \ref{subs:setup}. We think that the origin of the
unequal masses of the fragments
and the displacement of the barycentre can be explained by the development of an
asymmetrical random mass seed, that grow over time and are manifested mainly later, in the
more evolved binary systems. The fact that some binary  systems show upward displacement
while others downward, gives support to this explanation.

We notice that the mass fraction of the central primary configurations is
similar for all the models reported
in Table \ref{tab:propfrags2}; this is so because the number of SPH particles
entering into the central region is similar
for all the models; therefore, the mass of the central primary configuration does not
scale with the mass of the parent structure;see Fig.\ref{MassCenP}.

It is clear that the values calculated for the energy
ratios $\alpha_f$ and $\beta_f$, unfortunately depend on the values chosen for
$\rho_{min}$ and $r_{max}$, so that there inevitably is certain ambiguity in defining
the fragment's boundaries. Despite of this, when we calculated the $\alpha_f$ and $\beta_f$ values
we observe that some binary fragments do show a clear tendency to virialize, as it can be
appreciated in Fig.\ref{AlphavsBetaFrags}. We emphasize that a similar
conclusion can be drawn from the calculation of ~\citet{NuestroRMAA}, where plots
of the $\alpha_f$ and $\beta_f$ time evolution were also presented. On the contrary, for
the central primaries, we do not observe any trend to approach the virial line;
see Fig.\ref{AlphavsBetaCenP}.

%%%%%%%%%%%%%%%%%%%%%%%%%%%%%%%%%%%%%%
%%%%%%%%%%%%%%%%%%%%%%%%%%%%%%%%%%%%%%
%%%%%%%%%%%%%%%%%%%%%%%%%%%%%%%%%%%%%%
%%%%%%%%%%%%%%%%%%%%%%%%%%%%%%%%%%%%%%%
\section{Concluding Remarks}
\label{sec:conclu}

In this paper, we have considered the gravitational collapse of several rotating
spherical gas structures of very different size and mass, such that the collapse was
triggered initially by means of the same azimuthal mass perturbation, whose
amplitude has been allowed to take the values $a=$0.1 and $a=$0.25. For the sake of
comparison between our models, we have used the same set of
values of the $\alpha$ and $\beta$ ratios.

The range of $\alpha$ considered here, 0.1-0.3, was similar to that of \cite{riaz};
this range reflects the observational fact that the temperatures of
the gas structures are mostly in the range of 9-12 K; see \cite{bergin}. The range
of $\beta$ considered here, 0.1,0.3 and 0.48, was initially motivated by recent observations
of the virial parameters of clouds, so that the critical virial parameter
is within the chosen range, as we mentioned in Sect.\ref{intro}.

However, we have extended this range of $\beta$ in order to look for the last collapsing
configurations, for which purpose the new range has been increased to values of 0.9 for
the lowest $\alpha$ and around 0.4 for the highest $\alpha$.
This range of $\beta$ values is wider than those considered by other authors;
see for instance, \cite{tsuribe1}, who used a maximum $\beta$ of 0.3.

We have thus revisited many binary configurations already seen in other
works, but here the configurations change depending on the mass of the parent
structure. Based on these results, we present the following assessment of the effect
of parent mass on the collapse.

The resulting configurations for panels with $\alpha=$0.1 have a well-defined
pair of mass condensations connected by a dense filament for the
M$_T$= 1 M$_{\odot}$ model. The connecting filaments become weaker and almost disappear
when the total mass increases to M$_T$= 5 M$_{\odot}$ and M$_T$= 50 M$_{\odot}$,
respectively. The last collapsing configurations are of the same kind; that
is, of the face-to-face type; see Sect.\ref{sec:resultados}.

When $\alpha=$0.2, we have observed the formation of many orbiting
binary configurations for the M$_T$= 1 M$_{\odot}$ model; this tendency
becomes more pronounced when the mass perturbation increases from $a=$0.1
to $a=0.25$, so that the more massive the mass condensations, the greater the
probability of an orbiting binary configuration.

\begin{figure}
\begin{center}
\begin{tabular}{ccc}
\includegraphics[width=2.2in]{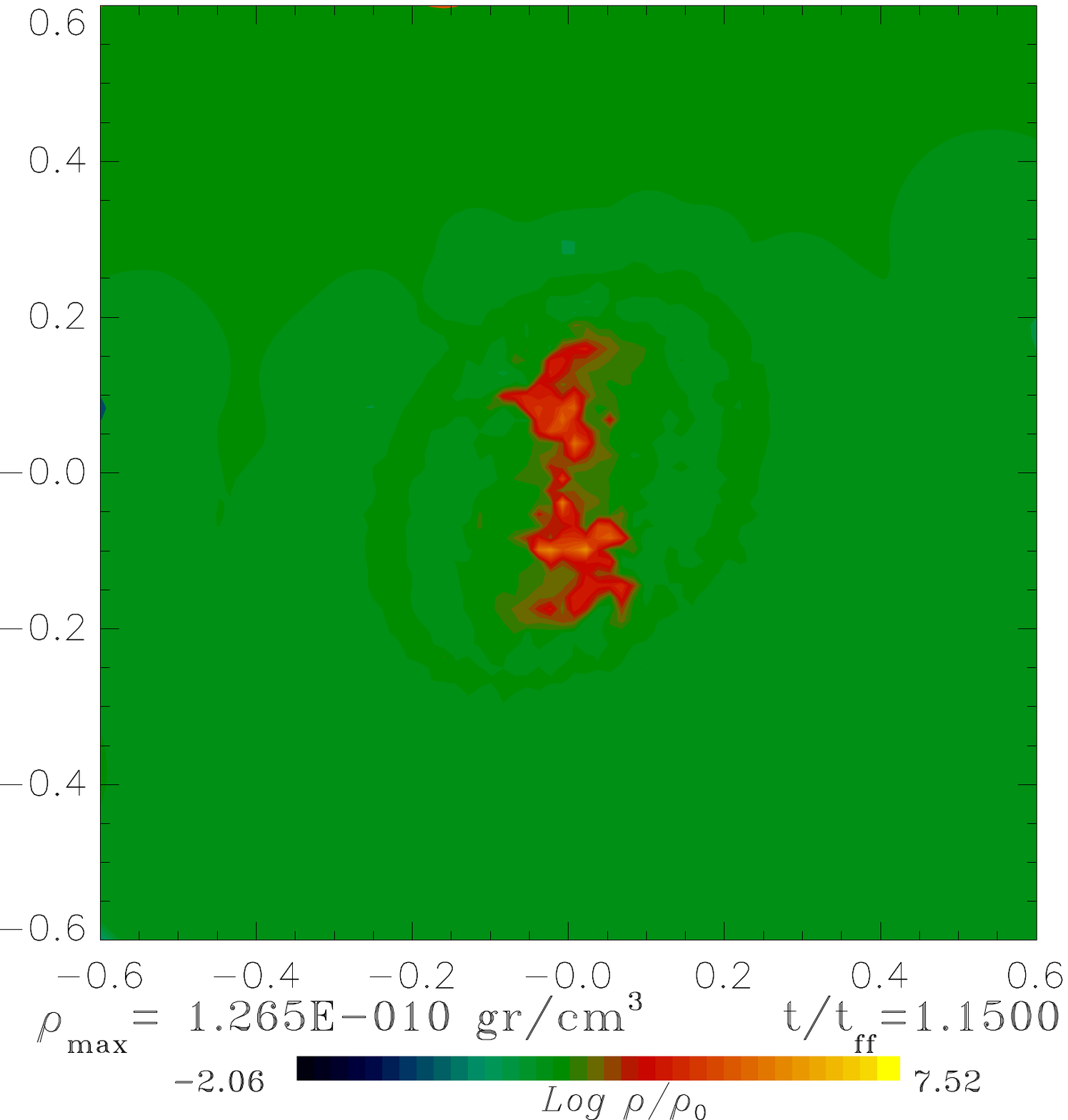} & \includegraphics[width=2.2in]{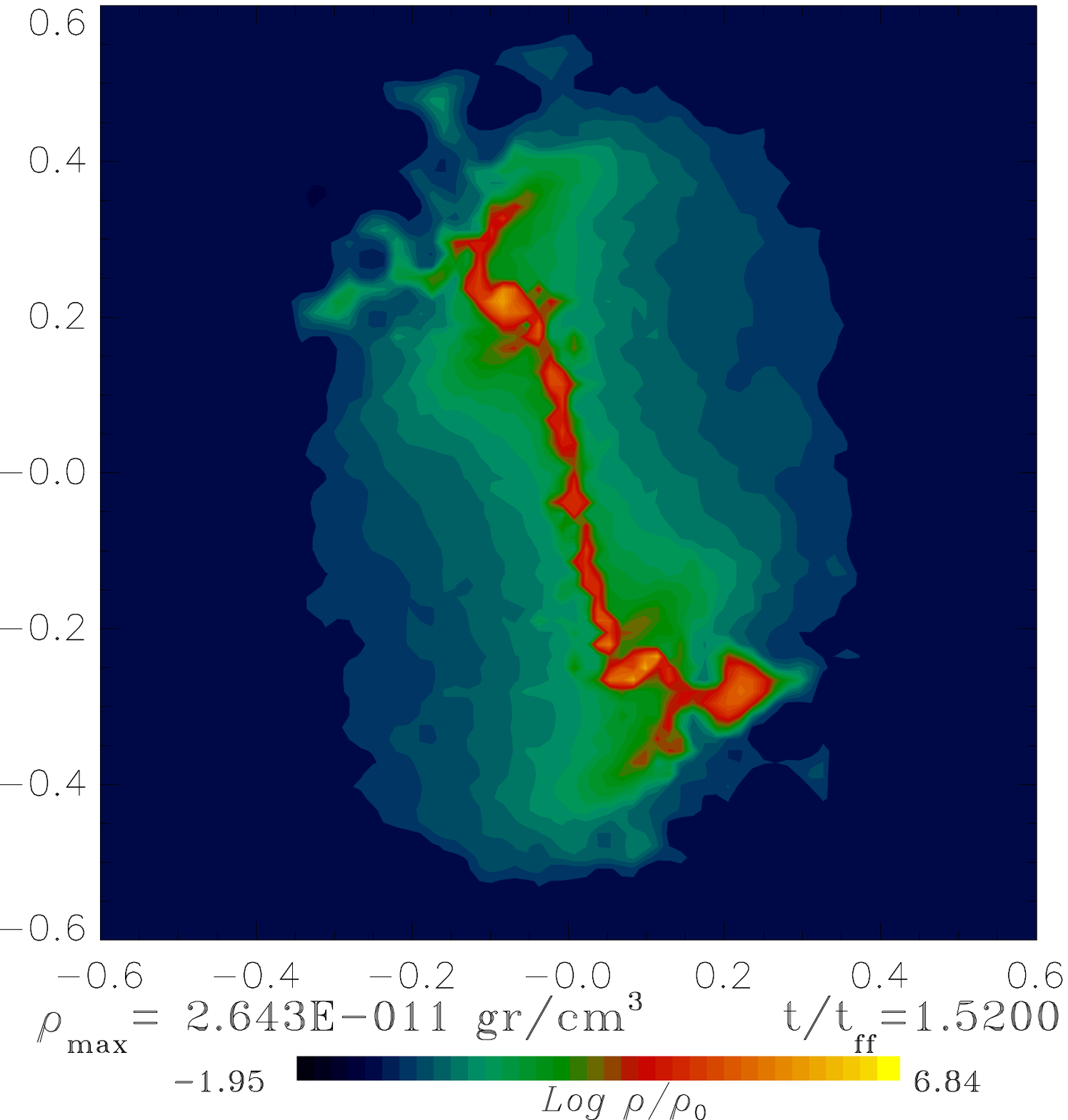}&
\includegraphics[width=2.2in]{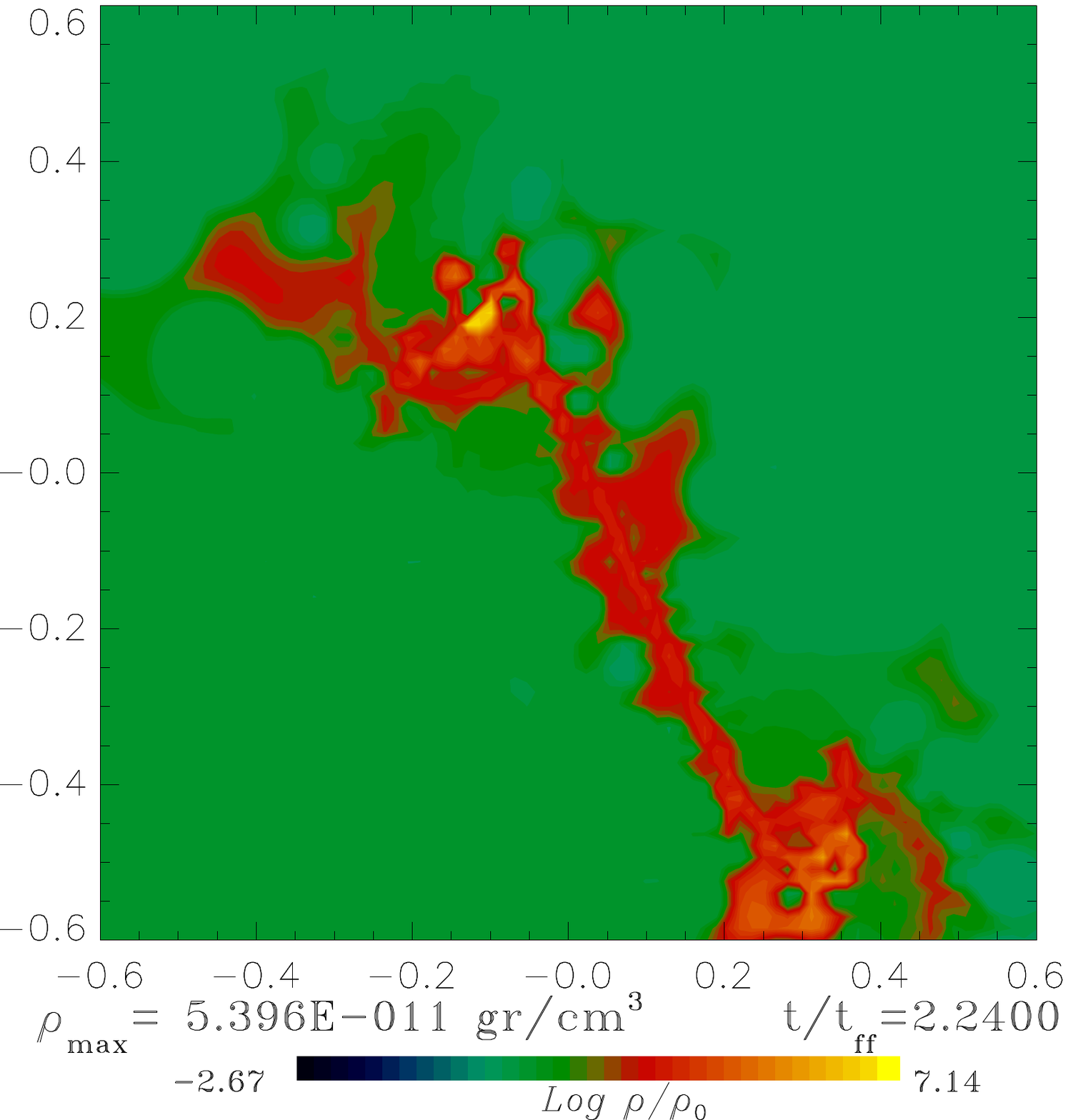}
\end{tabular}
\caption{\label{CPrueba60p1} Core models with M$_T$= 1 M$_{\odot}$, $a=$0.1
and $\alpha=$0.1; the corresponding $\beta$ are
(left) 0.1
(middle) 0.3
(right) 0.48.}
\end{center}
\end{figure}

\begin{figure}
\begin{center}
\begin{tabular}{ccc}
\includegraphics[width=2.2in]{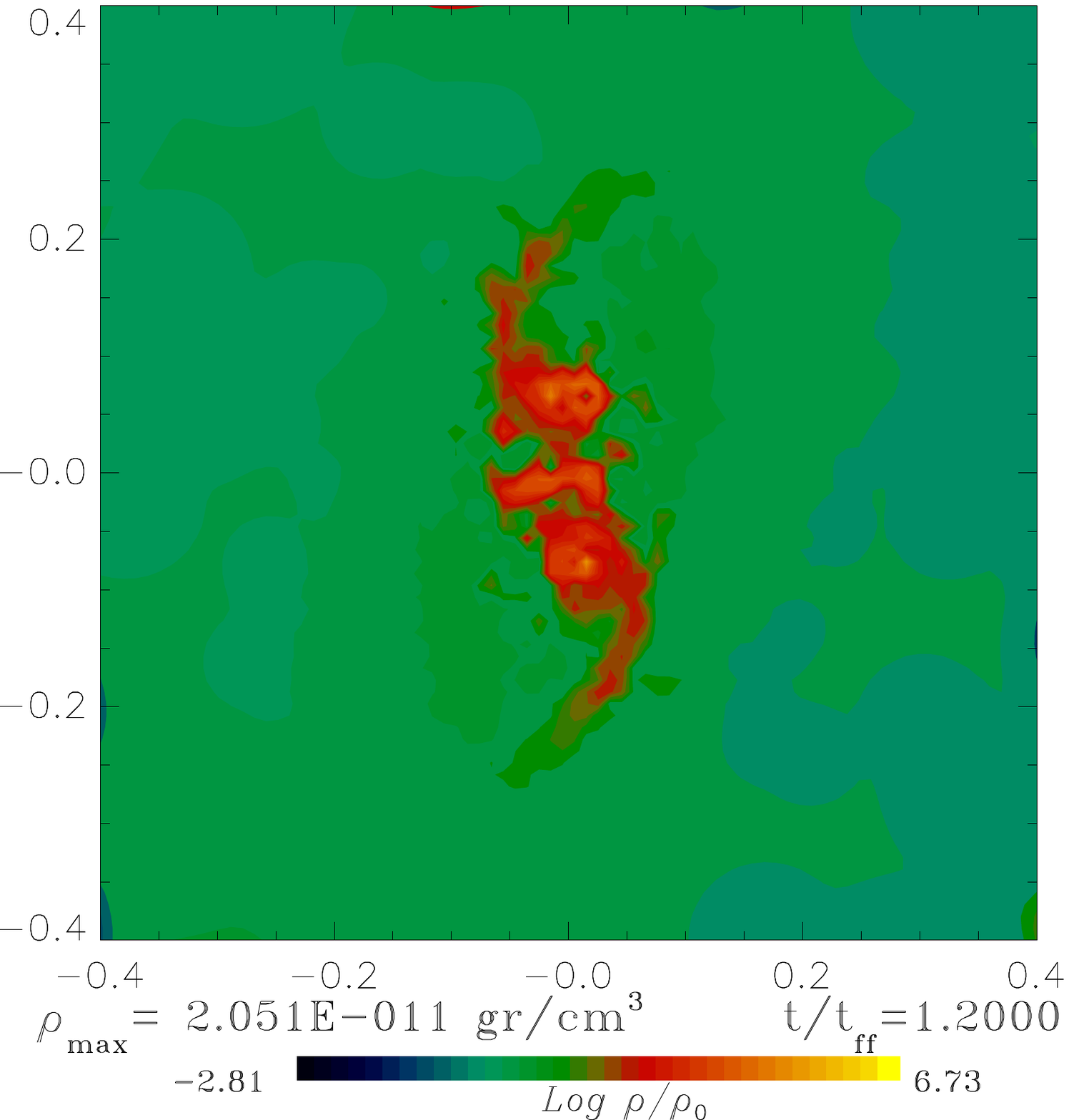} &\includegraphics[width=2.2in]{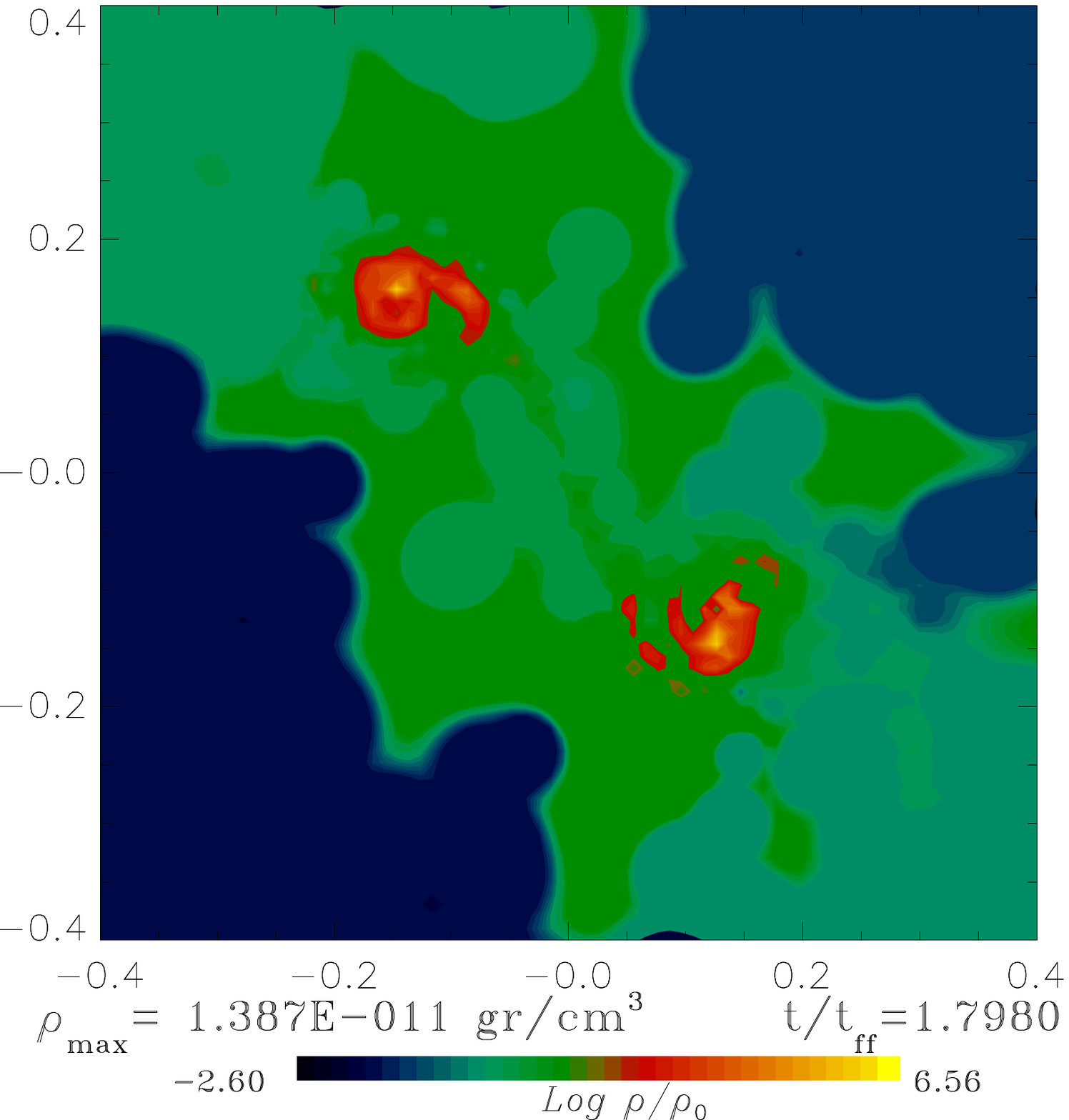} &
\includegraphics[width=2.2in]{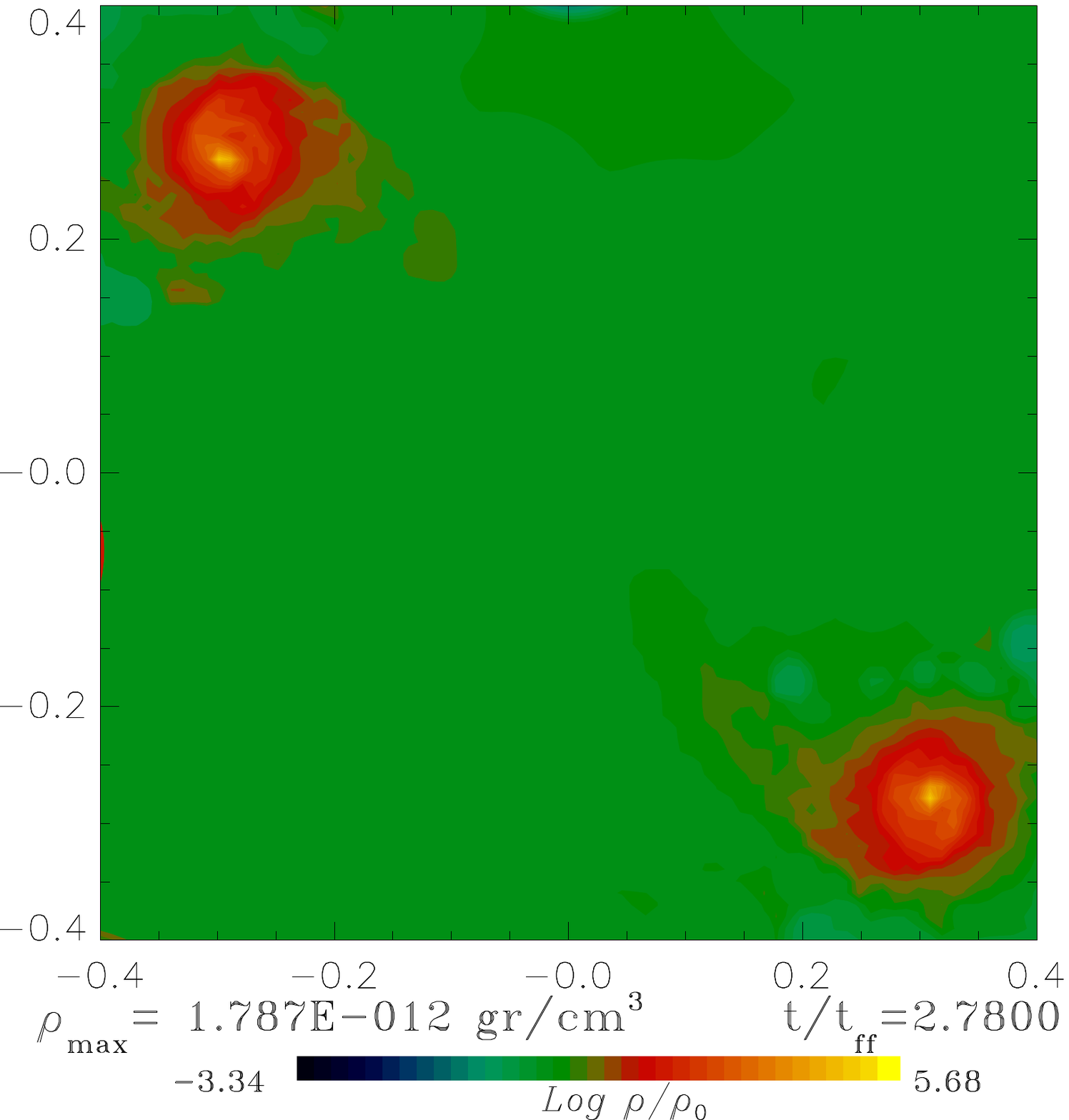}
\end{tabular}
\caption{\label{CPrueba70p2} Core models with M$_T$=1 M$_{\odot}$, $a=$0.25 and $\alpha=$0.2;
the corresponding $\beta$ are
(left) 0.1
(middle) 0.3
(right) 0.48.}
\end{center}
\end{figure}

\begin{figure}
\begin{center}
\begin{tabular}{ccc}
\includegraphics[width=2.2in]{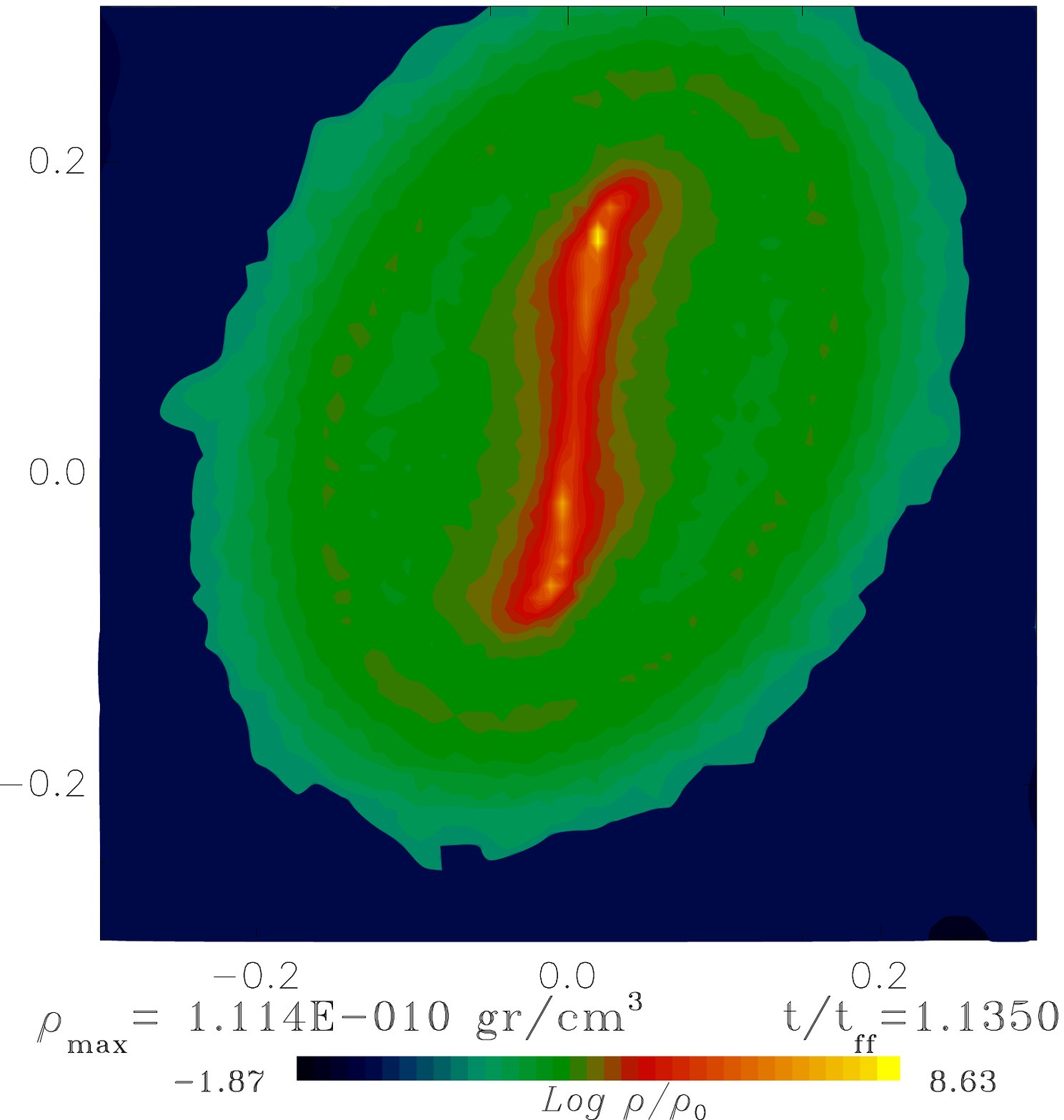} & \includegraphics[width=2.2in]{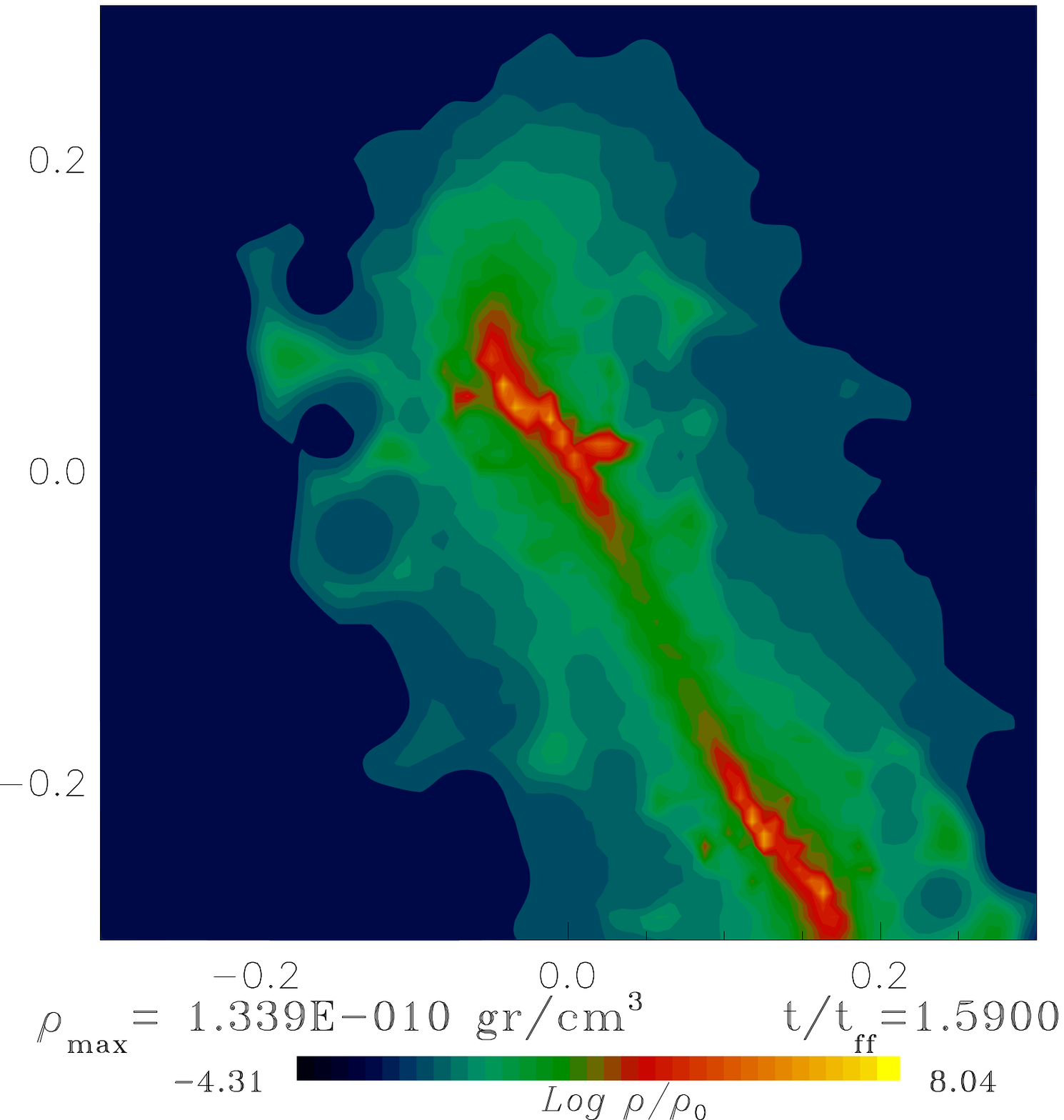} &
\includegraphics[width=2.2in]{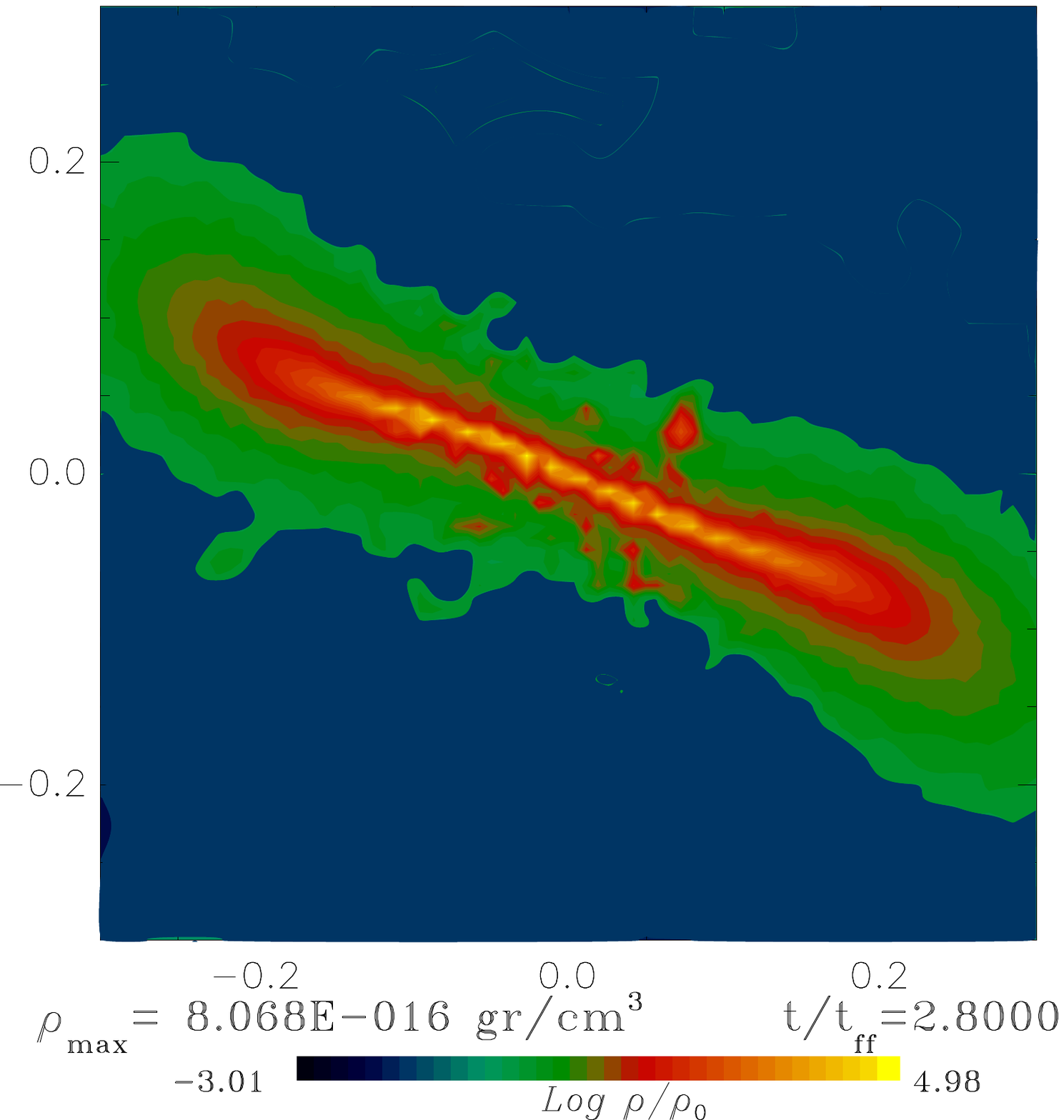} \\
\end{tabular}
\caption{\label{CPrueba30p2} Core models with M$_T$= 5 M$_{\odot}$, $a=$0.1 and $\alpha=$0.2;
the corresponding $\beta$ are
(left) 0.1
(middle) 0.3
(right) 0.48. This is the last collapsing configuration; therefore, there is a missing panel in
Fig.\ref{lastcolconfM5}.}
\end{center}
\end{figure}

\begin{figure}
\begin{center}
\begin{tabular}{ccc}
\includegraphics[width=2.2in]{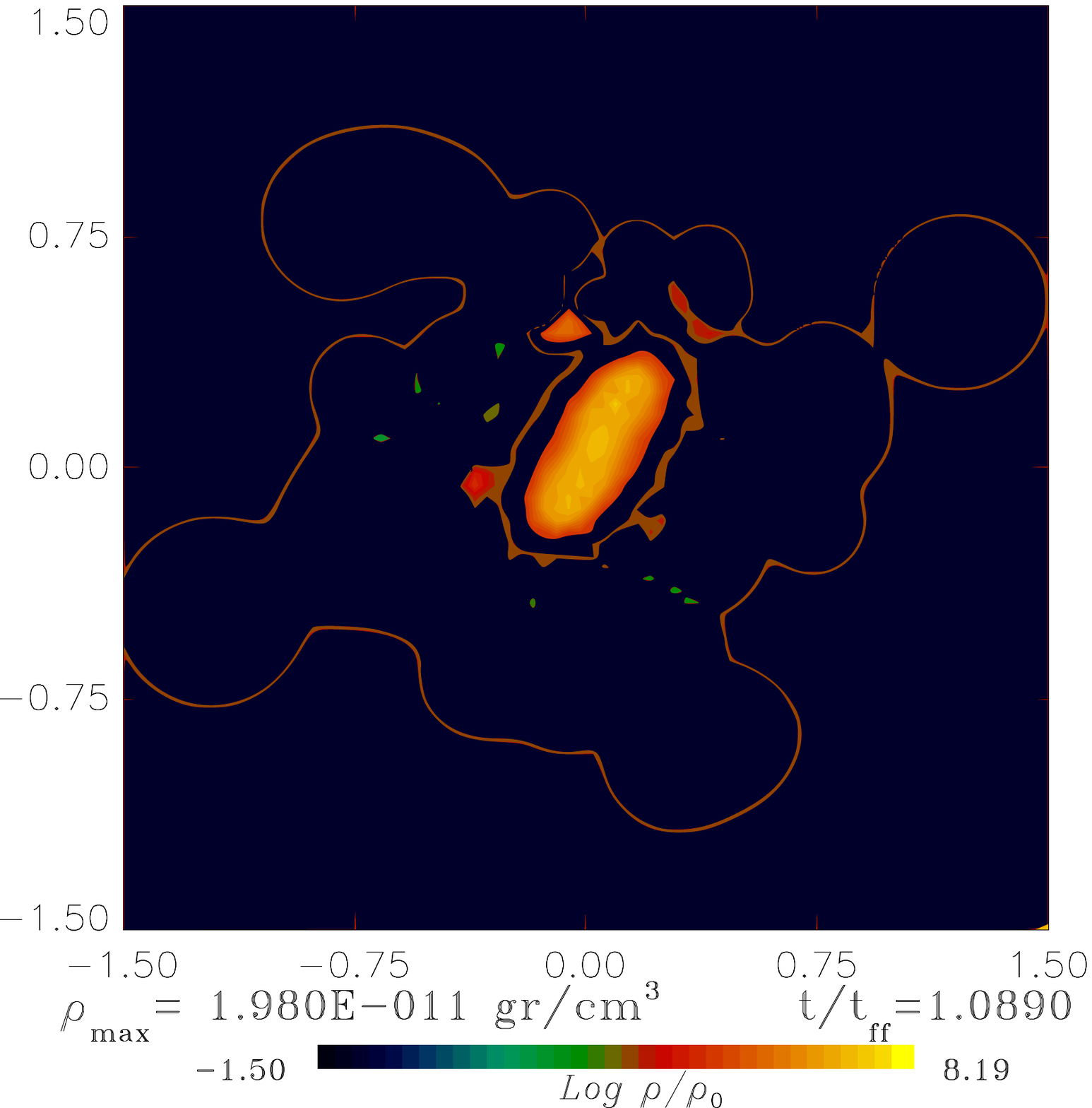} & \includegraphics[width=2.2in]{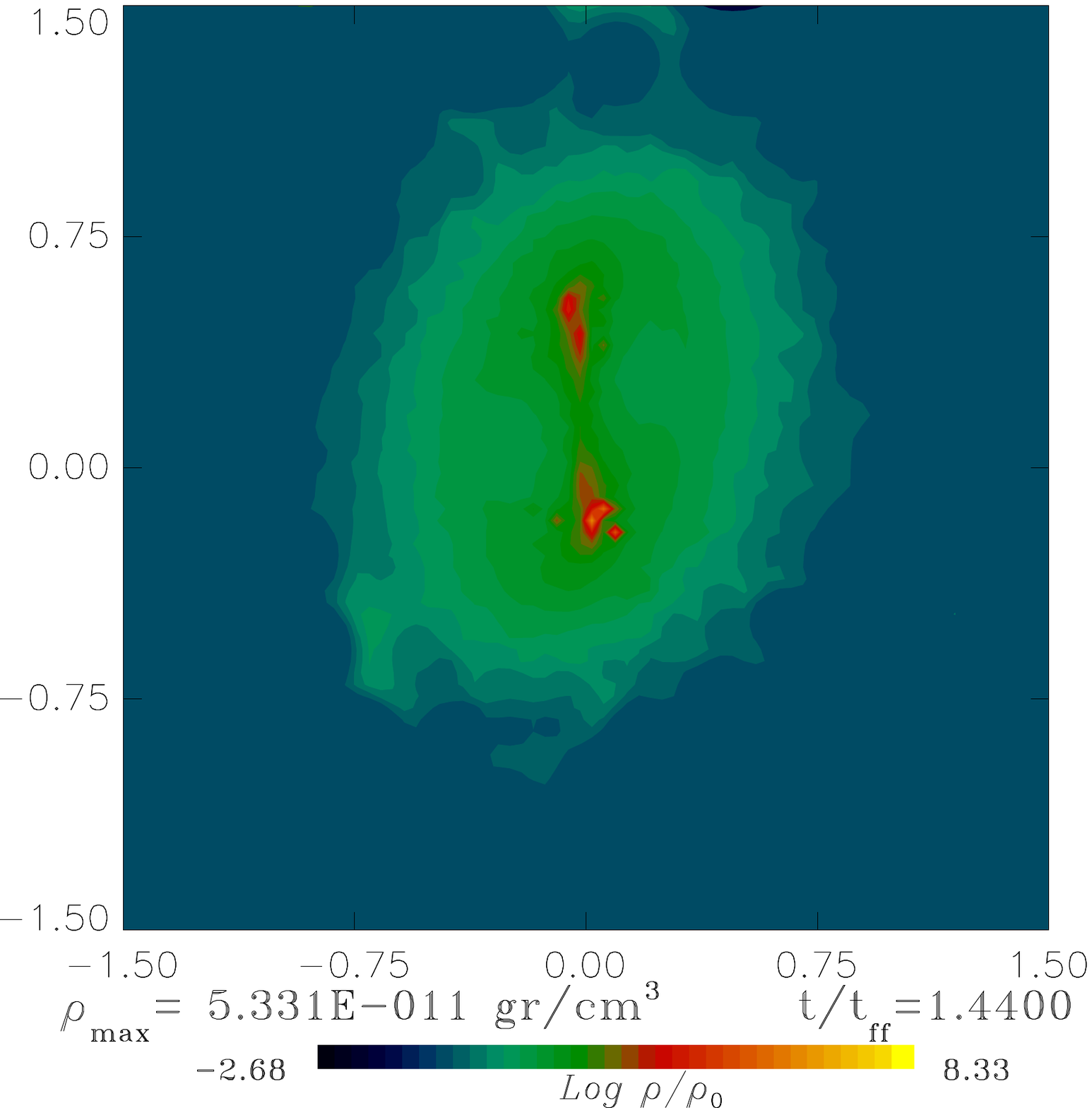} &
\includegraphics[width=2.2in]{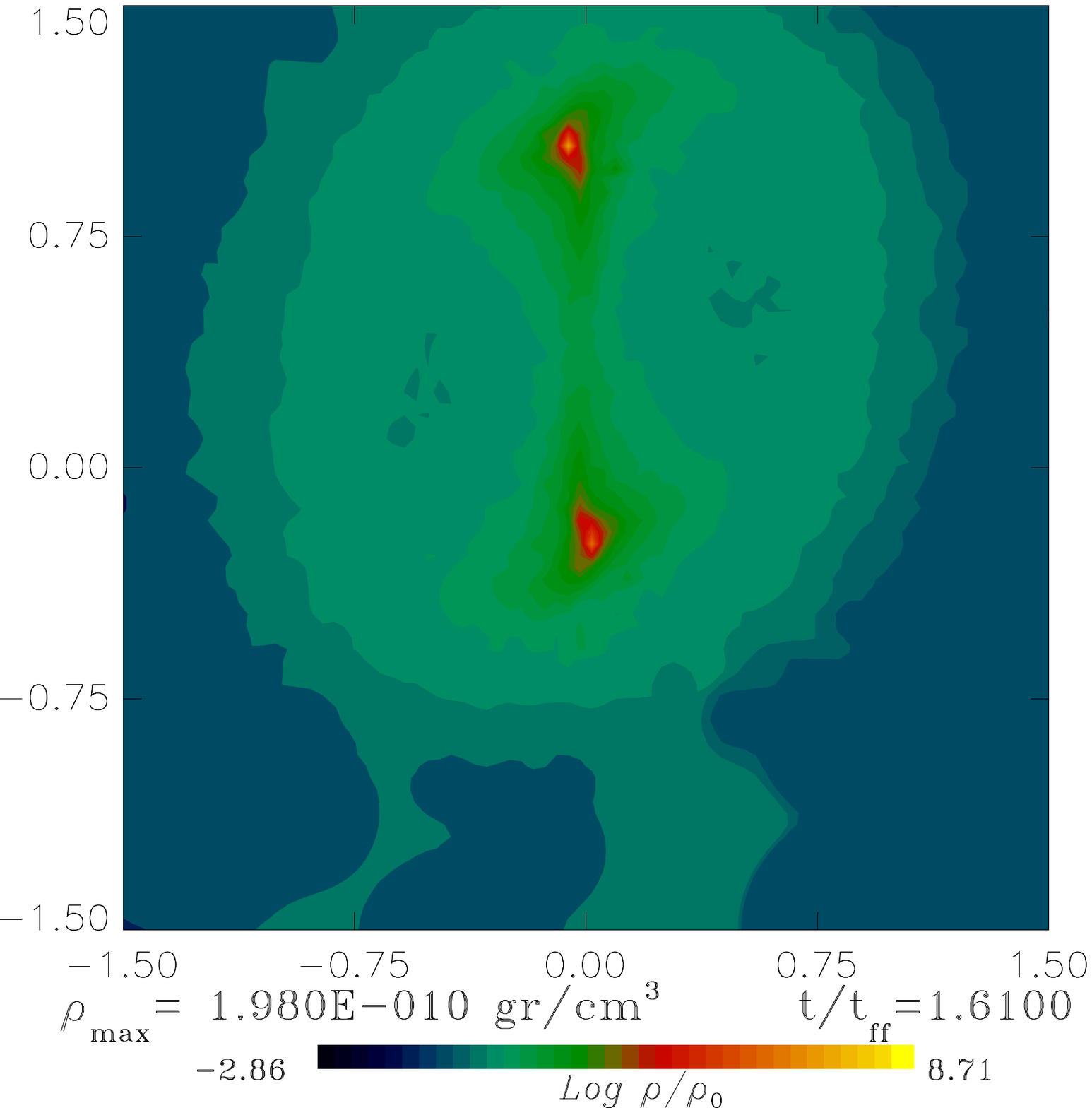} \\
\end{tabular}
\caption{\label{CPrueba40p2} Core models with M$_T$= 5 M$_{\odot}$, $a=$0.25 and $\alpha=$0.2;
the corresponding $\beta$ are
(left) 0.1
(middle) 0.3
(right) 0.48.}
\end{center}
\end{figure}

\begin{figure}
\begin{center}
\begin{tabular}{ccc}
\includegraphics[width=2.2in]{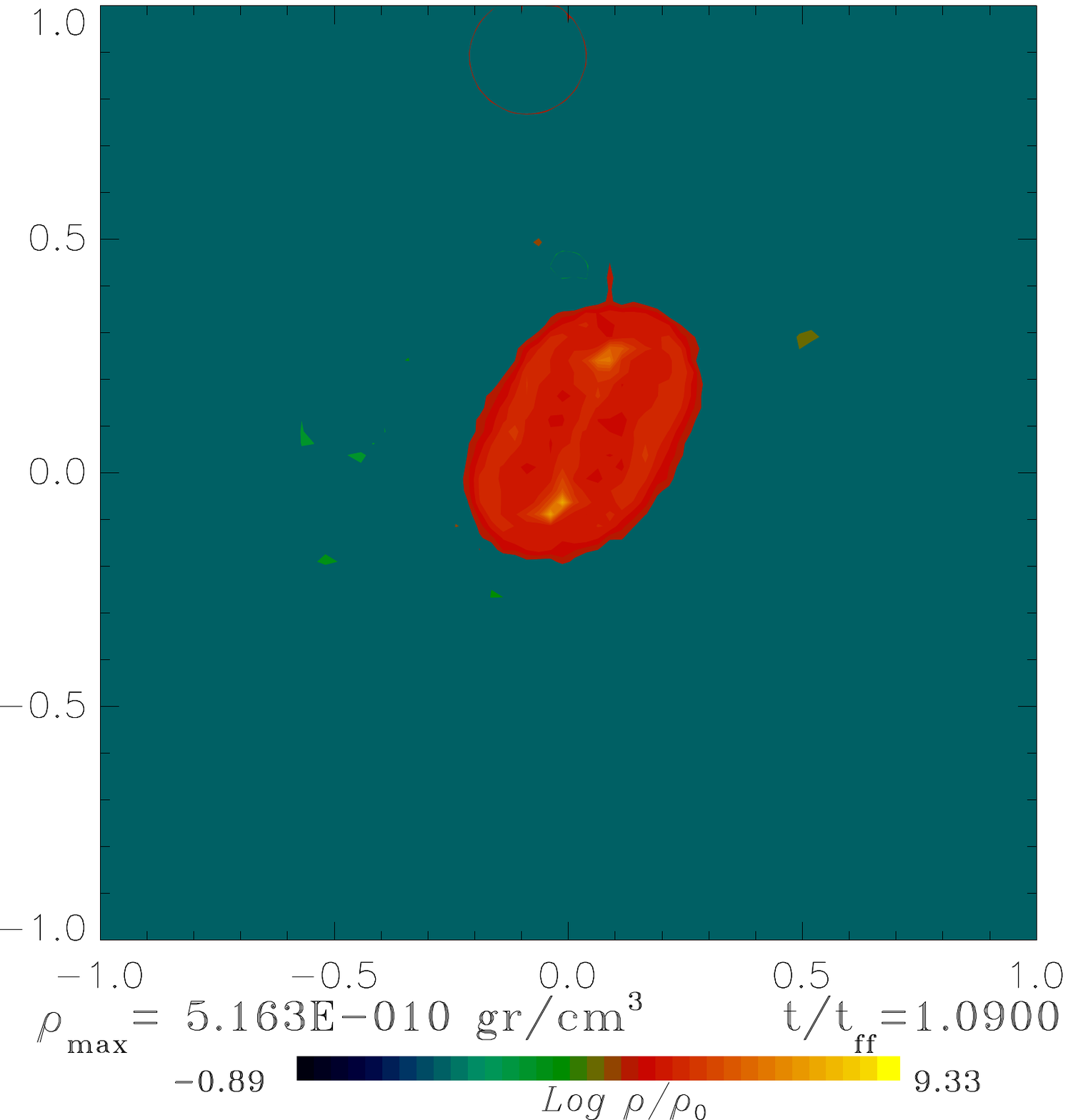} & \includegraphics[width=2.2in]{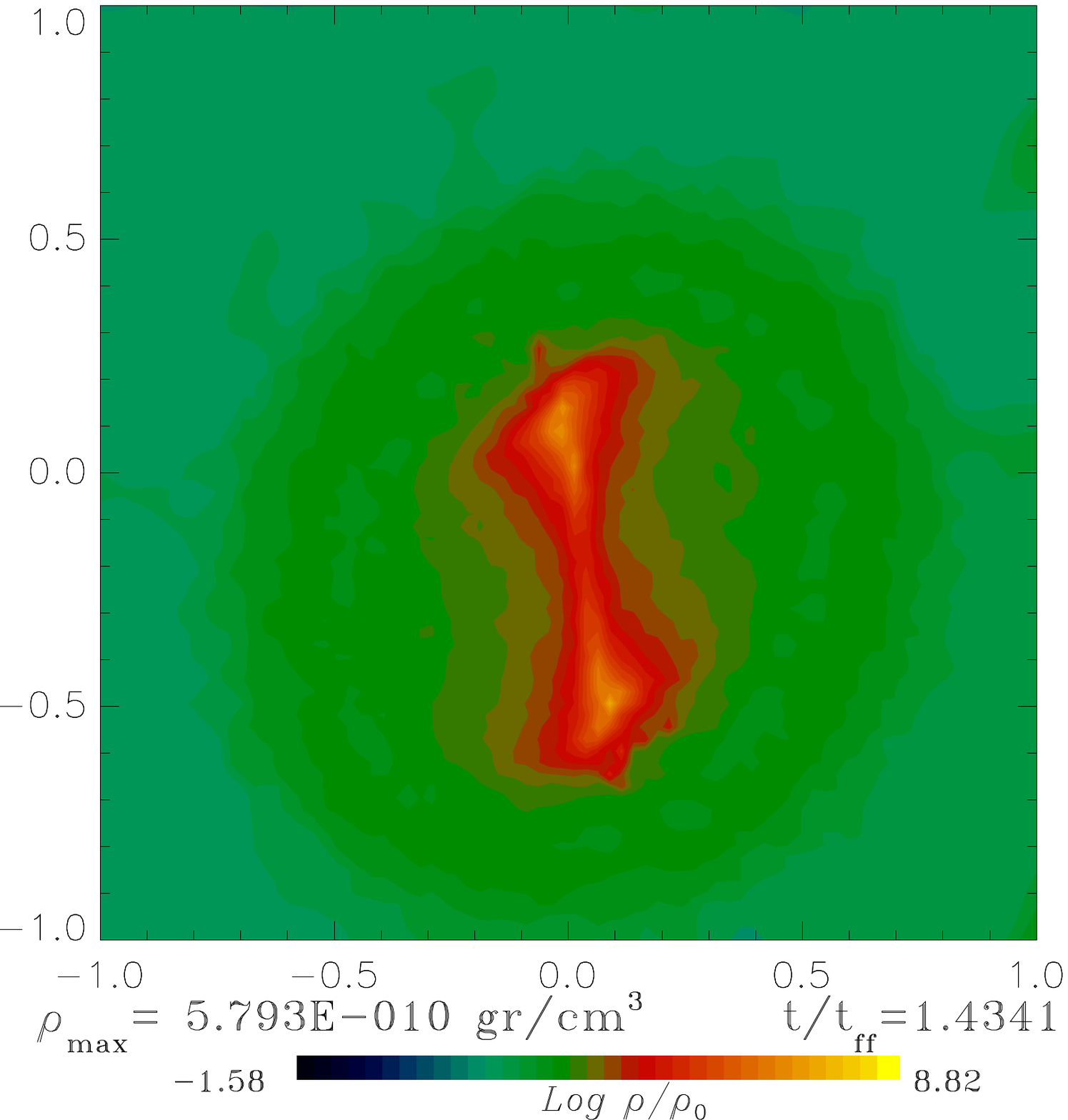} &
\includegraphics[width=2.2in]{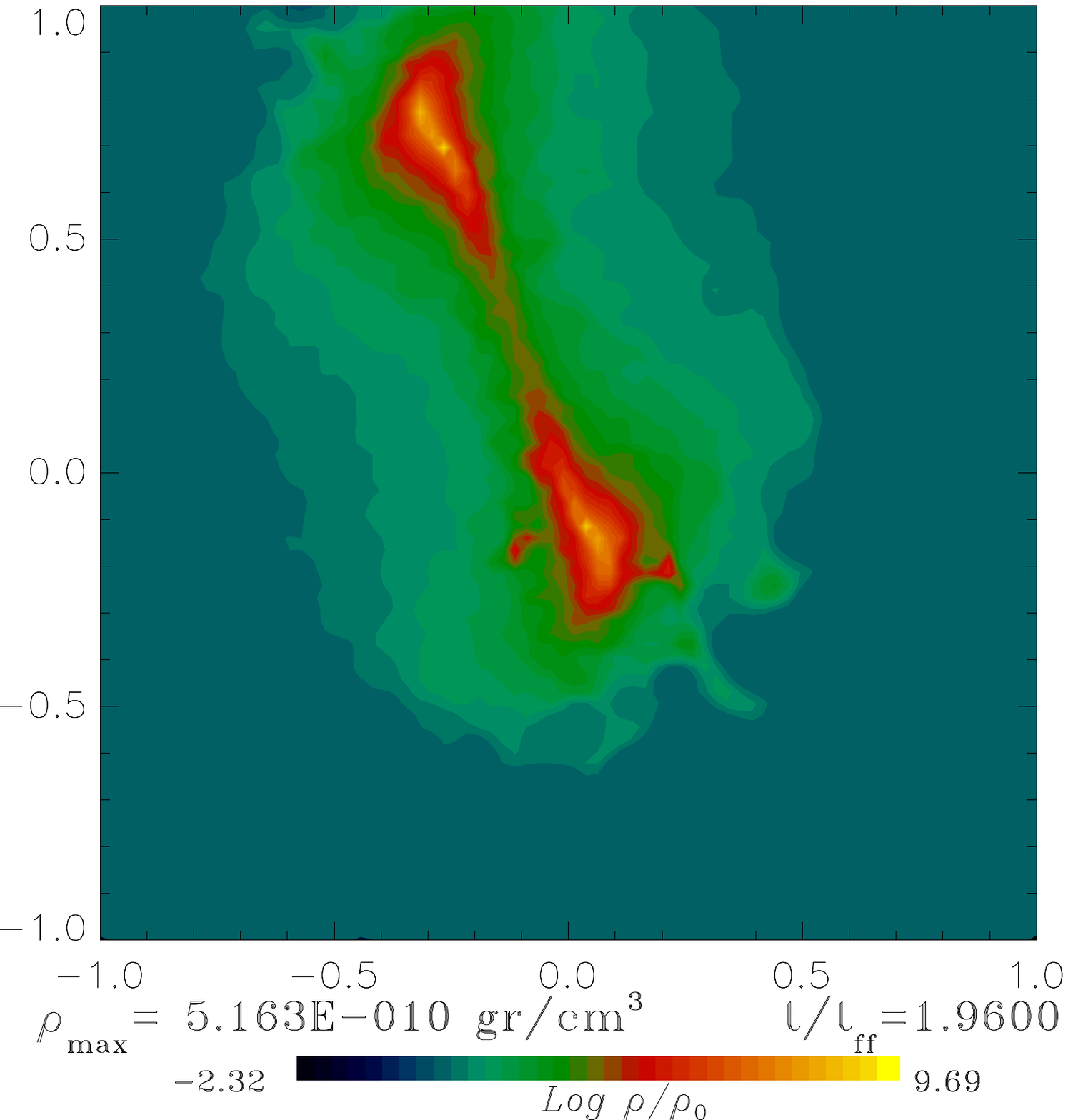} \\
\end{tabular}
\caption{\label{CPrueba0p1} Clump models with M$_T$ 400 M$_{\odot}$, $a=$0.1
and $\alpha=$0.1; the corresponding $\beta$ are
(left) 0.1
(middle) 0.3
(right) 0.48.}
\end{center}
\end{figure}

\begin{figure}
\begin{center}
\begin{tabular}{ccc}
\includegraphics[width=2.2in]{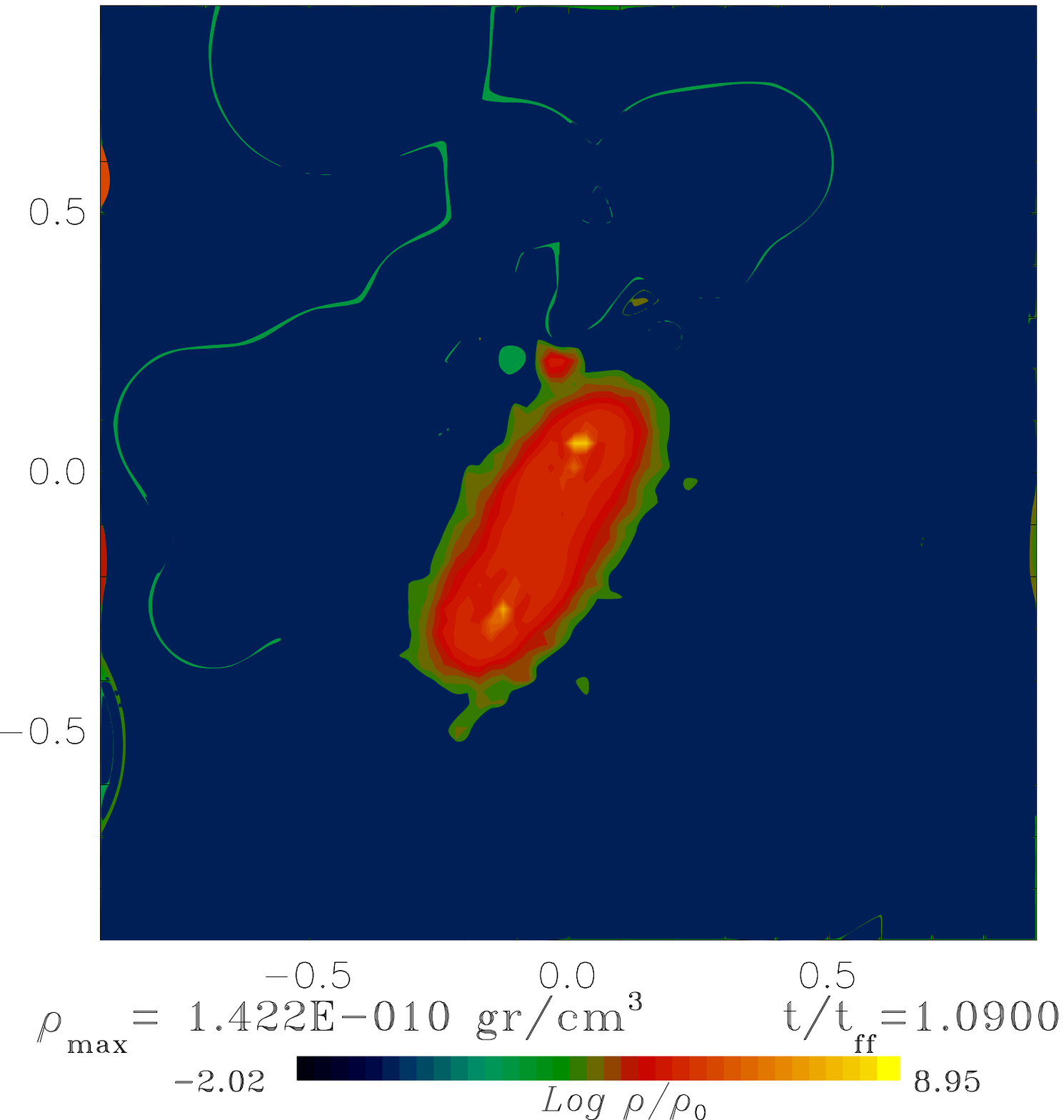} &\includegraphics[width=2.2in]{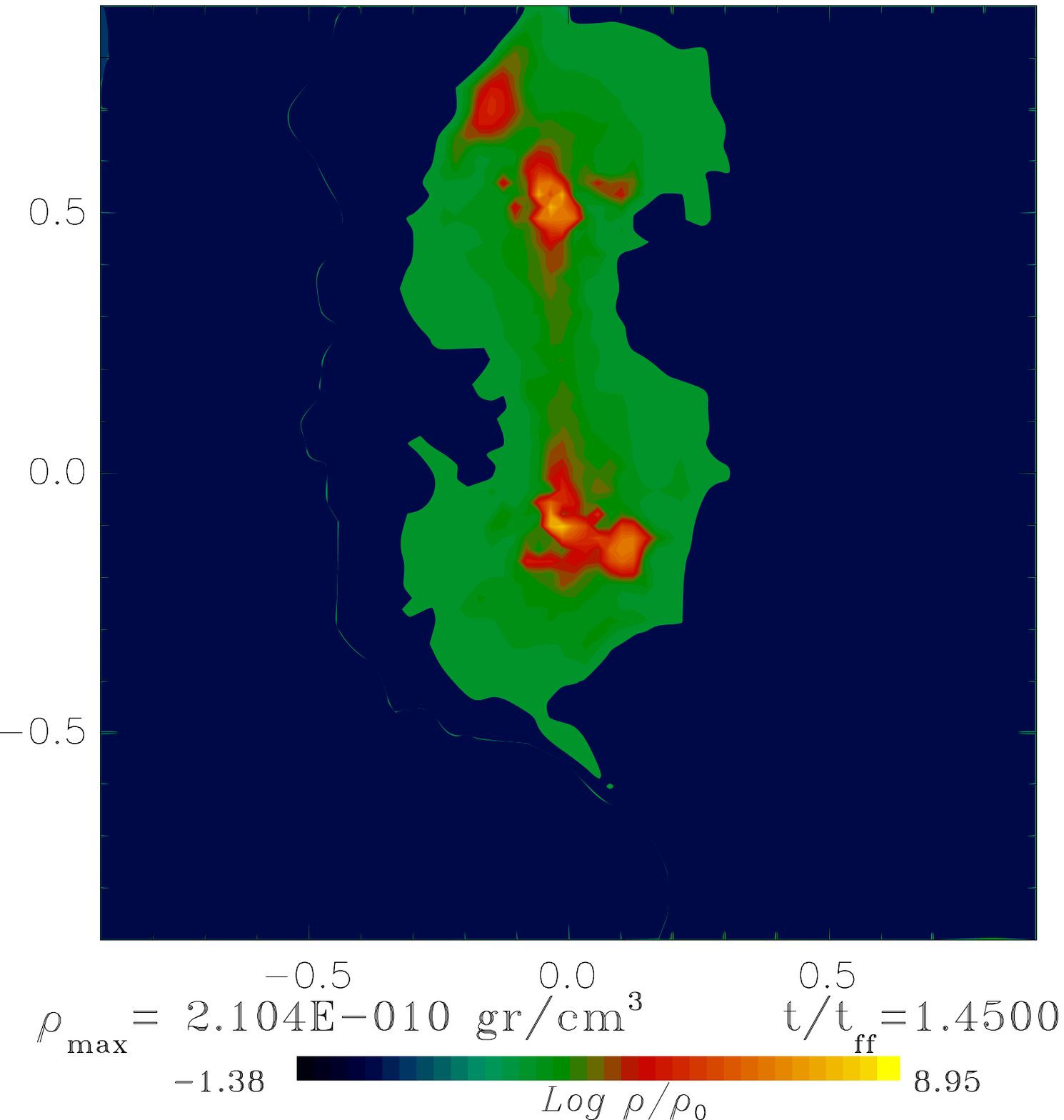} &
\includegraphics[width=2.2in]{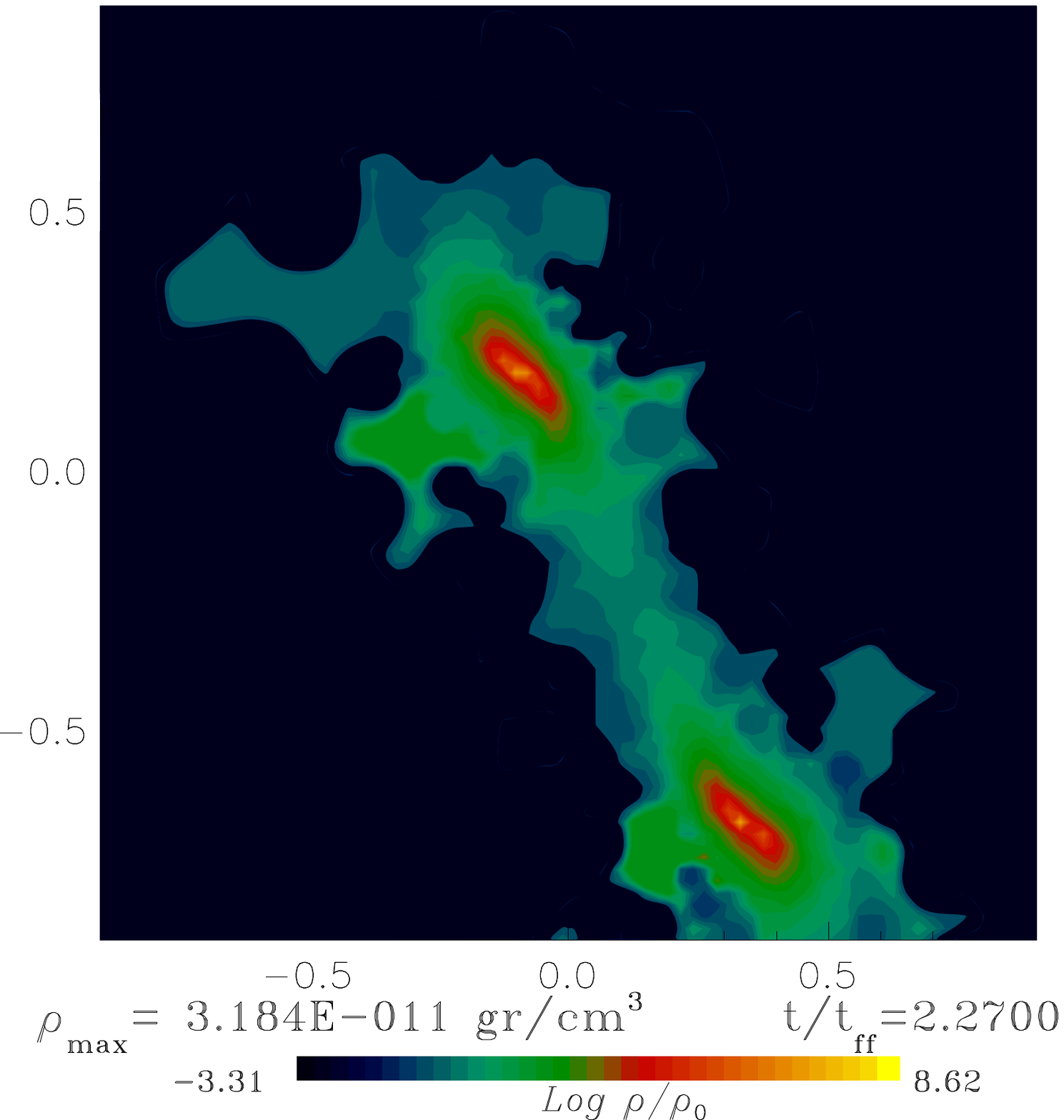}\\
\end{tabular}
\caption{\label{CPrueba50p2} Clump models with M$_T$= 400 M$_{\odot}$, $a=$0.25 and $\alpha=$0.2;
the corresponding $\beta$ are
(left) 0.1
(middle) 0.3
(right) 0.48.}
\end{center}
\end{figure}

For the M$_T$= 5 M$_{\odot}$ model, what has been observed primarily are
configurations where
thin and dense filaments connect two very small mass condensations. When
the mass of the parent gas structure is
M$_T$= 50 M$_{\odot}$, mainly face-to-face configurations are
formed, in which the mass condensations are predominant and the
filaments become thicker and less dense than those seen in the previous
M$_T$=5 M$_{\odot}$ model. The last collapsing configurations are of three
kinds; i.e., some runs finish with a central
primary (see Figs.\ref{CPrueba60p1} and \ref{CPrueba0p1}); other runs
finish with a binary (see Figs.\ref{CPrueba70p2}, Figs.\ref{CPrueba40p2} and
\ref{CPrueba50p2}), and finally, only one run finishes with a thin and
dense filament (see Fig.\ref{CPrueba30p2}).

When $\alpha=$0.3, we have seen the formation of most of the types of binary
configurations discussed in this paper; i.e., some runs show a central
primary, and other runs show a well-defined pair of mass condensations connected by
a filament that becomes weaker as the total parent mass increases. However, the
last collapsing configurations are of the type of
a central primary for all the models.

It must be emphasized that we have seen more differences between the outcomes of
the M$_T$=1 M$_{\odot}$ model and those obtained from the
M$_T$=5 M$_{\odot}$, M$_T$=50 M$_{\odot}$ and M$_T$=400 M$_{\odot}$ models. In addition, we have seen
many similarities in the outcomes of these last two models. These observations indicate
that there is a critical total mass not much larger than one
solar mass, for which the mass condensations to be formed will be too massive and
the centrifugal force will not be able to make them start to orbit around each
other.

As we saw in Sect.\ref{sec:dis}, the binary systems formed from
massive gas structures can be the origin of wide binaries, in which a
very large separation between the mass components can be reached for
large values of $\beta$. It is interesting to mention that a high frequency of
wide binaries has recently been observed in regions having very different physical
properties; see \cite{duch}.

Lastly, we mentioned that the mass perturbation shown in Eq.~\ref{masspert} was
invented to favor fragmentation of the so-called "standard isothermal test case"
of M$_T$=1 M$_{\odot}$. For the results outlined in this paper and those reported
in other papers, we conclude that this mass perturbation scheme works quite
well for the low-mass gas structure, but needs some improvement for the
higher mass case.

In order to calculate the integral properties
of the some configurations, we took particles
by applying selection criteria based on two parameters, a minimum density and
a maximum radius, whose values
were fixed in advance. Thus, one
would expect slight differences in the reported results as they are
definition-dependent. Nevertheless, we find that there is a clear correlation between
total mass of the parent gas structure and the separation of binary fragments; we do not observe a similar
correlation for the mass of the fragments.

It must be emphasized that the initial grid of spherical concentric shells, where SPH
particles were randomly located on all the available radial surface of each shell, appears to
be a well-suited setup for numerical simulations of binary formation as we have obtained
in a natural way, that the mass of the fragments in binary configurations are not always equal
one to the other.
%%%%%%%%%%%%%%%%%%%%%%%%%%%%%%%%%%%%%%%
%%%%%%%%%%%%%%%%%%%%%%%%%%%%%%%%%%%%%%%%

\acknowledgments

The author thankfully acknowledge the computer resources, technical expertise and support provided by the
Laboratorio Nacional de Superc\'omputo del Sureste de M\' exico through the grant number O-2016/047; 
I thank ACARUS-UNISON for the use of their computing facilities in the development of this 
manuscript. 
%%%%%%%%%%%%%%%%%%%%%%%%%%%%%%%%%%%%%%%%%%%%%%%%%%%%
%%%%%%%%%%%%%%%%%%%%%%%%%%%%%%%%%%%%%%%%%%%%%%%%%%%%%%%%%%%%

\clearpage
%%%%%%%%%%%%%%%%%%%%%%%%%%%%%%%%%%%%%%%%%%%%%%%%%%%%%%%%%%%%%
\begin{deluxetable}{lcccccc}
\tablecolumns{7} \tablewidth{0pc} \tablecaption{ \label{tab:modelos}
The models and main results} \tablehead{ \colhead{Model}  &
\colhead{M$_T$/M$_{\odot}$ }   &  \colhead{a} & \colhead{$\alpha$} &
\colhead{$\beta$}
    & \colhead{outcome} & \colhead{Figure}
}
\startdata
1  & 1 & 0.1 & 0.1 & 0.1    &  connecting filament          & \ref{CPrueba60p1} l\\
2  & 1 & 0.1 & 0.1 & 0.3    &   connecting filament         & \ref{CPrueba60p1} m\\
3  & 1 & 0.1 & 0.1 & 0.48   &  connecting filament          & \ref{CPrueba60p1} r\\
4  & 1 & 0.1 & 0.1 & 0.758  &  face-to-face                 & \ref{lastcolconfM1} 1\\
\hline
5  & 1 & 0.1 & 0.2 & 0.1    &   binary                      & \ref{CPrueba60p2} l\\
6  & 1 & 0.1 & 0.2 & 0.3    &    connecting filament        & \ref{CPrueba60p2} m\\
7  & 1 & 0.1 & 0.2 & 0.48   &    central primary with arms  & \ref{CPrueba60p2} r\\
8  & 1 & 0.1 & 0.2 & 0.5344 &    central primary            & \ref{lastcolconfM1} 2\\
\hline
9  & 1 & 0.1 & 0.3 & 0.1    &  central primary with disk    & \ref{CPrueba60p3} l\\
10 & 1 & 0.1 & 0.3 & 0.3    &  central primary with arms    & \ref{CPrueba60p3} m\\
11 & 1 & 0.1 & 0.3 & 0.3983 &  central primary              & \ref{lastcolconfM1} 3\\
\hline
\hline
12 & 1 & 0.25 & 0.1 & 0.1   &  connecting thin filament     & \ref{CPrueba70p1} l\\
13 & 1 & 0.25 & 0.1 & 0.3   &  connecting filament          & \ref{CPrueba70p1} m\\
14 & 1 & 0.25 & 0.1 & 0.48  &   connecting filament         & \ref{CPrueba70p1} r\\
15 & 1 & 0.25 & 0.1 & 0.8871&  face-to-face                            & \ref{lastcolconfM1} 4\\
\hline
16 & 1 & 0.25 & 0.2 & 0.1   &    binary         & \ref{CPrueba70p2} l\\
17 & 1 & 0.25 & 0.2 & 0.3   &    binary         & \ref{CPrueba70p2} m\\
18 & 1 & 0.25 & 0.2 & 0.48  &    binary         & \ref{CPrueba70p2} r\\
19 & 1 & 0.25 & 0.2 & 0.5411&    binary         & \ref{lastcolconfM1} 5\\
\hline
20 & 1 & 0.25 & 0.3 & 0.1   &    binary         & \ref{CPrueba70p3} l\\
21 & 1 & 0.25 & 0.3 & 0.3   &    binary         & \ref{CPrueba70p3} m\\
22 & 1 & 0.25 & 0.3 & 0.3983 &   central primary& \ref{lastcolconfM1} 6\\
\hline
\hline
23 & 5 & 0.1 & 0.1 & 0.1     &  connecting filament with disk & \ref{CPrueba30p1} l\\
24 & 5 & 0.1 & 0.1 & 0.3     &    connecting filament         & \ref{CPrueba30p1} m\\
25 & 5 & 0.1 & 0.1 & 0.48    &  face-to-face                  & \ref{CPrueba30p1} r\\
26 & 5 & 0.1 & 0.1 & 0.6118  &  face-to-face                  & \ref{lastcolconfM5} 1\\
\hline
27 & 5 & 0.1 & 0.2 & 0.1     &   thin filament   & \ref{CPrueba30p2} l\\
28 & 5 & 0.1 & 0.2 & 0.3     &   thin filament   & \ref{CPrueba30p2} m\\
29 & 5 & 0.1 & 0.2 & 0.48    &   thin filament   & \ref{CPrueba30p2} r\\
\hline
30 & 5 & 0.1 & 0.3 & 0.1     &  connecting filament        & \ref{CPrueba30p3} l\\
31 & 5 & 0.1 & 0.3 & 0.3     &   thin  filament            & \ref{CPrueba30p3} m\\
32 & 5 & 0.1 & 0.3 & 0.3594  &   central primary with arms & \ref{lastcolconfM5} 3\\
\hline
\hline
33 & 5 & 0.25 & 0.1 & 0.1    &   face-to-face     & \ref{CPrueba40p1} l\\
34 & 5 & 0.25 & 0.1 & 0.3    &   face-to-face     & \ref{CPrueba40p1} m\\
35 & 5 & 0.25 & 0.1 & 0.48   &   face-to-face     & \ref{CPrueba40p1} r\\
36 & 5 & 0.25 & 0.1 & 0.6989 &   face-to-face      & \ref{lastcolconfM5} 4\\
\hline
37 & 5 & 0.25 & 0.2 & 0.1    &   connecting filament   & \ref{CPrueba40p2} l\\
38 & 5 & 0.25 & 0.2 & 0.3    &   face-to-face          & \ref{CPrueba40p2} m\\
39 & 5 & 0.25 & 0.2 & 0.48   &   face-to-face          & \ref{CPrueba40p2} r \\
40 & 5 & 0.25 & 0.2 & 0.7467 &   face-to-face          & \ref{lastcolconfM5} 5\\
\hline
41 & 5 & 0.25 & 0.3 & 0.1    &   connecting thin filament    & \ref{CPrueba40p3} l\\
42 & 5 & 0.25 & 0.3 & 0.3    &   connecting filament         & \ref{CPrueba40p3} m\\
43 & 5 & 0.25 & 0.3 & 0.3918 &   central primary with arms   & \ref{lastcolconfM5} 6\\
\hline
\hline
44 & 50 & 0.1 & 0.1 & 0.1  & connecting filament  & \ref{CPrueba80p1} l \\
45 & 50 & 0.1 & 0.1 & 0.3  & face-to-face         & \ref{CPrueba80p1} m \\
46 & 50 & 0.1 & 0.1 & 0.48 & face-to-face         & \ref{CPrueba80p1} r \\
\hline
47 & 50 & 0.1 & 0.2 & 0.1  & connecting thin filament  & \ref{CPrueba80p2} l \\
48 & 50 & 0.1 & 0.2 & 0.3  & connecting filament       & \ref{CPrueba80p2} m \\
49 & 50 & 0.1 & 0.2 & 0.48 & thin filament             & \ref{CPrueba80p2} r \\
\hline
50 & 50 & 0.1 & 0.3 & 0.1  & connecting filament & \ref{CPrueba80p3} l \\
51 & 50 & 0.1 & 0.3 & 0.3  & thin filament       & \ref{CPrueba80p3} m \\
\hline
52 & 50 & 0.25 & 0.1 & 0.1  & face-to-face       & \ref{CPrueba90p1} l \\
53 & 50 & 0.25 & 0.1 & 0.3  & face-to-face       & \ref{CPrueba90p1} m \\
54 & 50 & 0.25 & 0.1 & 0.48 & face-to-face       & \ref{CPrueba90p1}r \\
\hline
55 & 50 & 0.25 & 0.2 & 0.1  & connecting filament & \ref{CPrueba90p2} l \\
56 & 50 & 0.25 & 0.2 & 0.3  & face-to-face        & \ref{CPrueba90p2} m \\
57 & 50 & 0.25 & 0.2 & 0.48 & face-to-face        & \ref{CPrueba90p2} r \\
\hline
58 & 50 & 0.25 & 0.3 & 0.1  & connecting filament & \ref{CPrueba90p3} l \\
59 & 50 & 0.25 & 0.3 & 0.3  & face-to-face        & \ref{CPrueba90p3} m \\
\hline
\hline
60 &  400 & 0.1 & 0.1 & 0.1     &  connecting filament with disk    & \ref{CPrueba0p1} l\\
61 &  400 & 0.1 & 0.1 & 0.3     &  connecting filament              & \ref{CPrueba0p1} m\\
62 &  400 & 0.1 & 0.1 & 0.48    &  connecting filament              & \ref{CPrueba0p1} r\\
63 & 400 & 0.1 & 0.1 & 0.7      &  face-to-face                     & \ref{lastcolconfM400} 1\\
\hline
64 &  400 & 0.1 & 0.2 & 0.1     &    filament                       & \ref{CPrueba0p2} l\\
65 &  400 & 0.1 & 0.2 & 0.3     &    connecting filament            & \ref{CPrueba0p2} m\\
66 &  400 & 0.1 & 0.2 & 0.48    &    filament                       & \ref{CPrueba0p2} r\\
67 &  400  & 0.1 & 0.2 & 0.55   &     central primary with arms     & \ref{lastcolconfM400} 2 \\
\hline
68 & 400  & 0.1 & 0.3 & 0.1     & connecting thin filament          & \ref{CPrueba0p3} l\\
69 & 400  & 0.1 & 0.3 & 0.3     & filament                          & \ref{CPrueba0p3} m\\
70 & 400  & 0.1 & 0.3 & 0.38    & central primary with arms         & \ref{lastcolconfM400} 3\\
\hline
\hline
71 &  400 & 0.25 & 0.1 & 0.1    & connecting filament               & \ref{CPrueba50p1} l\\
72 &  400 & 0.25 & 0.1 & 0.3    & connecting filament with disk     & \ref{CPrueba50p1} m\\
73 &  400 & 0.25 & 0.1 & 0.48   & face-to-face                      & \ref{CPrueba50p1} r\\
74 & 400 & 0.25 & 0.1 & 0.81    & face-to-face                      & \ref{lastcolconfM400} 4\\
\hline
75 &  400 & 0.25 & 0.2 & 0.1    &  connecting filament with disk    & \ref{CPrueba50p2} l\\
76 &  400 & 0.25 & 0.2 & 0.3    &  connecting weak filament         & \ref{CPrueba50p2} m\\
77 &  400 & 0.25 & 0.2 & 0.48   &  face-to-face                     & \ref{CPrueba50p2} r\\
78 & 400  & 0.25 & 0.2 & 0.55   & face-to-face                      & \ref{lastcolconfM400} 5 \\
\hline
79 & 400  & 0.25 & 0.3 & 0.1    &   connecting filament with disk   & \ref{CPrueba50p3} l\\
80 & 400  & 0.25 & 0.3 & 0.3    &   connecting weak filament        & \ref{CPrueba50p3} m \\
81 & 400  & 0.25 & 0.3 & 0.385  & central primary with arms         & \ref{lastcolconfM400} 6 \\
\enddata
\end{deluxetable}
%%%%%%%%%%%%%%%%%%%%%%%%%%%%%%%%%%%%%%%%%%%%%%%%%%%%%%%%%%%%%
\clearpage
%%%%%%%%%%%%%%%%%%%%%%%%%%%%%%%%%%%%%%%%%%%%%%%%%%%%%%%%%%%%
\begin{deluxetable}{lccccccccc}
\tablecolumns{10} \tablewidth{0pc} \tablecaption{
\label{tab:propfrags} Physical properties of fragments of the binary
and face-to-face configurations.} \tablehead{ \colhead{Model}  &
\colhead{M$_T$/M$_{\odot}$ }   &  \colhead{r$_{max}$/R$_0$} &
\colhead{M$_{f1}$/M$_T$}    & \colhead{M$_{f2}$/M$_T$} &
\colhead{sep [au] } & $\alpha_{f1}$ & $\beta_{f1}$  & $\alpha_{f2}$
& $\beta_{f2}$  } \startdata
4    &  1   & 0.225    & 5.5e-02    & 4.9e-02   & 6877.1   &  0.11  &  0.28   &  0.12  & 0.22  \\
5    &  1   & 0.02     & 1.4e-01    & 1.32e-01  & 133.5    &  0.15  &  0.32   &  0.15  & 0.34  \\
6    &  1   & 0.04     & 8.3e-02    & 1.46e-02  & 1062.3   &  0.13  &  0.30   &  0.07  & 0.72  \\
15   &  1   & 1.0      & 3.8e-02    & 3.4e-02   & 19004.2  &  0.15  &  0.24   &  0.19  & 0.13  \\
17   &  1   & 0.04     & 1.3e-01    & 5.6e-03   & 1856.3   &  0.16  &  0.31   &  0.09  & 0.86   \\
18   &  1   & 0.1      & 1.0e-01    & 1.0e-01   & 2720.3   &  0.12  &  0.31   &  0.12  & 0.31  \\
19   &  1   & 0.02     & 5.4e-02    & 5.1e-02   & 324.5    &  0.17  &  0.30   &  0.16  & 0.30  \\
20   &  1   & 0.015    & 1.3e-01    & 1.2e-01   & 333.6    &  0.19  &  0.26   &  0.19  & 0.26  \\
21   &  1   & 0.025    & 9.8e-02    & 9.8e-02   & 333.8    &  0.17  &  0.29   &  0.17  & 0.29 \\
\hline
25   &  5   & 0.1      & 5.0e-02    & 4.99e-02  & 3206.4   &  7.7   &  0.23   &  6.40  & 0.20  \\
26   &  5   & 0.15     & 5.3e-02    & 5.25e-02  & 55284.9  &  0.28  &  0.14   &  0.27  & 0.14   \\
33   &  5   & 0.15     & 2.0e-01    & 1.6e-01   & 2002.2   &  2.24  &  0.50   &  0.50  & 0.58  \\
34   &  5   & 0.15     & 1.2e-01    & 1.2e-01   & 2848.5   &  5.18  &  0.30   &  5.10  & 0.31  \\
35   &  5   & 0.15     & 8.4e-02    & 2.2e-02   & 5005.9   &  5.90  &  0.26   &  0.08  & 0.58  \\
36   &  5   & 0.15     & 5.36e-02   & 5.29e-02  & 86903.3  &  0.14  &  0.23   &  0.07  & 0.22  \\
38   &  5   & 0.1      & 8.3e-02    & 4.96e-02  & 2502.3   &  5.6   &  0.24   &  0.12  & 0.47 \\
39   &  5   & 0.1      & 5.7e-02    & 1.04e-02  & 5005.6   & 14.4   &  0.24   &  0.16  & 0.6  \\
40   &  5   & 0.15     & 4.8e-02    & 4.8e-02   & 99934.4  &  0.40  &  0.20   &  0.17  & 0.22  \\
\hline
45   & 50   & 0.062    & 4.77e-02   & 5.09e-02  & 37642.66 &  0.05  &  0.29   & 0.06   & 0.36 \\
46   & 50   & 0.125    & 6.74e-02   & 5.35e-02  & 55592.61 &  0.10  &  0.21   & 0.06   & 0.30 \\
52   & 50   & 0.0625   & 8.19e-02   & 7.87e-02  & 33227.5  &  0.03  &  0.45   & 0.02   & 0.45 \\
53   & 50   & 0.1      & 7.69e-02   & 7.41e-02  & 53690.4  &  0.05  &  0.31   & 0.04   & 0.28 \\
54   & 50   & 0.125    & 6.91e-02   & 6.66e-02  & 77490.5  &  0.08  &  0.28   & 0.05   & 0.23 \\
56   & 50   & 0.05     & 4.63e-02   & 4.38e-02  & 40819.7  &  0.17  &  0.26   & 0.07   & 0.26 \\
57   & 50   & 0.070    & 4.35e-02   & 4.05e-02  & 61478.6  &  0.14  &  0.28   & 0.11   & 0.21 \\
59   & 50   & 0.065    & 6.46e-02   & 6.30e-02  & 21099.1  &  0.14  &  0.26   & 0.12   & 0.24 \\
\hline
63   & 400  & 0.25     & 1.28e-02   & 7.17e-05  & 308823.5 &  0.13  &  0.04   &  0.15  & 0.35 \\
73   & 400  & 0.4      & 2.05e-01   & 2.06e-01  & 259714.3 &  0.2   &  0.2    &  0.15  & 0.23  \\
74   & 400  & 0.5      & 6.8e-02    & 6.8e-02   & 627298.3 &  0.45  &  0.068  &  0.29  & 0.18  \\
77   & 400  & 0.1      & 8.78e-02   & 8.65e-02  & 205888.0 &  0.15  &  0.23   &  0.14  & 0.22 \\
78   & 400  & 0.085    & 1.2e-02    & 4.1e-03   & 124480.4 &  0.17  &  0.03   &  0.23  & 0.23 \\
\hline
\hline
\enddata
\end{deluxetable}
\clearpage
%%%%%%%%%%%%%%%%%%%%%%%%%%%%%%%%%%%%%%%%%%%%%%%%%%%%%%%%%%%%
\begin{deluxetable}{lccccc}
\tablecolumns{6} \tablewidth{0pc} \tablecaption{
\label{tab:propfrags2} Physical properties of the fragment of the
central primary configurations.} \tablehead{ \colhead{Model}  &
\colhead{M$_T$/M$_{\odot}$ }   &  \colhead{r$_{max}$/R$_0$} &
\colhead{M$_{F}$/M$_T$}  & $\alpha_F$ & $\beta_F$  } \startdata
8     &  1     &  0.03    &  4.6e-02    &  0.18   &  0.20 \\
9     &  1     &  0.01    &  1.0e-01    &  0.22   &  0.18 \\
10    &  1     &  0.02    &  7.8e-02    &  0.17   &  0.22 \\
11    &  1     &  0.035   &  5.37e-02   &  0.18   &  0.21  \\
22    &  1     &  0.015   &  1.14e-01   &  0.16   &  0.31 \\
\hline
32    &  5     &  0.015   &  2.15e-02   &  0.15   &  0.26 \\
43    &  5     &  0.02    &  2.13e-02   &  0.19   &  0.24 \\
\hline
67    & 400    &  0.008   &  9.77e-03   &  0.13   &  0.27 \\
70    & 400    &  0.005   &  1.20e-02   &  0.18   &  0.23\\
81    & 400    &  0.01    &  1.82e-02   &  0.18   &  0.27 \\
\enddata
\end{deluxetable}
\clearpage
%%%%%%%%%%%%%%%%%%%%%%%%%%%%%%%%%%%%%%%%%%%%%%%%%%%%%%%%%%%%
%%%%%%%%%%%%%%%%%%%%%%%%%%%%%%%%%%%%%%%%%%%%%%%%%%%%%%%%%%%%%
%\clearpage
%%%%%%%%%%%%%%%%%%%%%%%%%%%%%%%%%%%%%
%%%%%%%%%%%%%%%%%%%%%%%%%%%%%%%%%%%%%%%
%\clearpage
%%%%%%%%%%%%%%%%%%%%%%%%%%%%%%%%%%%%%%%%
%%%%%%%%%%%%%%%%%%%%%%%%%%%%%%%%%%%%%%
%%%%%%%%%%%%%%%%%%%%%%%%%%%%%%%%%%%%%%%
%%%%%%%%%%%%%%%%%%%%%%%%%%%%%%%%%%%%%%%%
%\clearpage
%%%%%%%%%%%%%%%%%%%%%%%%%%%%%%%%%%%%%%%%
%%%%%%%%%%%%%%%%%%%%%%%%%%%%%%%%%%%%%%
%\newpage
%\clearpage
%%%%%%%%%%%%%%%%%%%%%%%%%%%%%%%%%%%%%%%
%%%%%%%%%%%%%%%%%%%%%%%%%%%%%%%%%%%%%%
%%%%%%%%%%%%%%%%%%%%%%%%%%%%%%%%%%%%%%%
%\clearpage
%%%%%%%%%%%%%%%%%%%%%%%%%%%%%%%%%%%%%%%
%%%%%%%%%%%%%%%%%%%%%%%%%%%%%%%%%%%%%%
%%%%%%%%%%%%%%%%%%%%%%%%%%%%%%%%%%%%%%%
%\clearpage
%%%%%%%%%%%%%%%%%%%%%%%%%%%%%%%%%%%%%%%
%%%%%%%%%%%%%%%%%%%%%%%%%%%%%%%%%%%%%%%%
\clearpage
%%%%%%%%%%%%%%%%%%%%%%%%%%%%%%%%%%%%%%%
%%%%%%%%%%%%%%%%%%%%%%%%%%%%%%%%%%%%%%%%%
%\newpage
\clearpage
%%%%%%%%%%%%%%%%%%%%%%%%%%%%%%%%%%%%%%%%%
%%%%%%%%%%%%%%%%%%%%%%%%%%%%%%%%%%%%%%
%%%%%%%%%%%%%%%%%%%%%%%%%%%%%%%%%%%%%%%
\clearpage
%%%%%%%%%%%%%%%%%%%%%%%%%%%%%%%%%%%%%%%
%%%%%%%%%%%%%%%%%%%%%%%%%%%%%%%%%%%%%%%%
%%%%%%%%%%%%%%%%%%%%%%%%%%%%%%%%%%%%%%%%%
%%%%%%%%%%%%%%%%%%%%%%%%%%%%%%%%%%%%%%%%%%
%%%%%%%%%%%%%%%%%%%%%%%%%%%%%%%%%%%%%%%%%
%%%%%%%%%%%%%%%%%%%%%%%%%%%%%%%%%%%%%%%%%
%\newpage
%\clearpage
%%%%%%%%%%%%%%%%%%%%%%%%%%%%%%%%%%%%%%%%
%%%%%%%%%%%%%%%%%%%%%%%%%%%%%%%%%%%%%%%
%%%%%%%%%%%%%%%%%%%%%%%%%%%%%%%%%%%%%%%%
%\clearpage
%%%%%%%%%%%%%%%%%%%%%%%%%%%%%%%%%%%%%%%%%
%%%%%%%%%%%%%%%%%%%%%%%%%%%%%%%%%%%%%%%%%%
%%%%%%%%%%%%%%%%%%%%%%%%%%%%%%%%%%%%%%
%%%%%%%%%%%%%%%%%%%%%%%%%%%%%%%%%%%%%%%%%%
%\clearpage
%%%%%%%%%%%%%%%%%%%%%%%%%%%%%%%%%%%%%
%%%%%%%%%%%%%%%%%%%%%%%%%%%%%%%%%%%%%%%
%\clearpage
%%%%%%%%%%%%%%%%%%%%%%%%%%%%%%%%%%%%%%%%
%%%%%%%%%%%%%%%%%%%%%%%%%%%%%%%%%%%%%%%%
%%%%%%%%%%%%%%%%%%%%%%%%%%%%%%%%%%%%%%%%
%%%%%%%%%%%%%%%%%%%%%%%%%%%%%%%%%%%%%%%%
%%%%%%%%%%%%%%%%%%%%%%%%%%%%%%%%%%%%%%%%
%%%%%%%%%%%%%%%%%%%%%%%%%%%%%%%%%%%%%%%%
%%%%%%%%%%%%%%%%%%%%%%%%%%%%%%%%%%%%%%%%
%%%%%%%%%%%%%%%%%%%%%%%%%%%%%%%%%%%%%%%%
%%%%%%%%%%%%%%%%%%%%%%%%%%%%%%%%%%%%%%%%
%%%%%%%%%%%%%%%%%%%%%%%%%%%%%%%%%%%%%%%%
%%%%%%%%%%%%%%%%%%%%%%%%%%%%%%%%%%%%%%%%
%%%%%%%%%%%%%%%%%%%%%%%%%%%%%%%%%%%%%%%%
%%%%%%%%%%%%%%%%%%%%%%%%%%%%%%%%%%%%%%%%
%%%%%%%%%%%%%%%%%%%%%%%%%%%%%%%%%%%%%%%%
%%%%%%%%%%%%%%%%%%%%%%%%%%%%%%%%%%%%%%%%
%%%%%%%%%%%%%%%%%%%%%%%%%%%%%%%%%%%%%%%%
%%%%%%%%%%%%%%%%%%%%%%%%%%%%%%%%%%%%%%%
%%%%%%%%%%%%%%%%%%%%%%%%%%%%%%%%%%%%%%%%%%%
%\clearpage
%%%%%%%%%%%%%%%%%%%%%%%%%%%%%%%%%%%%%%%%%
%%%%%%%%%%%%%%%%%%%%%%%%%%%%%%%%%%%%%%%%%%
\begin{figure}
\begin{center}
\begin{tabular}{ccc}
\includegraphics[width=2.2in]{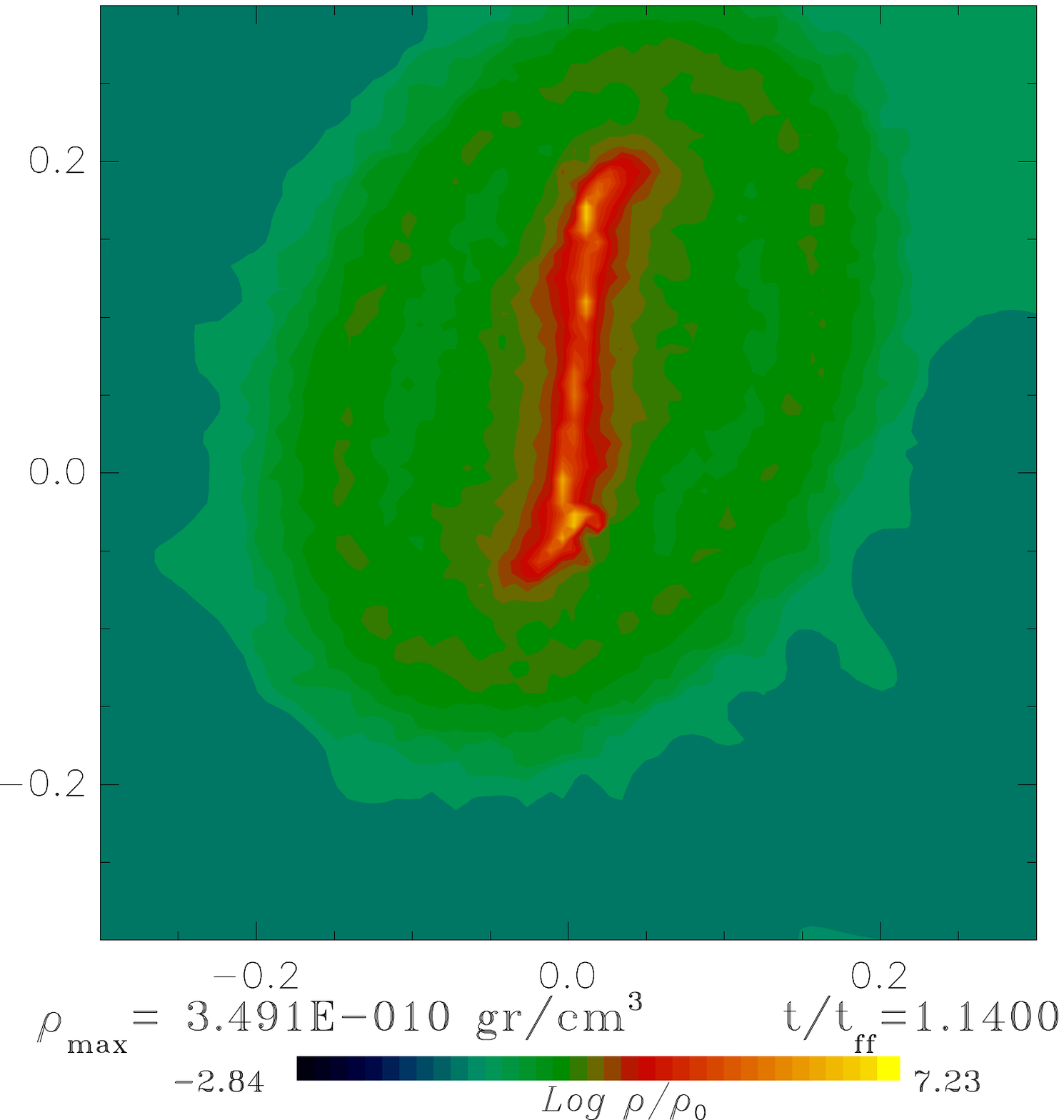} & \includegraphics[width=2.2in]{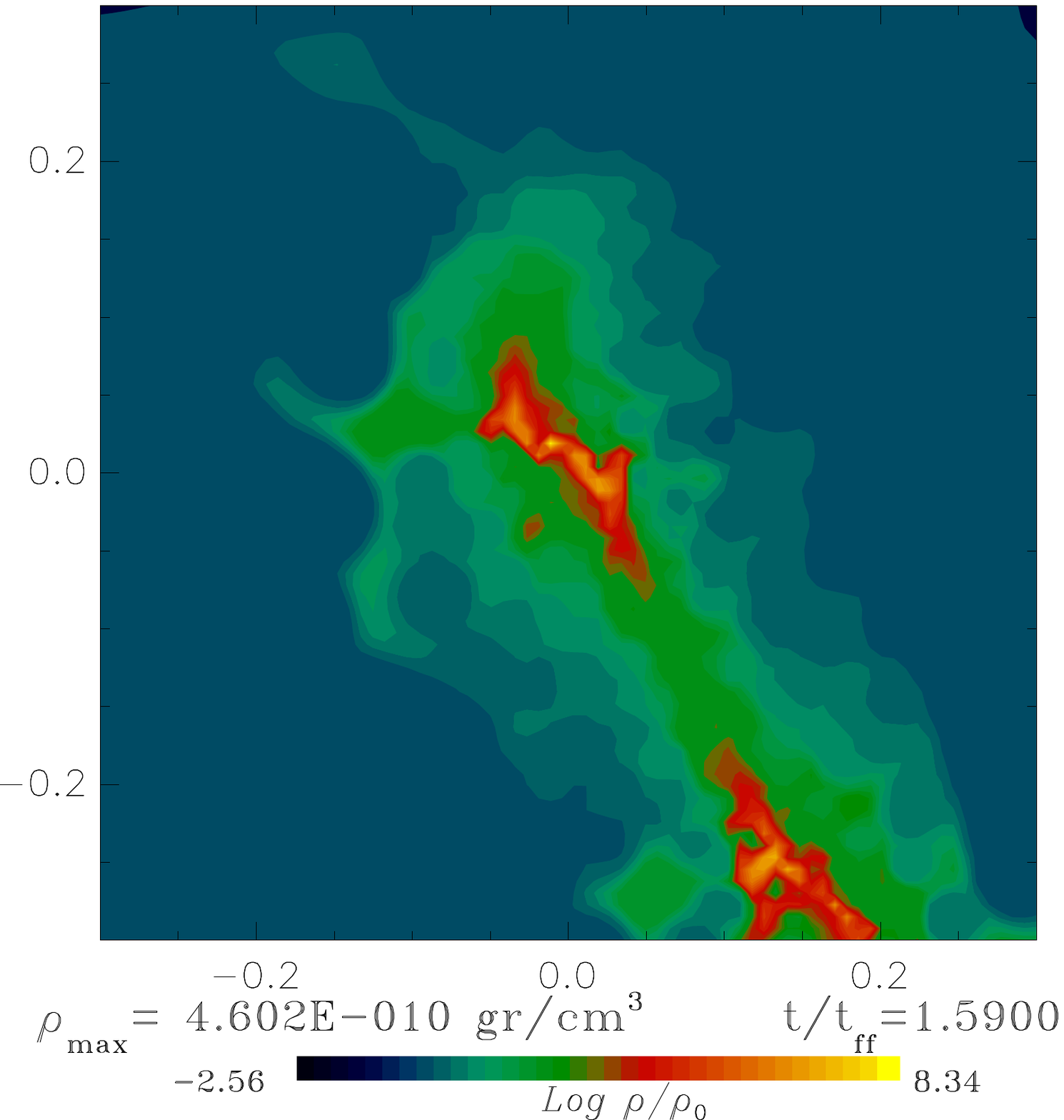} &
\includegraphics[width=2.2in]{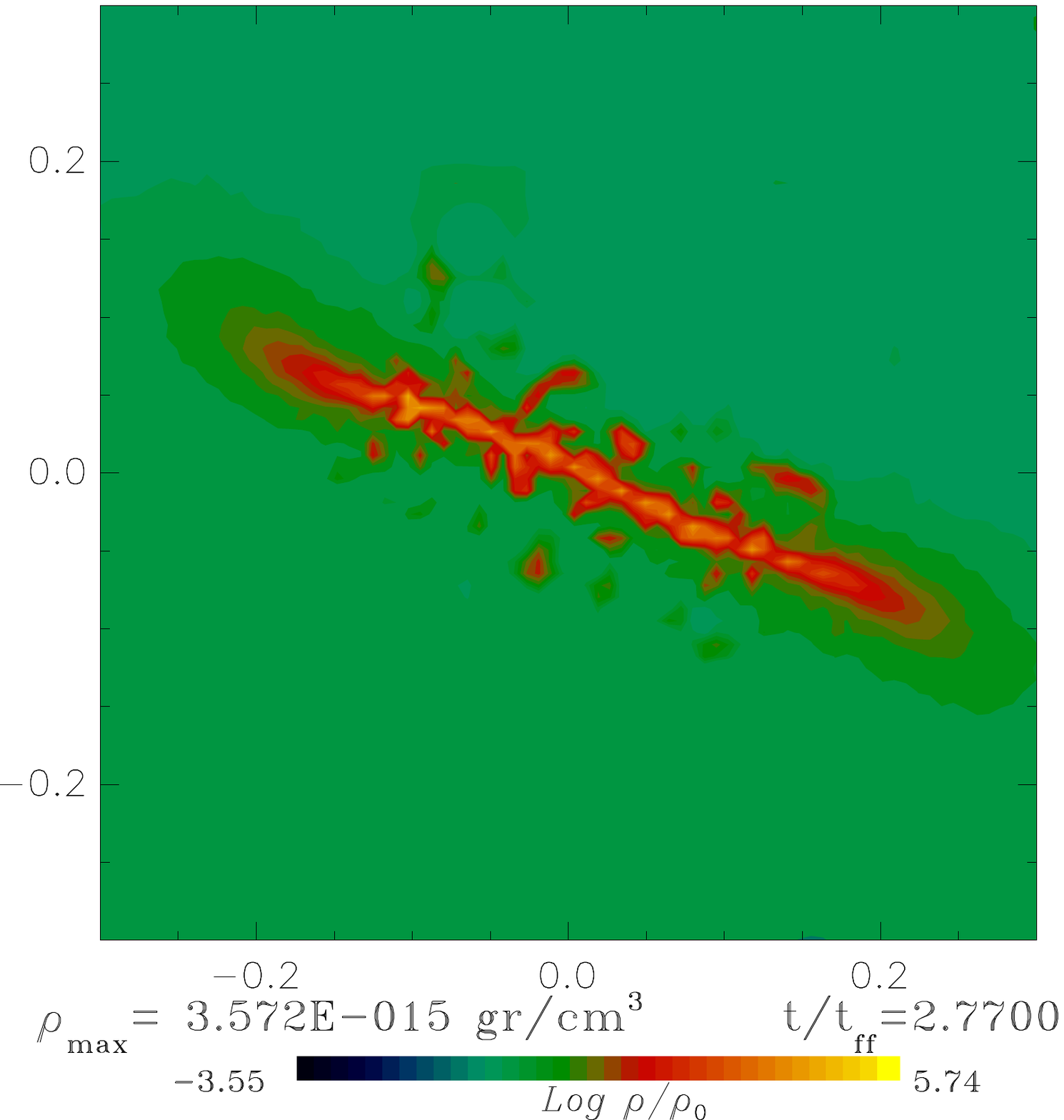} \\
\end{tabular}
\caption{\label{CPrueba0p2} Clump models with M$_T$= 400 M$_{\odot}$, $a=$0.1 and $\alpha=$0.2;
the corresponding $\beta$ are
(left) 0.1
(middle) 0.3
(right) 0.48.}
\end{center}
\end{figure}
%%%%%%%%%%%%%%%%%%%%%%%%%%%%%%%%%%%%%%%%%%%%
%%%%%%%%%%%%%%%%%%%%%%%%%%%%%%%%%%%%%%%%%%%%%%
%\clearpage
%%%%%%%%%%%%%%%%%%%%%%%%%%%%%%%%%%%%%%%%%%
%%%%%%%%%%%%%%%%%%%%%%%%%%%%%%%%%%%%%%%%%
%\newpage
%\clearpage
%%%%%%%%%%%%%%%%%%%%%%%%%%%%%%%%%%%%%%%%%
%%%%%%%%%%%%%%%%%%%%%%%%%%%%%%%%%%%%%%%%%%%
%\newpage
%\clearpage
%%%%%%%%%%%%%%%%%%%%%%%%%%%%%%%%%%%%%%%%%%%
%%%%%%%%%%%%%%%%%%%%%%%%%%%%%%%%%%%%%%%%%
%\clearpage
%%%%%%%%%%%%%%%%%%%%%%%%%%%%%%%%%%%%%%%%%
%%%%%%%%%%%%%%%%%%%%%%%%%%%%%%%%%%%%%%
\clearpage
%%%%%%%%%%%%%%%%%%%%%%%%%%%%%%%%%%%%%%%%
%%%%%%%%%%%%%%%%%%%%%%%%%%%%%%%%%%%%%%%%%

%%%%%%%%%%%%%%%%%%%%%%%%%%%%%%%%%%%%%%%%%
%%%%%%%%%%%%%%%%%%%%%%%%%%%%%%%%%%%%%%
%%%%%%%%%%%%%%%%%%%%%%%%%%%%%%%%%%%%%%
%\newpage
\clearpage
%%%%%%%%%%%%%%%%%%%%%%%%%%%%%%%%%%%%%%%%%%
%%%%%%%%%%%%%%%%%%%%%%%%%%%%%%%%%%%%%%%%%%
%%%%%%%%%%%%%%%%%%%%%%%%%%%%%%%%%%%%
%%%%%%%%%%%%%%%%%%%%%%%%%%%%%%%%%%%%%%%%
%\clearpage
%%%%%%%%%%%%%%%%%%%%%%%%%%%%%%%%%%%%%%%%
%%%%%%%%%%%%%%%%%%%%%%%%%%%%%%%%%%%%%%%%%%
%%%%%%%%%%%%%%%%%%%%%%%%%%%%%%%%%%%
%\clearpage
%%%%%%%%%%%%%%%%%%%%%%%%%%%%%%%%%%%%%%%%%%
%%%%%%%%%%%%%%%%%%%%%%%%%%%%%%%%%%%%
%%%%%%%%%%%%%%%%%%%%%%%%%%%%%%%%%%%%%%%%
%\clearpage
%%%%%%%%%%%%%%%%%%%%%%%%%%%%%%%%%%%%%%
%%%%%%%%%%%%%%%%%%%%%%%%%%%%%%%%%%%%%%%%
%%%%%%%%%%%%%%%%%%%%%%%%%%%%%%%%%%%%%%%%%
%%%%%%%%%%%%%%%%%%%%%%%%%%%%%%%%%%%%%%%%%%
%%%%%%%%%%%%%%%%%%%%%%%%%%%%%%%%%%%%%%%%%%%%%
%%%%%%%%%%%%%%%%%%%%%%%%%%%%%%%%%%%%%%%%%%%%%
\end{document}